\documentclass{JHEP3}

\usepackage{amsfonts,amsmath,amssymb}
\usepackage{epsfig,multicol,bbm}
\usepackage{float}
\usepackage{subfigure}
\usepackage{caption}
\usepackage{cite}
\usepackage{tabularx}
\newcommand\fverb{\setbox\fverbbox=\hbox\bgroup\verb}
\newcommand\fverbdo{\egroup\medskip\noindent
			\fbox{\unhbox\fverbbox}\ }
\newcommand\fverbit{\egroup\item[\fbox{\unhbox\fverbbox}]}
\newbox\fverbbox

\newcommand{\defeq}{\stackrel{\textup{\tiny def}}{=}}
\newcommand{\calN}{\mathcal{N}}

\newcommand{\calM}{\mathcal{M}}

\newcommand{\calR}{\mathcal{R}}
\newcommand{\ft}{\mathfrak{t}}
\newcommand{\fq}{\mathfrak{q}}

\def\be{\begin{equation}}
\def\ee{\end{equation}}
\def\bea{\begin{eqnarray}}
\def\eea{\end{eqnarray}}

\title{Non-Lagrangian Theories from Brane Junctions}

\preprint{DESY 13 - 176
\\
HU-Mathematik - 17 - 13
\\
 HU-EP-13/50
\\
KIAS-P13058
\\
RIKEN-MP-78}

\author{Ling Bao$^{a}$\footnote{Email: lingbao.work@gmail.com}
$\,$, Vladimir Mitev$^{b}$\footnote{Email: mitev@math.hu-berlin.de}
$\,$, Elli Pomoni$^{c}$\footnote{Email: elli.pomoni@desy.de}
$\,$, Masato Taki$^{d}$\footnote{Email: taki@riken.jp}
$\,$  and Futoshi Yagi$^{e,f}$\footnote{Email: fyagi@kias.re.kr}
\\
\\
\it $^a$ Chalmers University of Technology, SE-41296 G\"oteborg, Sweden
\\
\it $^b$ Institut f\"ur Mathematik und Institut f\"ur Physik, Humboldt-Universit\"at zu Berlin, IRIS Haus, Zum Gro{\ss}en Windkanal 6,  12489 Berlin, Germany
\\
\it $^c$ DESY Theory Group, Notkestra{\ss}e 85, 22607 Hamburg, Germany
\\
\it $^d$ Mathematical Physics Lab., RIKEN Nishina Center, Saitama 351-0198, Japan
\\
\it $^e$ International School of Advanced Studies (SISSA) and INFN, Sezione di Trieste,
via Bonomea 265, 34136 Trieste, Italy 
\\
\it $^f$ Korea Institute for Advanced Study (KIAS)
85 Hoegiro Dongdaemun-gu, 130-722, Seoul, Korea
}

\abstract{

\bigskip

In this article we use 5-brane junctions to study the 5D $T_N$ SCFTs corresponding to the 5D $\calN=1$ uplift of the 4D $\calN=2$ strongly coupled gauge theories, which are obtained by compactifying $N$ M5 branes on a sphere with three full punctures. Even though these theories have no Lagrangian description, by using the 5-brane junctions proposed by Benini, Benvenuti and Tachikawa, we are able to derive their Seiberg-Witten curves and Nekrasov partition functions. We cross-check our results with the 5D superconformal index proposed by Kim, Kim and Lee. Through the AGTW correspondence, we discuss the relations between 5D superconformal indices and $n$-point functions of the $q$-deformed $W_N$ Toda theories.
}

\keywords{Gauge theory, Topological strings, CFT}

\begin{document}


\section{Introduction}

In the seminal article~\cite{Gaiotto:2009we} Gaiotto argued that a large class, called class $\mathcal{S}$, of $\mathcal{N}=2$ superconformal field theories (SCFT) in four dimensions (4D) can be obtained by a twisted compactification of the 6D $(2,0)$ SCFT on a Riemann surface of genus $g$ with $n$ punctures.  The building blocks of the class $\mathcal{S}$ theories are tubes and pairs of pants that correspond to gauge groups and matter multiplets, respectively. Subsequently, the authors of \cite{Alday:2009aq, Wyllard:2009hg} proposed a relation between the partition functions of the $\mathcal{N}=2$ $SU(N)$ gauge theories and the correlation functions of the 2D $A_{N-1}$ Toda CFTs. If one can compute all the 2- and 3-point functions in a CFT, it would in principle lead to complete understanding of the $n$-point functions. Analogously, if we are able to find all the partition functions corresponding to the 2- and 3-point functions\footnote{As we will see below, the $T_N$ partition functions corresponding to the 3-point functions of the $W_N$ Toda theory have empty Young diagrams at their external legs \eqref{eq:TNjunctionpartitionfunction}. In order to obtain the $n$-point functions from the Nekrasov partition functions that  correspond to the 3-point functions, one has to use the partition functions \eqref{eq:TNjunctionpartitionfunction} but with non-empty Young diagrams for their external legs.}, we will be able to write down all the partition functions of theories that admit a pair of pants decomposition.

It is important to note that there is a fundamental difference between the $SU(2)$ and the $SU(N)$, $N>2$, cases. For the $SU(2)$ quiver gauge theories from \cite{Alday:2009aq} that are related to 2D Liouville CFT, there is only one type of puncture on the Riemann surface and hence has only one class of 2D 3-point functions to be calculated. On the other hand, the $SU(N)$ case with $N>2$ has more than one kind of punctures, containing $U(1)$ punctures and full $SU(N)$ ones \cite{Gaiotto:2009we}. So far, the case with three full $SU(N)$ punctures $T_N$ (also referred to as Triskelion \cite{Benini:2009mz}) has remained elusive, since neither the $T_N$ Nekrasov partition functions nor the Toda three-point correlators are known. The situation is further aggravated by the fact that the corresponding 4D theories do not posses a Lagrangian description. Even though there is no known Lagrangian description of the 4D $T_N$ theories, we are able to write down the partition functions for their 5D uplift \cite{Benini:2009gi} using topological strings on the dual geometry of the 5-brane junctions.

The simplest non-trivial example is the 4D $T_3$ theory that was found by Argyres and Seiberg in~\cite{Argyres:2007cn}. It was realized there as the strong coupling limit of an $SU(3)$ gauge theory with six flavors in the fundamental representation. This theory is known to enjoy global $E_6$ symmetry. The main focus of our study is the 5D uplift of this $T_3$ theory, which is quite special since it can be seen from a different point of view. Seiberg argued in \cite{Seiberg:1996bd} that the 5D $SU(2)$ gauge theory with $N_f$ fundamental flavors has a UV fixed point with $E_{N_f+1}$ symmetry, where the $SO(N_f)$ flavor and $U(1)$ instanton symmetries enhance to an $E_{N_f+1}$ global symmetry\footnote{These theories also appear in the context of the E-string theory, see \cite{Ganor:1996mu, Seiberg:1996vs, Witten:1996qb, Morrison:1996na, Morrison:1996pp, Ganor:1996gu, Klemm:1996hh, Ganor:1996xd, Ganor:1996pc, Minahan:1998vr}.} \cite{Seiberg:1996bd}.
The above mentioned 5D $T_3$ theory is in fact identical to this $E_6$ CFT \cite{Benini:2009gi}.

Powerful tools became available when one realizes gauge theories in string theory. One was suggested by Hanany and Witten in \cite{Hanany:1996ie}, who constructed gauge theories living on a system of NS5 and D$p$ branes in type IIA/B string theories. This approach provided a geometric way to realize and compute the Seiberg-Witten curves for 4D gauge theories by uplifting to M-theory \cite{Witten:1997sc}. For 5D gauge theories, the brane configuration is a type IIB $(p,q)$ 5-brane system, which is referred to as the \emph{web diagram} \cite{Aharony:1997ju, Aharony:1997bh} from which the SW curves are obtained \cite{Kol:1997fv,Aharony:1997bh,Brandhuber:1997ua,Bao:2011rc}. Another way to realize $\mathcal{N}=2$ gauge theories, known as geometric engineering \cite{Katz:1996fh,Katz:1997eq}, is to compactify type II string/M-theory on Calabi-Yau threefolds. This route has the advantage of allowing one to obtain the Nekrasov partition functions of the gauge theories by computing the partition functions of topological strings living on these backgrounds. Magically, the dual to the Calabi-Yau toric diagram turns out to be exactly equal to the web diagram used in the Hanany-Witten approach \cite{Leung:1997tw, Gorsky:1997mw}.

Flavors can be added in the Hanany-Witten setup by introducing D7-branes \cite{Witten:1997sc}. In particular, the $E_n$ SCFTs are obtained by adding $(n+3)$ 7-branes inside a 5-brane loop \cite{DeWolfe:1999hj}. In this configuration, the flavor symmetry becomes manifest, since it comes from the gauge symmetry of the 7-branes \cite{Gaberdiel:1997ud}. The web diagram with 7-branes is equivalent to local del Pezzo compactifications of M-theory \cite{Douglas:1996xp}. The del Pezzo surfaces are roughly speaking multi-point blow-ups of $\mathbb{CP}^2$, where the blow-up procedure corresponds to the inclusion of the 7-branes inside the 5-brane loop. The flavor symmetry is also manifest in the del Pezzo construction, and it corresponds to the rotations of the blow-up points. The SW curves of the $E_n$ theories are derived based on the del Pezzo compactification \cite{Minahan:1997ch,Eguchi:2002fc,Eguchi:2002nx}\footnote{The 4D SW curve was found in \cite{Minahan:1996fg,Minahan:1996cj}.}. The enhancement of the flavor symmetry then becomes manifest on the level of the SW curves. The partition function of the del Pezzo compactification was studied in \cite{Minahan:1997ct,Diaconescu:2005mv,Diaconescu:2005tr,Konishi:2006ya,Mohri:2001zz,Sakai:2011xg,Sakai:2012ik,Huang:2013yta}.

For $E_n$ with $n \le 3$, we can pull all the D7-branes outside of the 5-brane loop by employing the rules from \cite{Gaberdiel:1998mv,DeWolfe:1998zf,Iqbal:1998xb,DeWolfe:1998bi,DeWolfe:1998eu,DeWolfe:1998pr} without changing the theory\footnote{It is known that SW curve does not change by this procedure also for $n=4,5$ \cite{Kol:1997fv,Aharony:1997bh,Brandhuber:1997ua,Bao:2011rc}.}. Since the five dimensional theory is insensitive to the size of the regularized external 5-brane, we can move the 7-branes to infinity. It is easier to obtain the SW curve in this setup, by uplifting to M-theory following \cite{Witten:1997sc}. Moreover, the computation of the Nekrasov partition function is also fairly simple, since the corresponding Calabi-Yau manifold is known to be toric. We can thus use the topological vertex method, which is much easier than computing the topological string partition function for non-toric Calabi-Yau geometries.

However, a subtlety with this strategy is that this brane construction is problematic if we try to move all the 7-branes to infinity for $E_n$ with $n \ge 4$ \cite{Aharony:1997ju}. This is because parallel\footnote{For $n \ge 7$, crossing or jumping also appear \cite{Benini:2009gi}, but we will not discuss these issues in this paper.} external 5-branes appear in the naive brane web for the realization of higher flavor symmetries. They include extra degrees of freedom due to the fact that 1-branes can propagate along the parallel 5-branes. No desirable theory can arise from such an ``untidy" brane system. Moreover, the flavor symmetry is not manifest in the web system. The corresponding toric M-theory compactification also includes these extra degrees of freedom due to M2-branes propagating along the flat direction of the Calabi-Yau. Nevertheless, in this paper we demonstrate that for all the flavor cases up to $N_f=5$, the above mentioned computation methods for the SW curve and the Nekrasov partition function are still applicable. The $N_f=5$ case corresponds precisely to the $T_3$ 5-brane multi-junction \cite{Benini:2009gi}.

In this paper, we first compute the SW curve of the $T_N$ junction using the method of \cite{Witten:1997sc}. We demonstrate that the previously derived SW curve can be reproduced from this brane setup and we show the way the $E_6$ symmetry is realized, even though in this setup the full global symmetry is not manifest. We also compute the Nekrasov partition functions of the $T_N$ junctions as refined topological string partition functions \cite{Iqbal:2007ii,Taki:2007dh}.
At this point we make use of the quite recent  conjecture of Iqbal and Vafa \cite{Iqbal:2012xm}, that says that the 5D superconformal index,  which is the partition function on $S^4\times S^1$, can be obtained from the 5D Nekrasov partition function and thus from the topological string partition function
\be
\mathcal{I}^{5D}= \int d a \, |Z_{\textrm{Nek}}^{5D} (a)|^2\propto \int d a \, |Z_{\textrm{top}} (a)|^2.
\ee
This provides a way to test our results against the 5D superconformal index computed via localization on $S^4\times S^1$ by Kim, Kim and Lee in  \cite{Kim:2012gu}. The $E_6$ superconformal index is obtained from the $T_3$ Nekrasov partition function\footnote{As discussed in \cite{Iqbal:2012xm}, this is a property of odd dimensions. In even dimensions, the superconformal index cannot possibly be derived from the Nekrasov partition function, see for a 4D example \cite{Gadde:2009dj}.} by using the idea presented in \cite{Iqbal:2012xm} and we find that the results coincide with those of \cite{Kim:2012gu}. When parallel external 5-brane legs appear in the toric web diagram, the corresponding topological string partition functions contain extra degrees of freedom. In contrast to the massive spectrum in 5D which forms a representation of the Wigner little group $SU(2)\times SU(2)$, referred to as the \textit{full spin content representation}, these extra states do not transform as a correct representation under the Poincar\'e symmetry. Therefore, we call them \textit{non-full spin content} contributions. Based on the discussion in \cite{Iqbal:2012xm, Gopakumar:1998jq}\footnote{The non-compact modes of the moduli of the wrapped M2-branes do not form a correct representation of the Poincar\'e symmetry.}, we interpret this part as the contribution to the extra degrees of freedom appearing from the parallel 5-branes explained above. It should therefore be removed. To obtain the superconformal index from the topological string partition function, we have to eliminate all the non-full spin content from the partition function. Schematically, the partition function can be expressed as a sum of Young diagrams assigned to the product of strip geometries as
\begin{equation}
 Z_{T_N}=\frac{1}{Z_{\text{non-full spin}}}\sum_{\boldsymbol{Y}}\prod_{i=1}^N Z^{\text{strip}}_{i}(\boldsymbol{Y})  \, .
\end{equation}
The factor $Z_{\text{non-full spin}}$ is the BPS spectrum which does not form a representation of the Poincar\'e symmetry, and $Z^{\text{strip}}$ is the partition function of the strip geometry. A factor similar to $Z_{\text{non-full spin}}$ was also responsible for a mismatch observed in \cite{Bergman:2013ala}.

Finally, the 5D version of the AGTW relation \cite{Awata:2009ur,Awata:2010yy,Mironov:2011dk,Itoyama:2013mca}, which suggests that the 5D Nekrasov partition functions are equal to the conformal block of $q$-deformed $W_N$ Toda, implies the following relation between the superconformal index and the correlation functions of the corresponding $q$-deformed Toda field theory \cite{Nieri:2013yra}:
\begin{align}
\mathcal{I}^{5D}(x,y)=\int [da] \Big|
Z_{\textrm{Nek}}^{\textrm{5D}}(a,m,\beta,\epsilon_{1,2})\Big|^2
\propto  \langle
V_{\boldsymbol{\alpha}_1}(z_1)\cdots V_{\boldsymbol{\alpha}_n}(z_n)
\rangle_{q\textrm{-Toda}}.
\end{align}
This is an important entry in the dictionary of the 5D/2D AGTW correspondence. The partition functions of the $T_N$ brane junctions predict, up to an overall coefficient, the corresponding DOZZ formula for the three-point functions. 

The organization of the paper is as follows: In section~\ref{sec:7brane}, we review the procedure for obtaining the 5-brane junction of \cite{Benini:2009gi}. This is done by pulling the 7-branes out of the 5-brane loop. In section~\ref{sec:seibergwitten}, we compute the SW curve for the 5-brane junction. Although the full $E_6$ global symmetry is not immediately manifest, we discuss how it is realized. We also analyze the 4D limit of this 5D SW curve, and show that it reproduces the known 4D curve. In section~\ref{sec:topologicalstrings}, we compute the 5D superconformal index of $SU(2)$ gauge theories with $N_f =\{0,\dots,5\}$ from the topological string partition functions and show that they coincides with the results in \cite{Kim:2012gu}. Furthermore, we discuss the generalization to $T_N$. In section~\ref{sec:correlationfunctions}, we discuss applications of our result in the context of the AGTW relation. We also discuss how the Nekrasov partition function of the 5D $T_N$ CFT is related to the three-point functions of the $q$-deformed $W_N$ Toda theory and how in general its $n$-point functions can be obtained from the 5D superconformal index. Lastly, section~\ref{sec:conclusions} is devoted to conclusions and discussions.

We would like to remark that there is a certain overlap between the results of section~\ref{sec:topologicalstrings} and those of the article \cite{Hayashi:2013qwa}  that appeared on the same day as ours.

\section{7-branes and the toric web diagram of $T_3$}
\label{sec:7brane}

In this section we review the brane construction of 5D gauge theories. In particular, we see that the 5D $SU(2)$ gauge theory with $N_f=5$ flavors is obtained\footnote{F.Y.~thanks Dan Xie for helpful discussions on this point.} from the 5D $T_3$ multi-junction \cite{Benini:2009gi}.

Many 5D gauge theories are realized as the world volume theories of the type IIB $(p,q)$ 5-brane webs \cite{Aharony:1997bh},
which is a 5D uplift of the well-known Hanany-Witten brane construction of four dimensional $\mathcal{N}=2$ gauge theories \cite{Hanany:1996ie}. This web construction can be generalized by introducing 7-branes \cite{DeWolfe:1999hj}
without further breaking any supersymmetry. Strings can stretch between the D5 branes and the D7 branes rendering quark hypermultiplets in the fundamental representation of the color group. Thus, adding D7-branes leads to an extra hypermultiplet in the fundamental representation.

\begin{figure}[h!]
\centering
\includegraphics[height=3.2cm]{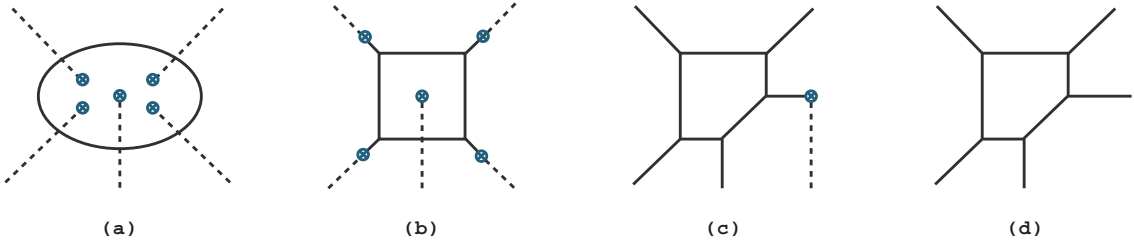}
\caption{\it  In this figure, the possible ways of constructing $SU(2)$ gauge theory with $N_f=1$ are depicted. The blue circles denote the 7-brane, the solid lines the $(p,q)$ 5-branes and the dashed lines the branch cuts that start from the 7-branes and extend to infinity. In part (a) we have a 5-brane loop with five 7-branes in it. In part (b) we pulled four of the 7-branes out of the 5-brane loop. In part (c) we also pull out the fifth 7-brane. In part (d) we pulled the 7-branes to infinity and are left with a 5-brane web.}
\label{fig:7rule}
\end{figure}

The 7-branes allow the construction of $SU(2)$ gauge theories with up to $N_f=8$ flavors \cite{DeWolfe:1999hj}. An elementary example of how it works is depicted in figure~\ref{fig:7rule}. Consider the $(p,q)$ web configuration of the pure $SU(2)$ gauge theory and add a 7-brane at the center of the web-toric diagram. The 5-branes fill the $01234$ directions and form webs in the $56$ plane. The 7-branes fill the $01234789$ directions and are point like in the $56$ plane. Adding multiple 7-branes inside the 5-brane loop increases the number of the hypermultiplets, and the gauge symmetry of the 7-branes becomes the flavor symmetry of the 5D theory under consideration, see \cite{Gaberdiel:1997ud, DeWolfe:1999hj}. The 5-brane web on a generic point of the Coulomb branch leads to a 5D theory at intermediate energy scales. The fixed point theory is realized nicely as a collapsed limit of the 5-brane loop \cite{Aharony:1997bh}. We can thus understand the $E_n$ SCFTs from the perspective of the type IIB brane configuration.

The 7-branes are filling (7+1) dimensions while being pointlike in the remaining two ($56$ plane). They are the magnetic sources of the dilaton-axion scalar $\tau = \chi + i e^{-\phi}$. This complex scalar experiences a monodromy around each 7-brane, which is accounted for by introducing a branch cut associated with each 7-brane \cite{Gaberdiel:1998mv,DeWolfe:1998zf,Iqbal:1998xb,DeWolfe:1998bi,DeWolfe:1998eu,DeWolfe:1998pr}. The branch cut that starts from a 7-brane and goes to infinity is depicted in figure~\ref{fig:7rule} by a dashed line. A generic 7-brane setup is specified by listing the 7-branes in the order in which their branch cuts are crossed when encircling them in a counterclockwise direction.

A 7-brane is labeled by two co-prime integers\footnote{This up to a sign, since a $(p,q)$ 7-brane is the same object as a $(-p,-q)$ 7-brane.} $(p,q)$. The dilaton-axion scalar $\tau$ transforms with the $SL(2,\mathbb{Z})$ monodromy matrix $K_{(p,q)}$ when crossing the cut of a $(p,q)$-brane. Combining the two charges as a vector $(p\  q)$, the monodromy matrix $K_{(p,q)}$ reads
\be
\label{eq:rule}
K_{(p,q)} = 1 + \left( \begin{array}{c}
     p \\
    q \\
   \end{array} \right)\left( \begin{array}{cc}
     p &
    q
   \end{array} \right) S =  \left( \begin{array}{cc}  1+pq & -p^2
\\
q^2 & 1-pq \\
 \end{array} \right)
\,,
\ee
where $S \defeq \left( \begin{array}{cc}  0 &-1 \\ 1 & 0\\
 \end{array} \right)
$.

In a generalized $(p,q)$-web containing both 5- and 7-branes, when a $(p,q)$ 5- or 7- brane crosses the branch cut of a $(P,Q)$ 7-brane it changes into a $(p',q')$ 5- or 7-brane according to the rule
\be
\label{eq:transformpq}
\left(
\begin{array}{c}
p' \\
q'
\end{array}
\right)
= \left(
\begin{array}{cc}
1 + PQ & - P^2 \\
Q^2 & 1 - PQ
\end{array}
\right)
\left(
\begin{array}{c}
p \\
q
\end{array}
\right)
\, .
\ee
In the example depicted in figure~\ref{fig:7rule} (b), the branch cut attached to an $(-1,0)$ 7-brane generates a monodromy around the same brane given by the matrix in \eqref{eq:transformpq}, which here reads
\be
K_{(-1,0)} = \left(
\begin{array}{cc}
1 & 1 \\
0 & 1
\end{array}
\right).
\ee
At the same time, the 5-branes that are swapped by the branch cut change their charges according to
\begin{equation}
(0,1) \to (1,1) , \quad (-1,1) \to (0,1), \quad (-1,0) \to (-1,0)  .
\end{equation}
Due to the $(p,q)$ 5-brane charge conservation, when an NS5-brane is swapped by a branch cut, a new D5-brane attached to the D7-brane is generated.

\begin{figure}[h!]
\centering
\includegraphics[height=4cm]{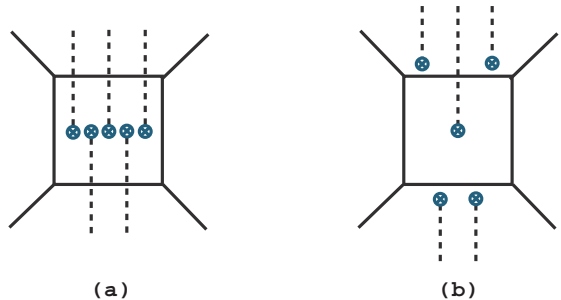}
\caption{\it To obtain the $SU(2)$ gauge theory with $N_f=5$, we begin with the pure $SU(2)$ web-toric diagram with five D7 branes inserted. In the first part of the figure all the flavors have the same mass, and the symmetry enhancement is manifest. In the second part of the figure, a mass deformation is performed by moving the 7-branes vertically.}
\label{fig:su2fiveflavor}
\end{figure}

\begin{figure}[h!]
\centering
\includegraphics[height=4cm]{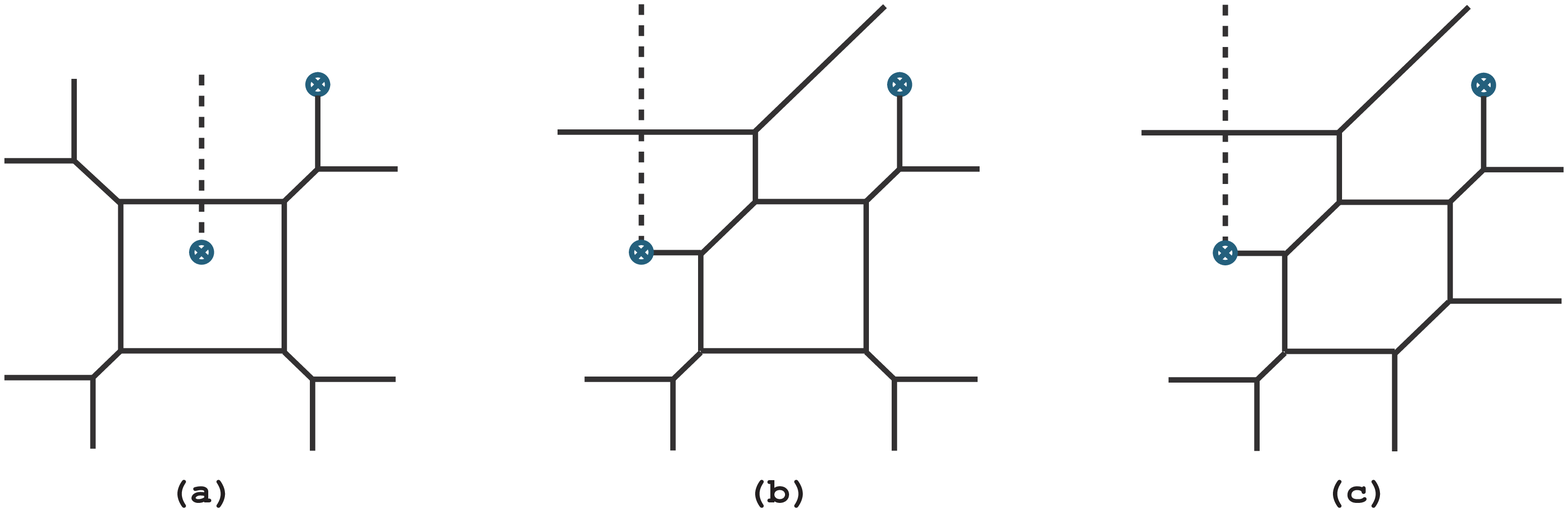}
\caption{\it
Then, from the mass deformed configuration, we  pull the 7-branes outside of the NS5-branes following the Hanany Witten effect. Lastly, after a flop on the lower right leg of the diagram we arrive at the $E_6$ web-toric diagram suggested by Benini, Benvenuti and Tachikawa.}
\label{fig:pullandflop}
\end{figure}

\begin{figure}[ht]
\centering
\includegraphics[height=5cm]{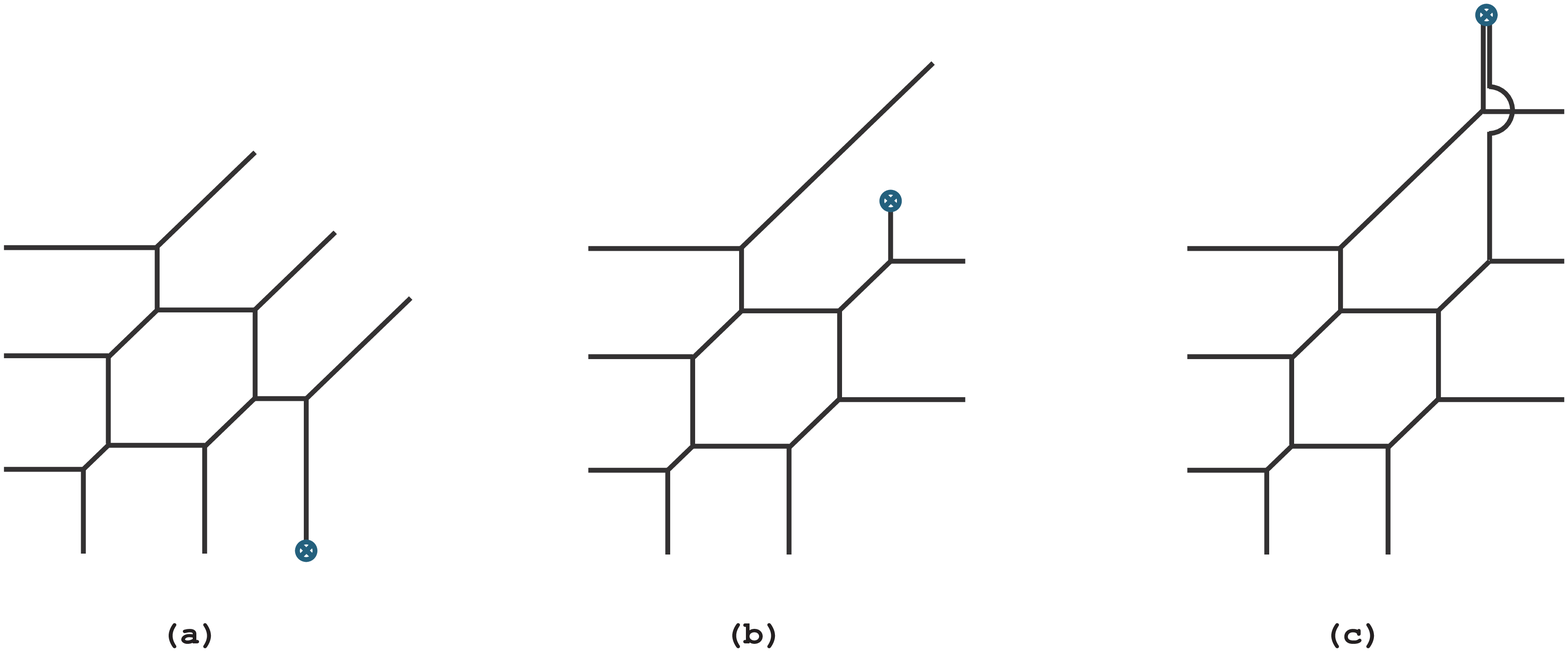}
\caption{\it Part (a) shows the brane setup for $T_3$, part (b) the setup for the $SU(2)$ gauge theory with five flavors and part (c) the setup for the $SU(3)$ gauge theory with six flavor. }
\label{fig:Brane-T3}
\end{figure}

Now that we have explained how the monodromy works, we will derive the $T_3$ multi-junction starting from the $SU(2)$ gauge theory with $N_f=5$ flavors. One possible brane setup is depicted in figure~\ref{fig:su2fiveflavor}, where all the five flavors are due to the 7-branes. Giving mass to a fundamental hypermultiplet corresponds to moving the 7-brane vertically. This is depicted in figure~\ref{fig:su2fiveflavor} (b). On contrary, moving the 7-branes horizontally will not change the theory. The next step is to move the four 7-branes horizontally out of the 5-brane loop and use the monodromy rule \eqref{eq:transformpq} to obtain figure \ref{fig:pullandflop} (a).
In figure~\ref{fig:pullandflop} (a) we insist on drawing the (0,1) 7-brane on which one of the NS5-branes ends (the upper right one) at a finite distance, so that we can pull out the remaining 7-brane to the left without intersecting any of the 5-branes (see figure~\ref{fig:pullandflop} (b)). After flopping, we obtain figure~\ref{fig:pullandflop} (c), which still keeps the natural interpretation of the $SU(2)$ theory with five flavors. This is also drawn in figure~\ref{fig:Brane-T3} (b), where the 7-brane with the branch cut has been pushed all the way to infinity.

We can now look at figure ~\ref{fig:Brane-T3} (b) and move the $(0,1)$ 7-brane downwards without altering the theory. By moving it downwards across the 5-brane and using the rule \eqref{eq:rule}, we obtain figure~\ref{fig:Brane-T3} (a), which is exactly the $T_3$ multi-junction introduced in \cite{Benini:2009gi}.

It is also interesting to demonstrate the other description. By moving the $(0,1)$ 7-brane in figure ~\ref{fig:Brane-T3} (b) upwards across the 5-brane and using the rule \eqref{eq:rule}, we obtain figure~\ref{fig:Brane-T3} (c). Here, two $(0,1)$ 5-branes are attached to a single $(0,1)$ 7-brane, and one of the $(0,1)$ 5-branes ``jumps over'' \cite{Benini:2009gi} the upper right D5-brane. It should be understood as a certain limit of the $SU(3)$ gauge theory with six flavors. This is the 5D uplift of the Argyres-Seiberg duality, which states that the four dimensional $E_6$ CFT is obtained at the strong coupling limit of the four dimensional $SU(3)$ gauge theory with six flavors. This interpretation is further investigated in section~\ref{sec:seibergwitten} at the level of the Seiberg-Witten curve. In particular, we show that the $E_6$ SW curve can be obtained form the curve for the $SU(3)$ gauge theory with six flavors by constraining its parameters as \eqref{eq:specialcurve}.

We conclude this section by noting that it is also possible to demonstrate, in a similar fashion, that the $T_N$ multi-junction is realized as a certain limit of the $SU(N)$ quiver gauge theory, which we also briefly discuss in section~\ref{sec:seibergwitten}.

\section{Seiberg-Witten curves}
\label{sec:seibergwitten}

In this section we derive the SW curve of the 5D $T_N$ junctions and take their 4D limit. We pay special attention to the $T_3$ junction that is the first non-trivial example and show how the $E_6$ Weyl symmetry is realized. We also show that the 5D $T_N$ junction curves
can be obtained from the curves of the $SU(N)^{N-2}$ quiver gauge theories if the gauge coupling constants and some of the Coulomb moduli parameters are tuned to certain values. 
In this section we follow closely \cite{Bao:2011rc} where our conventions were defined and the procedure was explained in great detail.
 
\subsection{Seiberg-Witten curves from M-theory}
\label{sec:seibergwittenMtheory}

\begin{figure}[htbp]
  \centering
  \includegraphics[height=4.7cm]{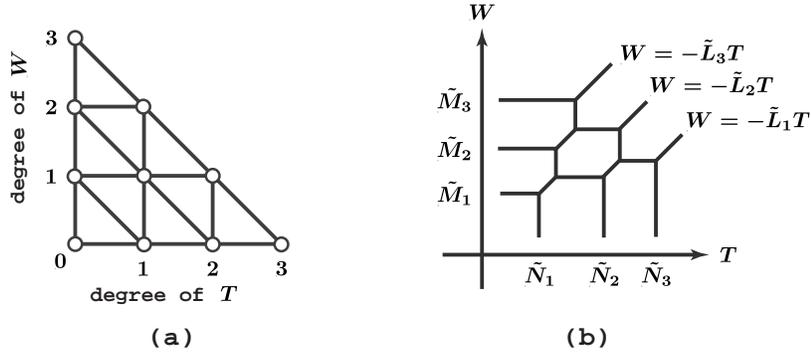}
  \caption{\it Part (a) shows the toric diagram and part (b) the brane setup for $E_6$ CFT.}
  \label{fig:brane_setup}
\end{figure}

We start from the brane setup of figure \ref{fig:brane_setup}.
The toric diagram associated with this configuration is depicted in part (a) of figure \ref{fig:brane_setup} and its associated SW curve is given by the polynomial equation \cite{Benini:2009gi, Brandhuber:1997ua}
\begin{equation}
\sum_{I,J \ge 0\atop{I+J \le 3}}^3 C_{IJ} T^I W^J = 0,
\end{equation}
where every vertex of the toric diagram corresponds to a coefficient $C_{IJ}$ \cite{Aharony:1997bh, Leung:1997tw}. The boundary condition depicted in part (b) of figure \ref{fig:brane_setup} leads to the following list of equations relating the parameters $C_{IJ}$:
\begin{equation}
\label{eq:condTW}
\begin{split}
\left(T \to 0\Rightarrow W \rightarrow \tilde{M}_I\right) & \Rightarrow \sum_{I=0}^3C_{0I}W^I=C_{03}\prod_{I=1}^3(W-\tilde{M}_I), \\
\left(W \to 0\Rightarrow T \rightarrow \tilde{N}_I\right) & \Rightarrow \sum_{I=0}^3C_{I0}T^I=C_{30}\prod_{I=1}^3(T-\tilde{N}_I), \\
\left(|T|\sim |W|\to \infty \Rightarrow W \rightarrow -\tilde{L}_IT\right) & \Rightarrow \sum_{I=0}^3C_{I3-I}T^IW^{3-I}=C_{03}\prod_{I=1}^3(W+\tilde{L}_IT).
\end{split}
\end{equation}
Compatibility of these three equations imposes the constraint:
\begin{equation}
\tilde{M}_1\tilde{M}_2\tilde{M}_3
= \tilde{N}_1\tilde{N}_2\tilde{N}_3
\tilde{L}_1\tilde{L}_2\tilde{L}_3.
\end{equation}
By rescaling the coordinates $W$ and $T$, we can choose the parameters of the curve to obey the relations
\begin{equation}
\label{eq:const_LMN}
\tilde{M}_1\tilde{M}_2\tilde{M}_3 = 1,
\qquad
\tilde{N}_1\tilde{N}_2\tilde{N}_3 = 1,
\qquad
\tilde{L}_1\tilde{L}_2\tilde{L}_3 = 1,
\end{equation}
which then implies $C_{30}=C_{03}=-C_{00}$.
Thus, putting the conditions \eqref{eq:condTW} and \eqref{eq:const_LMN} together, we find that the SW curve for the brane configuration of figure~\ref{fig:Brane-T3} is given by the equation
\begin{multline}
 W^3 - \left( \sum_I \tilde{M}_I \right) W^2 + \left( \sum_K \tilde{L}_K \right) W^2 T  + \left( \sum_I \tilde{M}_I^{-1} \right) W + U W T \\+ \left( \sum_K \tilde{L}_K^{-1} \right) W T^2  - 1 + \left( \sum_{J} \tilde{N}_J^{-1} \right) T - \left( \sum_{J} \tilde{N}_{J} \right) T^2 + \ T^3 = 0,
\label{eq:E6_SWcurve}
\end{multline}
where $U$ is a free parameter.

We would like to obtain also the expression of the curve 
based on the brane setup (c) in figure~\ref{fig:Brane-T3},
which is obtained from (a) in figure~\ref{fig:Brane-T3} 
by moving one of the $(0,1)$ 7-branes upward across the three (1,1) 5-branes. In \cite{Brandhuber:1997ua}, the authors introduced the coordinate change  which moves the flavor D5-branes from one hand side to the other side in the brane setup for the 5D $SU(N)$ theory with $N_f$ flavor. We expect that (a) and (c) in figure~\ref{fig:Brane-T3} are also related by an analogous coordinate change.
For that, we first observe that the $W$-independent part of \eqref{eq:E6_SWcurve} can be written as
\begin{equation}
\prod_{I=1}^3(T-\tilde{N}_I)=\tilde{N}_3^3(t-\tilde{N}_1\tilde{N}_3^{-1})(t-\tilde{N}_2\tilde{N}_3^{-1})(t-1),
\end{equation}
where we have set $t=\tilde{N}_3^{-1}T$. 
Then, if we perform the coordinate change 
\begin{equation}
\begin{split}
W
= -\tilde{N}_1^{-\frac{1}{3}}
\tilde{N}_3^{\frac{1}{3}}
(t-1) w,
\qquad
T = \tilde{N}_3 t,
\end{split}
\label{eq:wtoW}
\end{equation}
we can factor out a piece of $(t-1)$ from the curve to obtain the compact expression
\begin{equation}
\label{eq:alternativeeq:E6_SWcurve}
\prod_{i=1}^3(w-\tilde{m}_i)t^2+\left[-2w^3+\sum_{i=1}^6\tilde{m}_iw^2+U_1w+\left(1+\prod_{i=1}^6\tilde{m}_i\right)\right]t+\prod_{i=4}^6(w-\tilde{m}_i)=0,
\end{equation}
where we have defined the new parameters $\{\tilde{m}_i\}_{i=1}^6$ and $U_1$ as
\begin{equation}
\label{eq:newmass}
\tilde{m}_i  = \left\{\begin{array}{ll}\left(\tilde{N}_1^{\frac{1}{3}} \tilde{N}_3^{\frac{2}{3}}\right)
 \tilde{L}_i & \text{ for } i=1,2,3\\ \left(
\tilde{N}_1^{\frac{1}{3}} \tilde{N}_3^{-\frac{1}{3}}
\right) \tilde{M}_{i-3} & \text{ for } i=4,5,6 \end{array}\right. \quad \text{ and }  \quad U_1 = \tilde{N}_1^{\frac{2}{3}}\tilde{N}_3^{\frac{1}{3}} U.
\end{equation}
We interpret that the expression \eqref{eq:alternativeeq:E6_SWcurve} corresponds exactly to (c) in figure \ref{fig:Brane-T3}.

We can easily see that the curve described by \eqref{eq:newmass} is a special case of the \emph{general} 5D $SU(3)$ $N_f=6$ SW curve \cite{Nekrasov:1996cz, Bao:2011rc}
\begin{multline}
\label{eq:su3_6_general}
\prod_{i=1}^3(w-\tilde{m}_i)t^2+\left[-\left(1+q\prod_{i=1}^6\tilde{m}_i^{-\frac{1}{2}}\right)w^3+U_2w^2\right.\\\left.+U_1w+\left(1+q \prod_{i=1}^6\tilde{m}_i^{\frac{1}{2}}\right)\right]t+q  \prod_{i=1}^6 \tilde{m}_i ^{-\frac{1}{2}} \prod_{i=4}^6(w-\tilde{m}_i)=0,
\end{multline}
if we constrain the additional parameters as
\begin{equation}
\label{eq:specialcurve}
q = \prod_{i=1}^6 \tilde{m}_i^{\frac{1}{2}}, \qquad
U_2 = \sum_{i=1}^6 \tilde{m}_i.
\end{equation}
This is the 5D uplift of the statement in \cite{Argyres:2007cn} that the 4D $E_6$ CFT is obtained as the strong coupling limit of the 4D SU(3) theory with six flavors.

We end this subsection, by noting that  equation \eqref{eq:alternativeeq:E6_SWcurve} defines a curve of genus one. This is seen by solving \eqref{eq:alternativeeq:E6_SWcurve} for $t$ in terms of $w$. For illustration, the discriminant of this quadratic equation for the general curve in equation \eqref{eq:su3_6_general} reads
\begin{multline}
\Delta=\left(1-q\prod_{i=1}^6\tilde{m}_i^{-\frac{1}{2}}\right)^2w^6+2\left(2q\prod_{i=1}^6\tilde{m}_i^{-\frac{1}{2}}\sum_{j=1}^6\tilde{m}_i-U_2\left(1+q\prod_{i=1}^6\tilde{m}_i^{-\frac{1}{2}}\right)\right)w^5\\ +\left(U_2^2-2\left(1+q\prod_{i=1}^6\tilde{m_i}^{-\frac{1}{2}}\right)U_1-4q\prod_{i=1}^6\tilde{m_i}^{-\frac{1}{2}}\sum_{i<j=1}^6\tilde{m}_i\tilde{m}_j\right)w^4+\cdots,
\end{multline}
which reduces to a degree four polynomial if equations \eqref{eq:specialcurve} hold, meaning that two branch points get sent to infinity and that we are dealing with a genus one curve.

\subsection{Discrete symmetries}

Due to the introduction of the mass parameters, the global $E_6$ symmetry is broken to a discrete symmetry, which consists of the Weyl symmetry together with the $S_3$ symmetry of the extended Dynkin diagram. It is straightforward to show this discrete symmetry group for the curve derived in the previous subsection.

\begin{figure}[htbp]
  \centering
  \includegraphics[height=4.3cm]{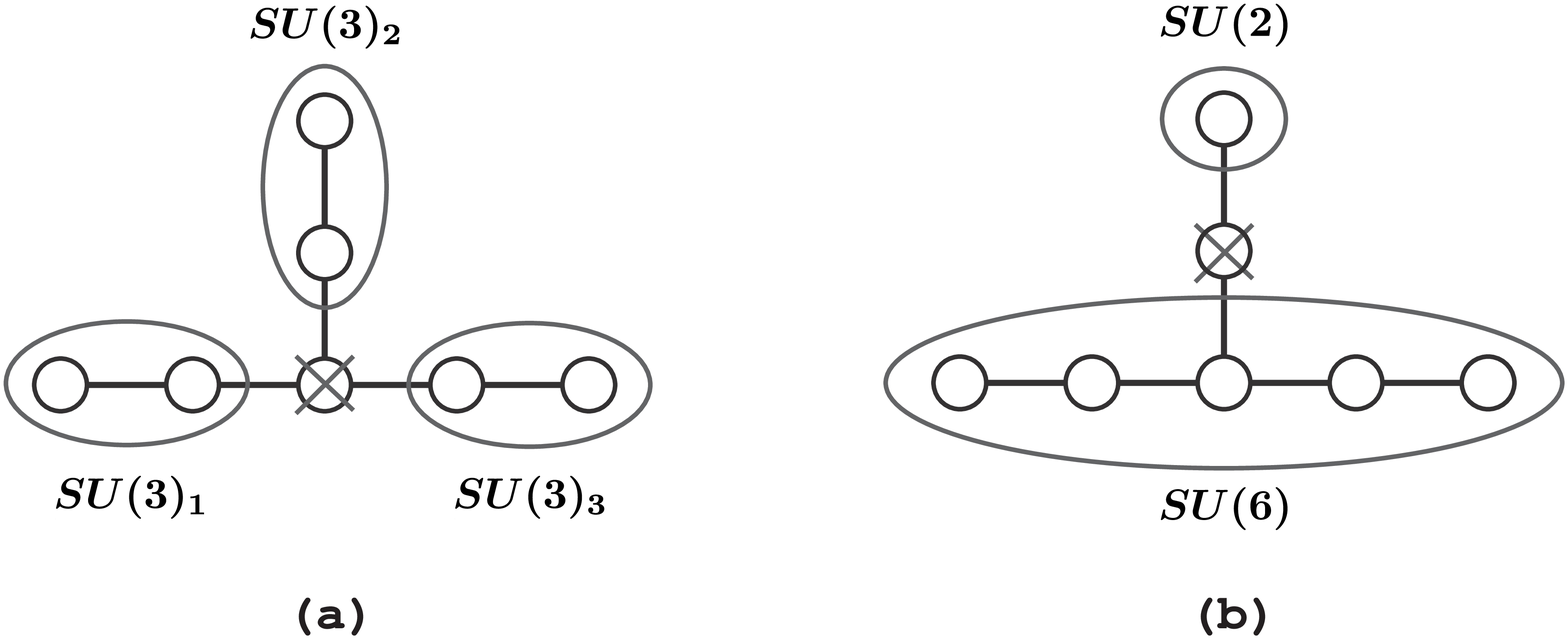}
  \caption{Part (a) of this figure shows the first maximal embedding $SU(3)^3 \subset E_6$, while part (b) depicts the second one \it $SU(6) \times SU(2) \subset E_6$.}
  \label{fig:E6subgroups}
\end{figure}

First, the Weyl symmetry of the $SU(3)^3$ subgroup of $E_6$ is manifest in the curve \eqref{eq:E6_SWcurve} as permutations of the mass parameters
\begin{equation}
\tilde{M}_I \leftrightarrow \tilde{M}_J ,
\qquad
\tilde{N}_I \leftrightarrow \tilde{N}_J ,
\qquad
\tilde{L}_I \leftrightarrow \tilde{L}_J.
\label{weylsu33}
\end{equation}
Notice that $SU(3)^3$ is a maximal subgroup of $E_6$, as depicted in the extended Dynkin diagram in part (a) of figure~\ref{fig:E6subgroups}. In addition, we can rewrite the curve in a way that makes another piece of the Weyl symmetry apparent. For this purpose, we set $\tilde{M}=\prod_{i=1}^6\tilde{m}_i$ and define
\begin{equation}
U_1=\tilde{M}^{\frac{1}{3}}U_1^{\prime}, \qquad \tilde{m}_i=\tilde{M}^{\frac{1}{6}}\tilde{m}_i^{\prime},  \qquad
w=\tilde{M}^{\frac{1}{6}}w^{\prime}, \qquad t = \frac{\tilde{M}^{\frac{1}{2}}t'}{\prod_{i=1}^3(w-\tilde{m}_i)}  .
\end{equation}
This transforms the alternative description of the curve \eqref{eq:alternativeeq:E6_SWcurve} into
\begin{equation}
\label{eq:su6xsu2_manifest}
\begin{split}
&t'{}^2
+ \left[
- 2 {w^{\prime}}^3
+ \left( \sum_{i=1}^6 \tilde{m}_i^{\prime} \right) {w^{\prime}}^2
+ U_1^{\prime} w^{\prime}
+ \left( \tilde{M}^{\frac{1}{2}}+\tilde{M}^{-\frac{1}{2}}\right)
\right] t'
+ \prod_{i=1}^6
(w^{\prime}-\tilde{m}_i^{\prime})
= 0.
\end{split}
\end{equation}
In this formulation, the Weyl symmetry of another maximal subgroup of $E_6$, namely $SU(6) \times SU(2)$ becomes manifest, with the explicit parameter transformations
\begin{equation}
\tilde{m}_i^{\prime} \leftrightarrow \tilde{m}_j^{\prime},
\qquad
\tilde{M} \leftrightarrow \tilde{M}^{-1}.
\label{weylsu6}
\end{equation}
The  $SU(6) \times SU(2)$ maximal subgroup of $E_6$  is depicted in figure~\ref{fig:E6subgroups}.
\begin{figure}[htbp]
\begin{center}
\includegraphics[width=5cm]{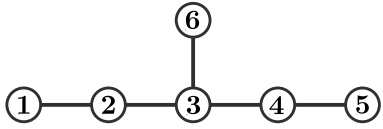}
\end{center}
  \caption{\it The Dynkin diagram of $E_6$}
  \label{fig:Dynkin_E6}
\end{figure}
The complete Weyl symmetry of $E_6$ is realized as follows. We recall that the non-vanishing entries of the Cartan matrix are $C_{ii}=2$ and $C_{ij}=-1$ if the node $i$ is connected to the node $j$ as illustrated in figure~\ref{fig:Dynkin_E6}. Denoting the simple roots of $E_6$ by $(\alpha_i)_{i=1}^6$, the action of the Weyl group generator associated to $\alpha_i$ on the set of simple roots is given by $\alpha_j\mapsto \alpha_j-C_{ij}\alpha_i$. If we now associate the simple roots to the parameters of the curve as
\begin{align}
&e^{-\beta\alpha_1}=\tilde{L}_1\tilde{L}_2^{-1},& &e^{-\beta\alpha_2}=\tilde{L}_2\tilde{L}_3^{-1},& &e^{-\beta\alpha_3}=\tilde{L}_3\tilde{N}_3\tilde{M}_1^{-1},&\nonumber\\
&e^{-\beta\alpha_4}=\tilde{M}_1\tilde{M}_2^{-1},& &e^{-\beta\alpha_5}=\tilde{M}_2\tilde{M}_3^{-1},& &e^{-\beta\alpha_6}=\tilde{N}_2\tilde{N}_3^{-1},&
\end{align}
we then get a well defined action of the Weyl group,
which is shown to be generated by the discrete symmetry \eqref{weylsu33} and \eqref{weylsu6}
after a lengthy calculation.
This indicates that the curve \eqref{eq:E6_SWcurve} is invariant under the whole 
$E_6$ Weyl symmetry up to coordinate transformations.
 
In addition to the $E_6$ Weyl symmetry already discussed, the original formulation \eqref{eq:E6_SWcurve} of the curve has an $S_3$ symmetry that is generated by two basic transformations $\sigma $ and $\tau$:
\begin{equation}
\label{ST_trans}
\begin{split}
\sigma:\  & T\leftrightarrow  W, \quad \tilde{M}_j\leftrightarrow \tilde{N}_j, \quad \tilde{L}_j\to \tilde{L}_{3-j}^{-1}, \quad U_1\to U_1, \\
\tau:\  & W \to  -W T^{-1}, \quad T\to  T^{-1}, \quad \tilde{M}_j\leftrightarrow \tilde{L}_j, \quad \tilde{N}_j\to \tilde{N}_{3-j}^{-1}, \quad U_1\to -U_1.
\end{split}
\end{equation}
We easily compute that $\sigma^2=\tau^2=\textbf{1}$ as well as $\sigma\tau\sigma=\tau\sigma\tau$. This implies that $\sigma$ and $\tau$ together generate an $S_3$ group, which we identify with the $S_3$ symmetry of the extended Dynkin diagram in figure \ref{fig:E6subgroups}.

Here, we comment that our curve for the 5D $T_3$ junction investigated above is actually equivalent to 
the curve for the 5D $E_6$ CFT obtained in \cite{Minahan:1997ch,Eguchi:2002fc,Eguchi:2002nx}.
In their expression, all of the coefficients can be
written in terms of $E_6$ characters, which makes the $E_6$ Weyl symmetry manifest.
By comparing the modular function called ``$j$-invariant'' for the two curves,
we can show that the periods of the curves are identical up to $SL(2,Z)$
modular transformation.
It indicates the existence of the coordinate change from one to the other
although its explicit form is too complicated to write down.
In appendix \ref{sec:Minahan-Nemeshanski}, we discuss this check in further detail.

\subsection{The 4D limit of the $E_6$ curve}

We will now consider the four dimensional limit of the SW curve \eqref{eq:alternativeeq:E6_SWcurve}. 
Denoting the $S^1$ circumference of the compactified 5D theory as $\beta$, we assume that the 4D coordinate $v$ is related to the 5D coordinate $w$ as\footnote{ To simplify the notations we use here a slightly different definition compared to our previous paper \cite{Bao:2011rc} in which $w = e^{- v/R_5}$. The difference is a rescaling of $v$ by the factor $2 \pi \alpha'$ due to the relation $\beta = \frac{2 \pi \alpha'}{R_5}$.}
\begin{align}
w = e^{- \beta v}.
\end{align}
The 4D coordinate $t$ is set equal to the 5D coordinate $t$. Accordingly, we also assume that the mass parameters of the 4D theory ($m_I$) are related to those of the 5D theory ($\tilde{m}_I$) as
\begin{align}
\label{eq:mm}
\tilde{m}_I = e^{- \beta m_I} , \qquad I=1,\dots,6.
\end{align}
Moreover, the following $\beta$ dependent variables are introduced for later convenience
\begin{align}
\label{eq:tildenontilde}
\tilde{M}_i = e^{- \beta M_i}, \qquad
\tilde{N}_i = e^{- \beta N_i}, \qquad
\tilde{L}_i = e^{- \beta L_i}.
\end{align}
We also expand the Coulomb moduli parameter $U_1$ in a power series
\begin{align}
U_1 = \sum_{k=0}^{\infty} u_k \beta^k.
\label{U_exp}
\end{align}
The expansion of the SW curve \eqref{eq:alternativeeq:E6_SWcurve} in terms of the radius $\beta$ is straightforward, in particular the first and the last term become
\begin{align}
\begin{split}
(w-\tilde{m}_1)(w-\tilde{m}_2)(w-\tilde{m}_3) t^2
& = - \beta^3 (v-m_1)(v-m_2)(v-m_3) t^2  + {\cal O}(\beta^4),
\cr
(w-\tilde{m}_4)(w-\tilde{m}_5)(w-\tilde{m}_6)
& = - \beta^3 (v-m_4)(v-m_5)(v-m_6) + {\cal O}(\beta^4).
\end{split}
\end{align}
These expressions indicate that a 4D SW curve will appear at the order of $\beta^3$ when we expand the full 5D curve in \eqref{eq:alternativeeq:E6_SWcurve}. Indeed, by setting
\begin{align}
u_0 = -6,
\qquad
u_1 = 2 \sum_{k=1}^6 m_k,
\qquad
u_2 = - \sum_{1 \le k \le \ell \le 6} m_k m_{\ell},
\end{align}
we find that up to order $\beta^2$ \eqref{eq:alternativeeq:E6_SWcurve} vanishes identically. Under these assumptions we can read off a 4D SW curve at the $\beta^3$ order of \eqref{eq:alternativeeq:E6_SWcurve}
\begin{align}
\begin{split}
(v-m_1)(v-m_2)(v-m_3) t^2
+ \left(
- 2 v^3 + \sum_{k=1}^6 m_k v^2
- \sum_{k < \ell} m_k m_{\ell} v
- u
\right) t &
\cr
+ (v-m_4)(v-m_5)(v-m_6) & = 0,
\end{split}
\label{eq:4DE6Curve}
\end{align}
where the parameter $u$ is defined as
\begin{align}
u \equiv u_3 - \frac{1}{6} \sum_{k=1}^6 m_k^3 - \frac{1}{6}  \left(\sum_{k=1}^6 m_k\right)^3.
\end{align}
This 4D curve can also be expressed as
\begin{align}
\begin{split}
(t-1)^2 v^3
- (t-1) (S_1 t - \tilde{S}_1) v^2
+ \left(
S_2 t^2
- \left(
S_2 + S_1 \tilde{S}_1 + \tilde{S}_2
\right) t
+ \tilde{S}_2
\right) v &
\cr
- \left(
S_3 t^2
+ u t
+ \tilde{S}_3
\right) &
= 0,
\end{split}
\end{align}
with the parameters $S_i$ and $\tilde{S}_i$ being defined as
\begin{align}
\label{eq:ChiToM}
&S_1 = m_1 + m_2 + m_3,
\quad
S_2 = m_1m_2 + m_2m_3 + m_3m_1,
\quad
S_3 = m_1m_2m_3,
\cr
&\tilde{S}_1= m_4 + m_5 + m_6,
\quad
\tilde{S}_2 = m_4m_5 + m_5m_6 + m_6m_4,
\quad
\tilde{S}_3 = m_4m_5m_6.
\end{align}
These parameters are the Casimirs \cite{Minahan:1996fg, Argyres:2007cn} of the $U(3)_1 \times 
U(3)_2$ flavor symmetries.

Furthermore, we can introduce a new coordinate $x$ which is related to the coordinate $v$ as
\begin{align}
v = xt + \frac{S_1 t - \tilde{S}_1}{3(t-1)}.
\end{align}
The 4D SW curve then takes the form
\begin{align}
\label{eq:E64D}
x^3 = \frac{P_2(t)}{t^2 (t-1)^2} x + \frac{P_3(t)}{t^3 (t-1)^3},
\end{align}
where $P_k(t)$ are polynomial functions with degree $k$ in $t$
\begin{equation}
\begin{split}
P_2(t)
= & - S_2' t^2
+ \left( S_2' + \tilde{S}_2'
+ \frac{1}{3} S_1^2
+ \frac{1}{3} S_1 \tilde{S}_1
+ \frac{1}{3} \tilde{S}_1^2
\right) t
- \tilde{S}_2',
\\
P_3(t) = &
S_3' t^3
- \left(
 S_3'
- \frac{2}{27} S_1^3
- \frac{1}{9} S_1^2 \tilde{S}_1
- u'
\right) t^2
+ \left(
\tilde{S}_3'
- \frac{2}{27} \tilde{S}_1^3
- \frac{1}{9} S_1 \tilde{S}_1^2
- u'
\right) t
- \tilde{S}_3'.
\end{split}
\end{equation}
Here, we have shifted the Coulomb moduli parameter as 
\begin{equation}
u' \equiv u + \frac{1}{3} (S_1 \tilde{S}_2 + \tilde{S}_1 S_2)
\end{equation}
and have introduced the $SU(2) \times SU(2)$ Casimirs:
\begin{align}
\begin{split}
&S_2' \equiv S_2 - \frac{1}{3} S_1^2
= L_1 L_2 + L_2 L_3 + L_3 L_1,
\cr
&S_3' \equiv S_3 - \frac{1}{3} S_1 S_2 + \frac{2}{27} S_1^3
= L_1 L_2 L_3,
\cr
&\tilde{S}_2' \equiv \tilde{S}_2 - \frac{1}{3} \tilde{S}_1^2
= M_1 M_2 + M_2 M_3 + M_3 M_1,
\cr
&\tilde{S}_3' \equiv
\tilde{S}_3 - \frac{1}{3} \tilde{S}_1 \tilde{S}_2 + \frac{2}{27} \tilde{S}_1^3
= M_1 M_2 M_3.
\end{split}
\end{align}
The expressions after the second equality signs follow from the relations \eqref{eq:const_LMN}, \eqref{eq:newmass} and \eqref{eq:ChiToM}. This form of the SW curve has previously been used by Gaiotto. For the massless case, \eqref{eq:E64D} coincides with Eq. (3.10) in \cite{Gaiotto:2009we}. This SW curve can be interpreted as the triple cover of a sphere with three punctures. It is straightforward to see that the curve has poles at $t=\{0,1,\infty\}$.

Let us look at the poles more closely starting at $t=0$. By looking at \eqref{eq:E64D} we see that around $t=0$
\begin{equation}
x = \frac{C}{t} + {\cal O} (1).
\end{equation}
After substituting it into \eqref{eq:E64D}, one obtains
\begin{equation}
C^3 =  P_2(0) C - P_3(0),
\end{equation}
which has the solutions
\begin{equation}
C = \left\{ M_1 , M_2 , M_3 \right\}.
\end{equation}
Observe that these are the mass parameters associated with the $SU(3)$ puncture at $t=0$.

Similarly, at $t=\infty$ we substitute
\begin{equation}
x = \frac{C'}{t} + {\cal O} (t^{-2})
\end{equation}
into \eqref{eq:E64D} to obtain
\begin{align}
C'^3
&=  \lim_{t\rightarrow \infty}\frac{P_2(t)}{t^2} C' + \lim_{t\rightarrow \infty} \frac{P_3(t)}{t^3}.
\end{align}
Solving for $C'$ yields
\begin{equation}
C' = \left\{ L_1 , L_2 , L_3 \right\}.
\end{equation}
These are the mass parameters associated with the $SU(3)$ puncture at $t=\infty$. Note that when computing the contour integral
\begin{equation}
\oint_{t=\infty} x dt = \oint_{t=\infty} \frac{C'}{t} dt = \oint_{s=0} \frac{C'}{s} ds \sim C',
\end{equation}
with $s=\frac{1}{t}$, the extra minus sign from $\frac{dt}{t} = - \frac{ds}{s} $ is canceled by changing the direction of the contour.

Lastly, at $t=1$
\begin{equation}
x = \frac{C''}{t-1} + {\cal O}(1)
\end{equation}
yields the equation
\begin{equation}
C''^3 =P_2(1)  C'' + P_3(1).
\end{equation}
Solving for $C''$, we find
\begin{eqnarray}
C'' = \left\{
- \frac{1}{3} S_1 - \frac{2}{3} \tilde{S}_1,
\frac{2}{3}S_1 + \frac{1}{3} \tilde{S}_1,
- \frac{1}{3} S_1 + \frac{1}{3} \tilde{S}_1
\right\}
 = - \left\{ N_1 ,N_2 ,N_3 \right\}
\end{eqnarray}
These are the mass parameters associated with the puncture at $t=1$. They are originally the mass parameters associated with
the $U(1)_1$ and the $U(1)_2$ puncture, respectively. However, $t=1$ corresponds to a third $SU(3)$ puncture, which is generated by decoupling the sphere with two $U(1)$ punctures from the two $SU(3)$ punctures. This interpretation is also consistent with the picture that the $E_6$ gauge theory corresponds to a sphere with three $SU(3)$ punctures.

As a side remark, the relations between the 4D coordinates $\{t,v\}$ and the 5D coordinates $\{T,W\}$ in \eqref{eq:wtoW}
\begin{equation}
\label{eq:relation4D5Dcoordinates}
T = \tilde{N}_3 t,\qquad
W = \tilde{N}_1^{-\frac{1}{3}} \tilde{N}_3^{\frac{1}{3}}
(t - 1) e^{-\beta v}
\end{equation}
are different from the corresponding ones given in \cite{Benini:2009gi}.

\subsection{Compatibility of the 4D and strong coupling limits}

Noticing that the constraint \eqref{eq:specialcurve} becomes $q \to 1$ in the 4D limit $\beta \to 0$, it is therefore natural to consider \eqref{eq:specialcurve} as being the 5D uplift of the strong coupling limit of the 4D SU(3) theory with six flavors, leading then to the 4D $E_6$ CFT \cite{Argyres:2007cn}. However, the 4D limit and the strong coupling limit may not commute a priori. In order to clarify this point, we consider here a more general way to take the 4D limit of the SW curve.

We start from the curve \eqref{eq:su3_6_general} for the general 5D SU(3) theory with six flavors. By expanding the Coulomb moduli parameters as
\begin{equation}
U_1 = \sum_{k=0}^{\infty} u_1^{(k)} \beta^k,\qquad
U_2 = \sum_{k=0}^{\infty} u_2^{(k)} \beta^k,
\end{equation}
it is shown below that the 4D limit does not depend on $u^{(k)}$ with $k \ge 4$, which we then can put to zero.

For the $E_6$ theory, one of the Coulomb moduli parameters obeys
\begin{align}
U_2 = \sum_{i=1}^6 \tilde{m}_i,
\end{align}
yielding
\begin{align}
u_2^{(0)} = 1,
\quad
u_2^{(1)} = - \sum_{i=1}^6 m_i,
\quad
u_2^{(2)} = \frac{1}{2} \sum_{i=1}^6 m_i^2,
\quad \dots
\label{eq:exp_U1_E6}
\end{align}
after a Taylor expansion.

It is worth emphasizing that the 4D limits of the $E_6$ CFT and of a generic $SU(3)$ six flavor theory are slightly different. In the $E_6$ case, we have the relation \eqref{eq:specialcurve} $q=\prod_{i=1}^{6} (\tilde{m}_i)^{1/2}$, which means that $q$ goes to 1 in a specific way when we take the 4D limit. On the other hand, for the $SU(3)$ theory, $q$ is regarded as being constant. Only after taking the 4D limit can we put $q \to 1$. Thus, it is not obvious that the two procedures give the same result. In order to compare the two 4D limits, we also expand $q$ in terms of $\beta$
\begin{align}
q = \sum_{k=0}^{\infty} q^{(k)} \beta^k.
\end{align}
For the $E_6$ theory, we have
\begin{align}
q^{(0)} = 1,
\quad
q^{(1)}= - \frac{1}{2} \sum_{i=1}^6 m_i,
\quad
q^{(2)}= \frac{1}{2} \left( \frac{1}{2} \sum_{i=1}^6 m_i \right)^2,
\quad
q^{(3)}= - \frac{1}{6} \left( \frac{1}{2} \sum_{i=1}^6 m_i \right)^3,
\quad \dots
\label{eq:exp_q_E6}
\end{align}
while for an $SU(3)$ theory, we have
\begin{align}
q^{(0)} = q_{4D},
\qquad
q^{(1)}= q^{(2)}=q^{(3)}=\dots=0.
\label{eq:exp_q_SU3}
\end{align}

By expanding the curve \eqref{eq:su3_6_general} in powers of $\beta$, the corresponding 4D SW curve is again obtained at the order $\beta^3$. Constraining the lower expansion coefficients ($\beta^0$, $\beta^1$, $\beta^2$) of the curve to vanish, we find the following relations among the parameters:
\begin{equation}
\begin{split}
u_1^{(0)} = -3-3q^{(0)}, & \qquad u_2^{(0)} = 3+3q^{(0)}, \\
u_1^{(1)} = \frac{1}{2} q^{(0)} \sum_{k=1}^6 m_k - 3 q^{(1)}, & \qquad
u_2^{(1)} = \frac{1}{2} q^{(0)} \sum_{k=1}^6 m_k + 3 q^{(1)}, \\
u_1^{(2)} = - u_2^{(2)} + \sum_{k=1}^6 m_k q^{(1)}. & \\
\end{split}
\end{equation}
Furthermore, the 4D SW curve becomes
\begin{align}
\begin{split}
0=&(v-m_1)(v-m_2)(v-m_3) t^2
\cr
& + \left(
- (1+q^{(0)}) v^3 + q^{(0)} \sum_{k=1}^6 m_k v^2
+ \tilde{u}_1 v
+ \tilde{u}_2
\right) t
\cr
& + q^{(0)} (v-m_4)(v-m_5)(v-m_6),
\end{split}
\label{eq:4D-su3-gen}
\end{align}
where we have defined
\begin{align}
\begin{split}
&\tilde{u}_1 \equiv u_2^{(2)}
- \frac{3}{8} q^{(0)} \left( \sum_{k=1}^6 m_k \right)^2
- \frac{1}{2} q^{(1)} \sum_{k=1}^6 m_k
- 3 q^{(2)},
\\
& \tilde{u}_2 \equiv -u_1^{(3)} - u_2^{(3)}
+ \frac{1}{24} q^{(0)} \left( \sum_{k=1}^6 m_k \right)^3
+ q^{(2)} \sum_{k=1}^6 m_k.
\end{split}
\label{eq:def_utilde}
\end{align}

This is the SW curve of the $SU(3)$ theory with six flavors, if we identify $q^{(0)} = q_{4D}$ as in \eqref{eq:exp_q_SU3}. It is remarkable that all the higher order terms in the expansion of $q$ are absorbed by the definition of the 4D Coulomb moduli parameters. In \eqref{eq:4D-su3-gen}, only the leading order $q^{(0)}$ appears explicitly. Therefore, the difference between the two expansions \eqref{eq:exp_q_E6} and \eqref{eq:exp_q_SU3} will not be seen in the 4D curve itself, if we further take the limit $q_0 = q_{4D} \to 1$. Thanks to this feature, the 4D limit and the strong coupling limit commute.

By inserting \eqref{eq:exp_U1_E6} and \eqref{eq:exp_q_E6} into \eqref{eq:def_utilde}, we obtain
\begin{equation}
q^{(0)}=1, \qquad \tilde{u}_1 = - \sum_{i<j} m_i m_j.
\label{q-u_1}
\end{equation}
The curve \eqref{eq:4D-su3-gen} then coincides with the 4D $E_6$ curve in \eqref{eq:4DE6Curve}.

\subsection{The curve for general $T_N$ junctions}
\label{subsec:curvegeneralN}

\begin{figure}[htbp]
\centering
  \includegraphics[height=5cm]{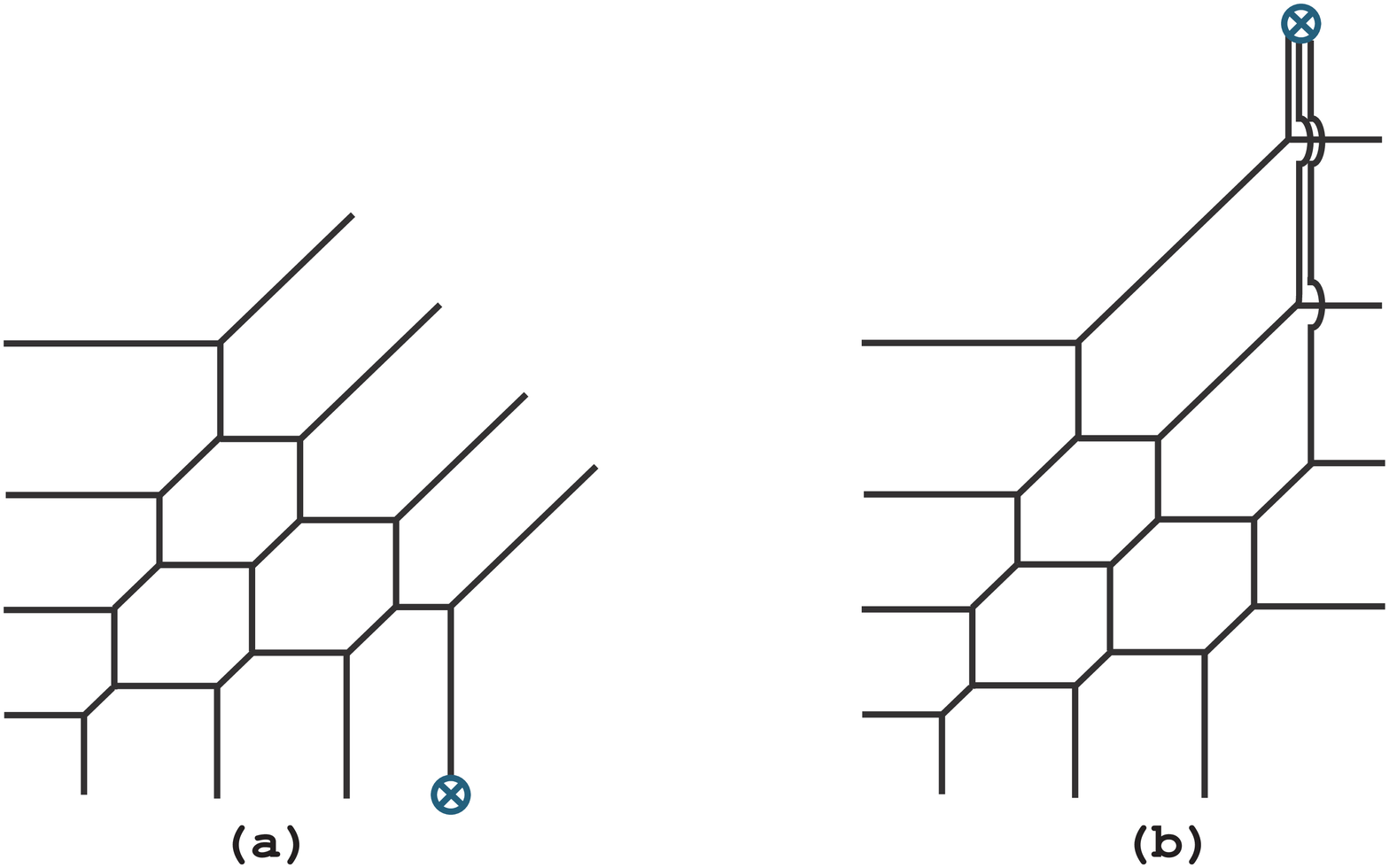}
  \caption{\it Part (a) of the figure shows the brane setup of $T_4$. Part (b) shows the brane setup of a special limit of the $SU(4) \times SU(4)$ quiver gauge theory.}
  \label{fig:Brane-T4}
\end{figure}

The analysis of the previous sections can be generalized to the $T_N$ theory for an arbitrary $N$. The SW curve for the $T_N$ theory has the form
\be
\sum_{i \geq 0\, , \, j\geq 0  \, , \, i + j \leq 0} C_{ij} T^i W^j = 0,
\ee
with the coefficients
\be
C_{0i} = (-1)^{N-i} S_{N-i}(\tilde{M}) \, , \qquad
C_{j0} = (-1)^{N-j} S_{N-j}(\tilde{N}) \, , \qquad
C_{i, N-i} = S_i(\tilde{L})\, .
\ee
We have introduced the elementary symmetric polynomials $S_n$ in the above expressions, in particular
\be
S_n(\tilde{M}) =  \sum_{1 \le i_1<i_2 < \cdots < i_n \le N} \tilde{M}_{i_1} \tilde{M}_{i_2} \cdots \tilde{M}_{i_n}\, ,
\ee
with $\tilde{M}_i$, $\tilde{N}_i$ and $\tilde{L}_i$ satisfying the constraints
\begin{align}
\prod_{i=1}^N \tilde{M}_i = \prod_{i=1}^N \tilde{N}_i = \prod_{i=1}^N \tilde{L}_i = 1.
\end{align}
The rest of the coefficients $C_{ij}$ corresponds to various Coulomb moduli $U_i$. There are $\frac{1}{2}(N-2)(N-1)$ in total. We now perform the coordinate transformation\footnote{In the previous section, we rescaled $W$ such that the center of mass of the color branes sits at $W=1$. Here, we omit this rescaling.}
\begin{equation}
\label{eq:rescalingWTTN}
W = -(t-1) w,
\qquad
T=\tilde{N}_N t,
\end{equation}
which we interpret as moving the $(0,1)$ 7-brane attached to one of the NS5-branes. An example for $N=4$ is given in figure~\ref{fig:Brane-T4}.
After factoring out $(t-1)$,  we obtain a polynomial of degree $N-1$ in $t$ and $N$ in $w$:
\begin{align}
0
&= (t-1)^{N-1} w^N
+ \left( S_1(\tilde{M}) - \tilde{N}_N S_1(\tilde{L}) t \right) (t-1)^{N-2} w^{N-1}
+ \cdots
+ \tilde{N}_N^{N} \prod_{i=1}^{N-1} \left( t - \frac{\tilde{N}_i}{\tilde{N}_N} \right)\nonumber\\
&= \left( \prod_{i=1}^N (w - \tilde{N}_N \tilde{L}_i) \right) t^{N-1}
+ \cdots + \prod_{i=1}^N (w - \tilde{M}_i).
\end{align}
This curve is consistent with figure~\ref{fig:Brane-T4} since $N-1$ NS5-branes coincide in the $w \to \infty$ limit.
This is the special case of the SW curve for the $SU(N)^{N-2}$ quiver gauge theory, where the gauge coupling constants and some of the Coulomb moduli parameters are tuned to certain specific values depending on the mass parameters and is the 5D uplift of the known statement that the 4D $T_N$ theory is constructed from the 4D $SU(N)^{N-2}$ quiver gauge theory \cite{Gaiotto:2009we}.

The $SU(N)^3$ symmetry is the full symmetry of the $T_N$ theory for $N>3$. The check of the Weyl symmetry would therefore be trivial.

The 4D limit can be obtained analogously by using the same coordinate transformation \eqref{eq:relation4D5Dcoordinates} up to an overall rescaling of $T$ and $W$ together with the parametrization \eqref{eq:tildenontilde}. The 4D curve appears at the order $\beta^N$
\begin{align}
0 = 
\left( \prod_{i=1}^N (v - (N_N + L_i) ) \right) t^{N-1}
+ \cdots + \prod_{i=1}^N (v - M_i).
\end{align}
In order to bring the above expression into Gaiotto's form, we first shift $v$ as
\begin{align}
v = v_{\rm new} + f(t)
\end{align}
so that the $v^{N-1}$ term disappears. Moreover, for a such term to vanish at $t \to 0$ and $t \to \infty$, the constraints
\begin{align}
f(0) = 0, \qquad f(\infty) = N_N
\end{align}
have to be satisfied. Under these conditions, we find the asymptotic behavior
\begin{align}
v_{\rm new} \sim M_i \quad (t \to 0),
\qquad \qquad
v_{\rm new} \sim L_i \quad (t \to \infty),
\end{align}
which imply that $x \equiv v_{\rm new} / t$ has poles at $t=0$ and $t=\infty$. The residues at these poles are $M_i$ and $L_i$, respectively. Furthermore, we expect the curve to have a pole at $t=1$ with the residue $N_i$ as well, but this is less trivial to show.

\section{Nekrasov partition functions from topological strings}
\label{sec:topologicalstrings}

The main result of this section is the derivation of the topological string partition functions for the $T_N$ junctions.
We read them off from the corresponding toric diagram using the refined topological vertex \cite{Iqbal:2007ii} (see also \cite{Taki:2007dh}). We begin the section by reviewing the rules of the topological vertex construction and apply them to the $SU(2)$ gauge theories with $N_f=0,\ldots, 4$. By comparing them with the corresponding 5D superconformal indices of \cite{Kim:2012gu}, we discover the presence of extra non-full spin content contributions in the topological string partition functions. In subsection \ref{subsec:Normalizing}, we argue that these non-full spin content contributions are of stringy nature and that they should be removed in order to obtain the correct Nekrasov partition functions that reproduce the 5D superconformal indices. With this knowledge at hand, we compute the topological string partition function for the $T_3$ junction and compare its index with the known results. We then generalize this by calculating the partition functions for the general junctions and finish by deriving a product formula for the partition function of the simplest junction $T_2$.

\subsection{From toric diagrams to partition functions}
\label{sec:rulestopologicalstring}

The purpose of this subsection is to review the rules for reading off the refined topological string partition function from a dual toric diagram. We begin by defining the main building blocks that will enter the partition function. Similar to Feynman diagrams, the procedure associates functions to the edges and to the vertices of the dual toric diagram, called \textit{edge} and \textit{vertex factors} respectively. The full partition function is then obtained as the product of all the edge and vertex factors, summed over all possible partitions associated with the internal edges. For a description of the computation of the topological partition function in the unrefined case, the reader is referred to \cite{Bao:2013wqa}.

First we will give a couple of preliminary definitions. The $\Omega$ background parameters $\ft$ and $\fq$ that enter the refined topological string partition function are
\begin{equation}
\fq=e^{-\beta \epsilon_1},\qquad \ft=e^{\beta \epsilon_2}.
\end{equation}
Letting $\lambda=(\lambda_1, \lambda_2, \ldots, )$ be a partition, we define
\begin{equation}
|\lambda|\defeq\sum_{(i,j)\in\lambda}1=\sum_{i=1}^{\ell(\lambda)}\lambda_i,\qquad ||\lambda||^2\defeq\sum_{(i,j)\in\lambda}\lambda_i=\sum_{i=1}^{\ell(\lambda)}\lambda_i^2.
\end{equation}
Given a toric diagram, the determination of the topological string partition function is done in several steps that we shall now describe carefully.
\begin{enumerate}
\item We begin by drawing the dual toric diagram and picking a preferred direction, which has to be chosen in a such a way that each vertex is connected to exactly one edge that points in the preferred direction. In our illustrations, the edges along the preferred direction are marked by two red strips.
\item Each edge is associated with a partition $\lambda$, an arrow pointing towards one of the two vertices of the edge, a vector $v$ with integer coefficients and an integer $\eta$. In addition, each edge is of a given length $\ell$. The partition $\lambda$ of an external line of the diagram has to be empty.
\begin{figure}[ht]
 \centering
  \includegraphics[height=2.5cm]{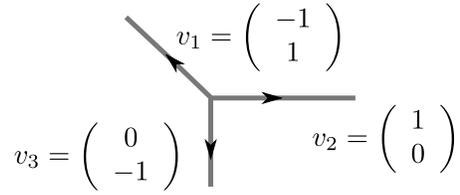}
  \caption{\it This figure illustrates the numbering of the edges.}
\label{fig:wedgesign}
\end{figure}
\item The arrows are chosen such that they are either all incoming or all outgoing at every vertex. The vectors $v$ with integer coefficients point in the direction of the arrows. At each vertex, we number the vectors of the adjacent edges \textit{clockwise} and impose the \textit{charge conservation condition}
\begin{equation}
\label{eq:chargeconservation}
\sum_{i=1}^3 v_i=0.
\end{equation}
Furthermore, we define the bilinear antisymmetric operation $\wedge$ as
\begin{equation}
v\wedge w\defeq  v_1w_2-v_2w_1,
\end{equation}
and require that
\begin{equation}
\label{eq:wedgesign}
v_i\wedge v_{i+1}=-1.
\end{equation}
We see in the example of figure~\ref{fig:wedgesign} that $v_1\wedge v_2=v_2\wedge v_3=v_3\wedge v_1=-1$.
\item To each edge with vector $v$, we associate an incoming vector $v_{\text{in}}$ and an outgoing vector $v_{\text{out}}$ as follows. First choose $v_{\text{in}}$ to be one of the two vectors that point towards one of the two vertices that the edge connects. Then $v_{\text{out}}$ is determined to be the one outgoing vector that has a positive scalar product with $v_{\text{in}}$.
\begin{figure}[h]
  \centering
  \includegraphics[ height=2.2cm]{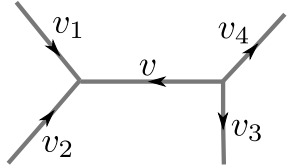}
\caption{\it This figure gives an example for the determination  $v_{\text{in}}$ and  $v_{\text{out}}$.}
\label{fig:framingdetermination}
\end{figure}
For example, in figure~\ref{fig:framingdetermination}, we can choose $v_1$ to be $v_{\text{in}}$, which then implies that $v_3$ is $v_{\text{out}}$. Alternatively, we can choose  $v_2$ to be $v_{\text{in}}$ in which case $v_4$ is $v_{\text{out}}$. We then associate a number $\eta$ to the edge of $v$ according to
\begin{equation}
\eta\defeq v_{\text{in}}\wedge v_{\text{out}}.
\end{equation}
Due to the constraints \eqref{eq:chargeconservation} and \eqref{eq:wedgesign}, the number $\eta$ does not depend on the choice of $v_{\text{in}}$.
\item Each end of an edge that is not oriented along the preferred direction is labeled by either $\ft$ or $\fq$ according to the following rules. We start by labeling the end of an arbitrary non-preferred segment by either $\ft$ or $\fq$. The other end of that segment is then labeled by the other variable.
As soon as one makes a choice for one edge, the associations for the rest of the diagram are uniquely determined as illustrated in figure \eqref{fig:diagramlabels}.
\begin{figure}[h]
\begin{center}
\includegraphics[height=4.5cm]{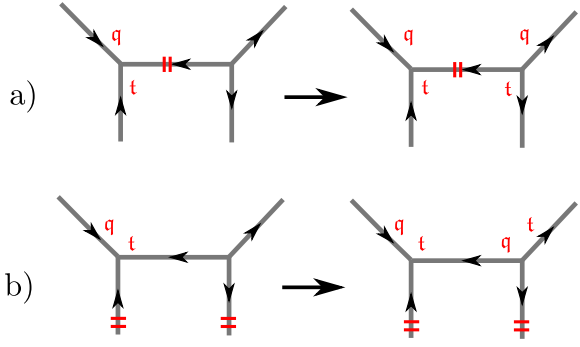}
\end{center}
\caption{\it This figure illustrates the way the $\ft$ and $\fq$ variables are placed.}
\label{fig:diagramlabels}
\end{figure}
However, the topological string partition function should not depend on the choice of the original placement of $\ft$ and $\fq$, see \cite{Iqbal:2008ra} for examples where this invariance has been checked.
\item We define the \textit{framing factor} functions, of which we use two\footnote{Our $\tilde{f}$ is different from the one of Iqbal and Koz{\c c}az \cite{Iqbal:2012mt}, specifically $\tilde{f}_{\nu}^{\text{our}}(\ft,\fq)=(\ft/\fq)^{|\nu|}\tilde{f}_{\nu}^{\text{IK}}(\ft,\fq)$. }, namely:
\begin{equation}
\label{eq:defframingfactors}
f_{\nu}(\ft,\fq)\defeq  (-1)^{|\nu|}\ft^{\frac{||\nu^t||^2}{2}}\fq^{-\frac{||\nu||^2}{2}}, \qquad \tilde{f}_{\nu}(\ft,\fq)\defeq  (-1)^{|\nu|}\ft^{\frac{||\nu^t||^2+|\nu|}{2}}\fq^{-\frac{||\nu||^2+|\nu|}{2}}.
\end{equation}
They obey the exchange relations
\begin{equation}
f_{\nu^t}(\fq,\ft)=f_{\nu}(\ft,\fq)^{-1}, \qquad \tilde{f}_{\nu^t}(\fq,\ft)=\tilde{f}_{\nu}(\ft,\fq)^{-1}.
\end{equation}
We then associate to the edge with partition $\lambda$ and number $\eta$ a \textit{framing factor}, which depends on the preferred direction and the association of $\ft$ and $\fq$ to the edges, as shown in figure~\ref{fig:framingfactors3}.
\begin{figure}[h]
\begin{center}
\includegraphics[height=2.1cm]{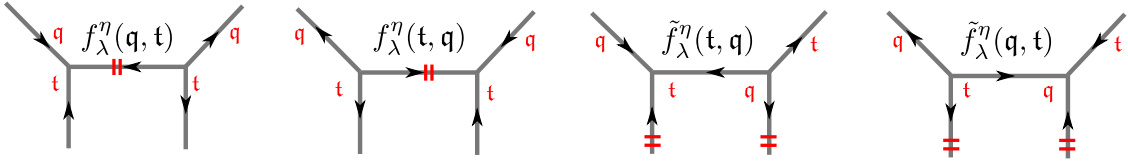}
\end{center}
\caption{\it This figure shows all the possible ways to associate a framing factor to an edge.}
\label{fig:framingfactors3}
\end{figure}

In addition, the K\"ahler moduli of an edge with vector $v=(p,q)$ is given by the equation
\begin{equation}
\label{eq:Qlength}
Q=\exp{\left(-\frac{\beta \ell}{2\pi \alpha^{\prime}\sqrt{p^2+q^2}} \right)},
\end{equation}
where $\ell$ is the length of the edge and $\beta$ is the circumference of the fifth dimensional circle.
The full \textit{edge factor} is then
\begin{equation}
\text{edge factor}\defeq (-Q)^{|\lambda|}\times \text{framing factor}.
\end{equation}
\item Within our conventions the \textit{refined topological vertex} reads:
\begin{equation}\label{eq:topvertex}
C_{\lambda\mu\nu}(\ft,\fq)\defeq \fq^{\frac{||\mu||^2+||\nu||^2}{2}}\ft^{-\frac{||\mu^t||^2}{2}}\tilde{Z}_{\nu}(\ft,\fq)\sum_{\eta}\left(\frac{\fq}{\ft}\right)^{\frac{|\eta|+|\lambda|-|\mu|}{2}}s_{\lambda^t/\eta}(\ft^{-\rho}\fq^{-\nu})s_{\mu/\eta}(\fq^{-\rho}\ft^{-\nu^t}),
\end{equation}
where we have used the functions $\tilde{Z}_{\nu}(\ft,\fq)$ defined in equation~\eqref{eq:mainfunctionsdefinitions} in the appendix.
The $s_{\lambda/\mu}(\textbf{x})$ are the skew-Schur functions of the possibly infinite vector $\textbf{x}=(x_1,\ldots)$. We remind that for a partition $\nu$, the vector  $\ft^{-\rho}\fq^{-\nu}$ is given by
\begin{equation}
\ft^{-\rho}\fq^{\nu}=(\ft^{\frac{1}{2}}\fq^{-\nu_1},\ft^{\frac{3}{2}}\fq^{-\nu_2},\ft^{\frac{5}{2}}\fq^{-\nu_3}, \ldots).
\end{equation}
To each vertex we associate a \textit{vertex factor}, namely a topological vertex function $C_{\lambda_1\lambda_2\lambda_3}$, where the $\lambda_i$ are the partitions associated to the lines connected to the vertex, counted \emph{clockwise} around the vertex such that \textit{the last edge is oriented along the preferred direction}. If the arrow of a segment is incoming towards the vertex, then the associated partition gets transposed in the topological vertex.
\begin{figure}[h]
\centering
\includegraphics[height=2.2cm]{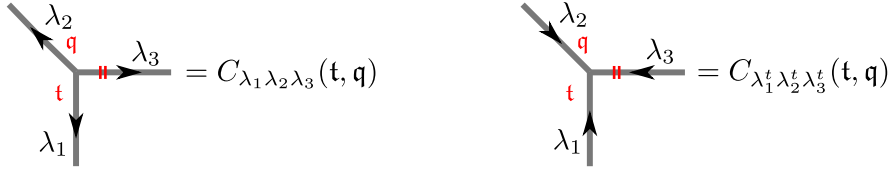}
\caption{\it The direction of the arrows determines whether the partitions or their transpose enter the vertex factor. Counting clockwise with the preferred direction being last, if $\ft$ is associated with the end of the first edge and $\fq$ with the end of the second, then the variables enter in the order $\ft,\fq$ in the vertex factor. }
\label{fig:clockwise}
\end{figure}
Whether we write $C_{\lambda_1\lambda_2\lambda_3}(\ft,\fq)$ or $C_{\lambda_1\lambda_2\lambda_3}(\fq,\ft)$ depends on the variables associated with the ends on the segments as illustrated in figure~\ref{fig:clockwise}.
\end{enumerate}

Having performed all the steps in the list above, we can finally write the topological string partition function as a sum over the partitions $\{\lambda_1,\cdots, \lambda_M\}$ of the $M$ internal edges of the toric diagram
\begin{equation}
Z=\sum_{\lambda_1,\cdots,\lambda_M}\  \prod_{\text{edges}}\text{edge factor}\times \prod_{\text{vertices}}\text{vertex factor} .
\end{equation}

\subsection{From zero to four flavors}

In this subsection, we shall compute the topological string partition function for SU(2) gauge theory with the number of flavors ranging from zero to four. Using the same approach as \cite{Iqbal:2012xm}, we shall then calculate the resulting index and compare with the results of \cite{Kim:2012gu}. In so doing, we find that it becomes necessary, starting at $N_f=2$, to divide out the topological string partition function by a factor that does not depend on the partitions entering the computation of the instantonic part.

\begin{figure}[ht]
\begin{center}
\includegraphics[height=4cm]{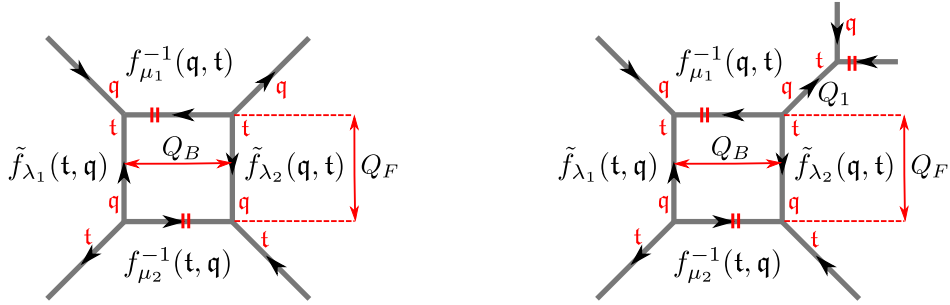}
\end{center}
\caption{\it The diagrams for $N_f=0$ and $N_f=1$. }
\label{fig:diagramNf01}
\end{figure}

\begin{figure}[ht]
\begin{center}
\includegraphics[height=4cm]{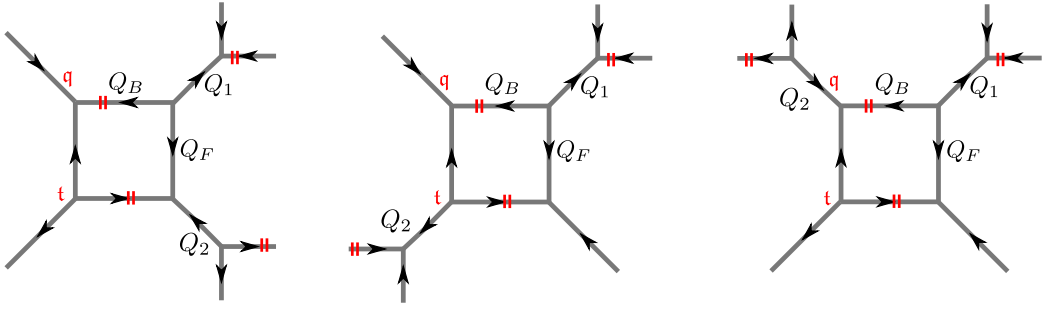}
\end{center}
\caption{\it The three different toric diagrams for SU(2) with two flavors. The non-trivial framing factors are identical to the one flavor case. To remove clutter, we denote only two of the positions of $\ft$ and $\fq$ in the diagram.}
\label{fig:diagramNf2}
\end{figure}

Let us warm up with the simple case of pure SU(2) gauge theory, for which the index was computed using topological string partition function in \cite{Iqbal:2012xm}. The diagram is depicted in figure \ref{fig:diagramNf01} and the partitions of the internal lines are named, starting from the left one and going clockwise, $\lambda_1$, $\mu_1$, $\lambda_2$ and $\mu_2$.
Instead of giving the lengths of the internal lines of the toric diagram, we prefer instead to directly give the K\"ahler parameter. Thus, for example, the edge factor of the top horizontal edge is $(-Q_B)^{|\mu_1|}f_{\mu_1}^{-1}(\fq,\ft)$. The topological string partition function is then given, according to the rules in section \ref{sec:rulestopologicalstring}, by the expression:
\begin{multline}
\label{eq:partfunction0before}
Z_{0}=\sum_{\boldsymbol{\lambda},\boldsymbol{\mu}}\left(-Q_F\right)^{|\lambda_1|}\tilde{f}_{\lambda_1}(\ft,\fq)\left(-Q_F\right)^{|\lambda_2|}\tilde{f}_{\lambda_2}(\fq,\ft)\left(-Q_B\right)^{|\mu_1|}f_{\mu_1}^{-1}(\fq,\ft)\left(-Q_B\right)^{|\mu_2|}f_{\mu_2}^{-1}(\ft,\fq)\\\times C_{\lambda_1^t\emptyset\mu_1^t}(\ft,\fq)C_{\emptyset\lambda_1\mu_2}(\ft,\fq)C_{\lambda_2^t \emptyset\mu_2^t}(\fq,\ft)C_{\emptyset\lambda_2\mu_1}(\fq,\ft) ,
\end{multline}
where $\boldsymbol{\lambda}$ is a vector of partitions $(\lambda_1,\lambda_2,\ldots)$ and similarly for $\boldsymbol{\mu}$ .
In a similar manner, we can also write down the partition function for $SU(2)$ $N_f=1$, whose diagram is depicted on the right in figure~\ref{fig:diagramNf01}. For one flavor, it does not matter on which external leg of the toric diagram we attach the flavor brane and furthermore, the framing factor of the corresponding edge is trivial. The resulting partition function reads
\begin{multline}
\label{eq:partfunction1before}
Z_{1}=\sum_{\boldsymbol{\lambda},\boldsymbol{\mu},\boldsymbol{\nu}}\left(-Q_F\right)^{|\lambda_1|}\tilde{f}_{\lambda_1}(\ft,\fq)\left(-Q_F\right)^{|\lambda_2|}\tilde{f}_{\lambda_2}(\fq,\ft)\left(-Q_B\right)^{|\mu_1|}f_{\mu_1}^{-1}(\fq,\ft)\left(-Q_B\right)^{|\mu_2|}f_{\mu_2}^{-1}(\ft,\fq) \\\times\left(-Q_1\right)^{|\nu_1|} C_{\lambda_1^t\emptyset\mu_1^t}(\ft,\fq)C_{\emptyset\lambda_1\mu_2}(\ft,\fq) C_{\lambda_2^t \emptyset\mu_2^t}(\fq,\ft)C_{\nu_1\lambda_2\mu_1}(\fq,\ft)C_{\nu_1^t\emptyset\emptyset}(\ft,\fq).
\end{multline}
Before simplifying the sums in \eqref{eq:partfunction0before} and \eqref{eq:partfunction1before}, let us consider the case of $N_f=2,3,4$. At two flavors, there are three a priori inequivalent toric diagrams as shown in figure \ref{fig:diagramNf2}.
The partition function corresponding to the leftmost diagram in figure~\ref{fig:diagramNf2} is given by
\begin{align}
\label{eq:partfunction2before}
Z_{2}=&\sum_{\boldsymbol{\lambda},\boldsymbol{\mu},\boldsymbol{\nu}}\left(-Q_F\right)^{|\lambda_1|}\tilde{f}_{\lambda_1}(\ft,\fq)\left(-Q_F\right)^{|\lambda_2|}\tilde{f}_{\lambda_2}(\fq,\ft)\left(-Q_B\right)^{|\mu_1|}f_{\mu_1}^{-1}(\fq,\ft)\left(-Q_B\right)^{|\mu_2|}f_{\mu_2}^{-1}(\ft,\fq) \nonumber\\&\times\left(-Q_1\right)^{|\nu_1|}\left(-Q_2\right)^{|\nu_2|} C_{\lambda_1^t\emptyset\mu_1^t}(\ft,\fq)C_{\emptyset\lambda_1\mu_2}(\ft,\fq) C_{\lambda_2^t \nu_2^t\mu_2^t}(\fq,\ft)C_{\nu_1\lambda_2\mu_1}(\fq,\ft)\\&\times C_{\nu_1^t\emptyset\emptyset}(\ft,\fq)C_{\emptyset \nu_2\emptyset}(\ft,\fq), \nonumber
\end{align}
the one for the center diagram in figure \ref{fig:diagramNf2} is
\begin{align}
\label{eq:partfunction2primebefore}
Z_{2}^{\prime}=&\sum_{\boldsymbol{\lambda},\boldsymbol{\mu},\boldsymbol{\nu}}\left(-Q_F\right)^{|\lambda_1|}\tilde{f}_{\lambda_1}(\ft,\fq)\left(-Q_F\right)^{|\lambda_2|}\tilde{f}_{\lambda_2}(\fq,\ft)\left(-Q_B\right)^{|\mu_1|}f_{\mu_1}^{-1}(\fq,\ft)\left(-Q_B\right)^{|\mu_2|}f_{\mu_2}^{-1}(\ft,\fq) \nonumber\\&\times\left(-Q_1\right)^{|\nu_1|}\left(-Q_2\right)^{|\nu_2|} C_{\lambda_1^t\emptyset\mu_1^t}(\ft,\fq)C_{\nu_2\lambda_1\mu_2}(\ft,\fq) C_{\lambda_2^t \emptyset\mu_2^t}(\fq,\ft)C_{\nu_1\lambda_2\mu_1}(\fq,\ft)\\&\times C_{\nu_1^t\emptyset\emptyset}(\ft,\fq)C_{\nu_2^t\emptyset \emptyset}(\fq,\ft),\nonumber
\end{align}
and the one for the rightmost diagram is
\begin{align}
\label{eq:partfunction2secondbefore}
Z_{2}^{\prime\prime}=&\sum_{\boldsymbol{\lambda},\boldsymbol{\mu},\boldsymbol{\nu}}\left(-Q_F\right)^{|\lambda_1|}\tilde{f}_{\lambda_1}(\ft,\fq)\left(-Q_F\right)^{|\lambda_2|}\tilde{f}_{\lambda_2}(\fq,\ft)\left(-Q_B\right)^{|\mu_1|}f_{\mu_1}^{-1}(\fq,\ft)\left(-Q_B\right)^{|\mu_2|}f_{\mu_2}^{-1}(\ft,\fq) \nonumber\\&\times\left(-Q_1\right)^{|\nu_1|}\left(-Q_2\right)^{|\nu_2|} C_{\lambda_1^t\nu_2^t\mu_1^t}(\ft,\fq)C_{\emptyset\lambda_1\mu_2}(\ft,\fq) C_{\lambda_2^t \emptyset\mu_2^t}(\fq,\ft)C_{\nu_1\lambda_2\mu_1}(\fq,\ft)\\&\times C_{\nu_1^t\emptyset\emptyset}(\ft,\fq)C_{\emptyset\nu_2\emptyset}(\ft,\fq) .\nonumber
\end{align}

\begin{figure}[ht]
\begin{center}
\includegraphics[height=4cm]{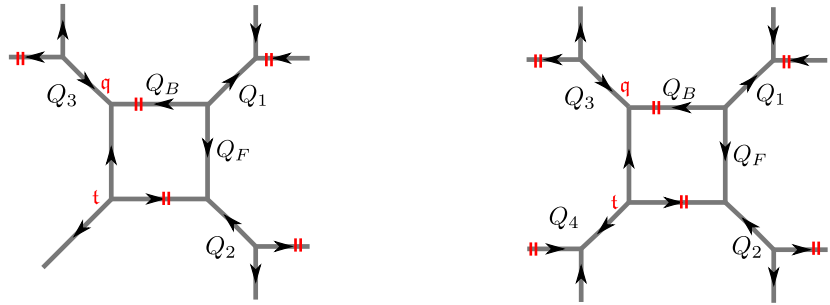}
\end{center}
\caption{\it The toric diagrams for SU(2) with $N_f=3$ and $N_f=4$. }
\label{fig:diagramNf34}
\end{figure}

The cases of three and four flavors are similar to the ones of zero and one flavor since there is essentially only one inequivalent diagram for each, as illustrated by figure \ref{fig:diagramNf34}. The partition functions read
\begin{align}
\label{eq:partfunction3before}
Z_{3}=&\sum_{\boldsymbol{\lambda},\boldsymbol{\mu},\boldsymbol{\nu}}\left(-Q_F\right)^{|\lambda_1|}\tilde{f}_{\lambda_1}(\ft,\fq)\left(-Q_F\right)^{|\lambda_2|}\tilde{f}_{\lambda_2}(\fq,\ft)\left(-Q_B\right)^{|\mu_1|}f_{\mu_1}^{-1}(\fq,\ft)\left(-Q_B\right)^{|\mu_2|}f_{\mu_2}^{-1}(\ft,\fq) \nonumber\\&\times\left(-Q_1\right)^{|\nu_1|}\left(-Q_2\right)^{|\nu_2|}\left(-Q_3\right)^{|\nu_3|} C_{\lambda_1^t\nu_3^t\mu_1^t}(\ft,\fq)C_{\emptyset\lambda_1\mu_2}(\ft,\fq) C_{\lambda_2^t \nu_2^t\mu_2^t}(\fq,\ft)\\&\times C_{\nu_1\lambda_2\mu_1}(\fq,\ft)C_{\nu_1^t\emptyset\emptyset}(\ft,\fq)C_{\emptyset \nu_2\emptyset}(\ft,\fq)C_{\emptyset \nu_3\emptyset}(\fq,\ft)\nonumber
\end{align}
and
\begin{align}
\label{eq:partfunction4before}
Z_{4}=&\sum_{\lambda\mu\nu}\left(-Q_F\right)^{|\lambda_1|}\tilde{f}_{\lambda_1}(\ft,\fq)\left(-Q_F\right)^{|\lambda_2|}\tilde{f}_{\lambda_2}(\fq,\ft)\left(-Q_B\right)^{|\mu_1|}f_{\mu_1}^{-1}(\fq,\ft)\left(-Q_B\right)^{|\mu_2|}f_{\mu_2}^{-1}(\ft,\fq) \nonumber\\&\times\left(-Q_1\right)^{|\nu_1|}\left(-Q_2\right)^{|\nu_2|}\left(-Q_3\right)^{|\nu_3|}\left(-Q_4\right)^{|\nu_4|} C_{\lambda_1^t\nu_3^t\mu_1^t}(\ft,\fq)C_{\nu_4\lambda_1\mu_2}(\ft,\fq)\\&\times C_{\lambda_2^t \nu_2^t\mu_2^t}(\fq,\ft)  C_{\nu_1\lambda_2\mu_1}(\fq,\ft)C_{\nu_1^t\emptyset\emptyset}(\ft,\fq)C_{\emptyset \nu_2\emptyset}(\ft,\fq)C_{\emptyset \nu_3\emptyset}(\fq,\ft)C_{ \nu_4^t\emptyset\emptyset}(\fq,\ft).\nonumber
\end{align}

To simplify the expressions \eqref{eq:partfunction0before} to \eqref{eq:partfunction4before}, it is convenient to define a function $\calR_{\lambda\mu}$ as
\begin{equation}
\calR_{\lambda\mu}(Q;\ft,\fq)\defeq \prod_{i,j=1}^{\infty}\left(1-Q \ft^{i-\frac{1}{2}-\lambda_j}\fq^{j-\frac{1}{2}-\mu_i}\right).
\end{equation}
We can use this function to express in a compact way two well known identities involving Schur functions, namely
\begin{multline}
\label{eq:schuridentity1}
\sum_{\lambda}Q^{|\lambda|}s_{\lambda/\mu_1}(\ft^{-\rho}\fq^{-\nu_1})s_{\lambda^t/\mu_2}(\fq^{-\rho}\ft^{-\nu_2})=\\=\calR_{\nu_2\nu_1}(-Q;\ft,\fq)\sum_{\lambda}Q^{|\mu_1|+|\mu_2|-|\lambda|}s_{\mu_2^t/\lambda}(\ft^{-\rho}\fq^{-\nu_1})s_{\mu_1^t/\lambda^t}(\fq^{-\rho}\ft^{-\nu_2}) ,
\end{multline}
as well as
\begin{multline}
\label{eq:schuridentity2}
\sum_{\lambda}Q^{|\lambda|}s_{\lambda/\mu_1}(\ft^{-\rho}\fq^{-\nu_1})s_{\lambda/\mu_2}(\fq^{-\rho}\ft^{-\nu_2})=\\=\calR_{\nu_2\nu_1}(Q;\ft,\fq)^{-1}\sum_{\lambda}Q^{|\mu_1|+|\mu_2|-|\lambda|}s_{\mu_2/\lambda}(\ft^{-\rho}\fq^{-\nu_1})s_{\mu_1/\lambda}(\fq^{-\rho}\ft^{-\nu_2}) .
\end{multline}
If either of the partitions $\mu_i$ is empty, equations \eqref{eq:schuridentity1} and \eqref{eq:schuridentity2} allow us to simplify the sums and replace them by an expression involving infinite products. Acting iteratively, the sums in \eqref{eq:partfunction0before} to \eqref{eq:partfunction1before} can be reduced to sums over the two partitions $\mu_1$ and $\mu_2$ that are associated with the internal lines oriented along the preferred direction.

In order to keep the formulas of the next part short and tidy, we introduce the shorthand
\begin{equation}
\tilde{Z}_{\mu_1\mu_2}(\ft,\fq)\defeq \ft^{||\mu_1||^2}\fq^{||\mu_2||^2}\tilde{Z}_{\mu_1}(\fq,\ft)\tilde{Z}_{\mu_2}(\ft,\fq)\tilde{Z}_{\mu_1^t}(\ft,\fq)\tilde{Z}_{\mu_2^t}(\fq,\ft).
\end{equation}
Furthermore, \textit{we will often drop the explicit dependence of $\ft$ and $\fq$}, i.e. we shall write
\begin{equation}
f(Q)\equiv f(Q;\ft,\fq).
\end{equation}
Taking now the topological string partition functions \eqref{eq:partfunction0before} to \eqref{eq:partfunction4before} and applying the identities  \eqref{eq:schuridentity1} and \eqref{eq:schuridentity2}, we obtain the result
\begin{equation}
\label{eq:partfunction0after}
Z_0=\sum_{\boldsymbol{\mu}}Q_B^{|\mu_1|+|\mu_2|}\frac{\tilde{Z}_{\mu_1\mu_2}}{\calR_{\mu_2^t\mu_1^t}(Q_F\sqrt{\frac{\fq}{\ft}})\calR_{\mu_2^t\mu_1^t}(Q_F\sqrt{\frac{\ft}{\fq}})},
\end{equation}
for pure SU(2),
\begin{equation}
\label{eq:partfunction1after}
Z_1=\sum_{\boldsymbol{\mu}}Q_B^{|\mu_1|+|\mu_2|}\frac{\tilde{Z}_{\mu_1\mu_2}\calR_{\mu_1\emptyset}(Q_1)\calR_{\mu_2^t \emptyset}(Q_1Q_F)}{\calR_{\mu_2^t\mu_1^t}(Q_F\sqrt{\frac{\fq}{\ft}})\calR_{\mu_2^t\mu_1^t}(Q_F\sqrt{\frac{\ft}{\fq}})},
\end{equation}
for $N_f=1$. In the $N_f=2$ case, we find
\begin{equation}
\label{eq:partfunction2after}
Z_{2}=\sum_{\boldsymbol{\mu}}Q_B^{|\mu_1|+|\mu_2|}\frac{\tilde{Z}_{\mu_1\mu_2}\calR_{\mu_1\emptyset}(Q_1)\calR_{\emptyset\mu_2}(Q_2)\calR_{\mu_2^t\emptyset}(Q_1 Q_F)\calR_{\emptyset\mu_1^t}(Q_2 Q_F)}{\calR_{\mu_2^t\mu_1^t}(Q_F\sqrt{\frac{\fq}{\ft}})\calR_{\mu_2^t\mu_1^t}(Q_F\sqrt{\frac{\ft}{\fq}})\calR_{\emptyset\emptyset}(Q_1 Q_2 Q_F\sqrt{\frac{\ft}{\fq}})},
\end{equation}
for the situation on the left in figure \ref{fig:diagramNf2},
\begin{equation}
\label{eq:partfunction2primeafter}
Z_{2}^{\prime}=\calR_{\emptyset\emptyset}(Q_1 Q_2 Q_F\sqrt{\frac{\ft}{\fq}})Z_{2}
\end{equation}
for the one depicted in the center and
\begin{equation}
\label{eq:partfunction2secondafter}
Z_{2}^{\prime\prime}=\sum_{\boldsymbol{\mu}}Q_B^{|\mu_1|+|\mu_2|} \frac{\tilde{Z}_{\mu_1\mu_2} \calR_{\mu_1\emptyset}(Q_1)\calR_{\mu_1\emptyset}(Q_2)\calR_{\mu_2^t\emptyset}(Q_1 Q_F)\calR_{\mu_2^t\emptyset}(Q_2 Q_F)}{\calR_{\mu_2^t\mu_1^t}\big(Q_F\sqrt{\frac{\fq}{\ft}}\big)\calR_{\mu_2^t\mu_1^t}\big(Q_F\sqrt{\frac{\ft}{\fq}}\big)}
\end{equation}
for the one on the right. Performing the sum in \eqref{eq:partfunction2secondafter} on a computer for partitions $\mu$ with $|\mu|<M$ for some large integer $M$, we experimentally uncover the identity
\begin{equation}
Z_{2}^{\prime}=\calR_{\emptyset\emptyset}\big(Q_1 Q_2 Q_B\sqrt{\frac{\fq}{\ft}}\big)Z_{2}^{\prime\prime}.
\end{equation}
For $N_f=3$, the partition function \eqref{eq:partfunction3before} becomes
\begin{equation}
\label{eq:partfunction3after}
Z_{3}=\sum_{\boldsymbol{\mu}}Q_B^{|\mu_1|+|\mu_2|} \frac{\tilde{Z}_{\mu_1\mu_2}\prod_{i=1}^2\calR_{\mu_1\emptyset}(Q_{2i-1})\calR_{\mu_2^t\emptyset}(Q_{2i-1} Q_F)      \calR_{\emptyset\mu_2}(Q_2)\calR_{\emptyset\mu_1^t}(Q_2 Q_F)}{\calR_{\mu_2^t\mu_1^t}(Q_F\sqrt{\frac{\fq}{\ft}})\calR_{\mu_2^t\mu_1^t}\big(Q_F\sqrt{\frac{\ft}{\fq}}\big)\calR_{\emptyset\emptyset}\big(Q_1 Q_2 Q_F\sqrt{\frac{\ft}{\fq}}\big)},
\end{equation}
while for four flavors we get the similar expression
\begin{equation}
\label{eq:partfunction4after}
Z_{4}=\sum_{\boldsymbol{\mu}}Q_B^{|\mu_1|+|\mu_2|}\frac{\tilde{Z}_{\mu_1\mu_2}\prod_{i=1}^2\calR_{\mu_1\emptyset}(Q_{2i-1})\calR_{\emptyset\mu_2}(Q_{2i})\calR_{\emptyset\mu_1^t}(Q_{2i}Q_F)\calR_{\mu_2^t\emptyset}(Q_{2i-1}Q_F)}{\calR_{\mu_2^t\mu_1^t}\big(Q_F\sqrt{\frac{\fq}{\ft}}\big)
\calR_{\mu_2^t\mu_1^t}\big(Q_F\sqrt{\frac{\ft}{\fq}}\big)\calR_{\emptyset\emptyset}\big(Q_1 Q_2 Q_F\sqrt{\frac{\ft}{\fq}}\big)\calR_{\emptyset\emptyset}\big(Q_3Q_4Q_F\sqrt{\frac{\fq}{\ft}}\big)}.
\end{equation}

We would now like to use equations \label{eq:partfunction0after} to \label{eq:partfunction4after} in order to extract the 5D superconformal index of protected states for the relevant gauge theory, as done in \cite{Iqbal:2012xm} for the zero flavor case. For this purpose, it is necessary to replace the variables $\ft$ and $\fq$ through $x$ and $y$ via
\begin{equation}
\ft=\frac{y}{x}, \qquad \fq = x y,
\end{equation}
and then carefully expand the partition functions for small $x$. This is problematic due to the presence of infinite products in the definition of $\calR_{\lambda\mu}$. To simplify the issue, we separate the functions $\calR_{\lambda\mu}$ into a finite product $\calN_{\lambda\mu}$ that depends on the partitions and an infinite product $\calM$ that doesn't according to equation \eqref{eq:mainfunctionsdefinitions}.
It is now trivial to replace $\ft$ and $\fq$ by $x$ and $y$ in $\calN_{\lambda\mu}(Q;\ft,\fq)$ and expand for small $x$. On the other hand, the function $\calM$ needs to be reformulated so that a power series expansion for small $x$ becomes possible. This is done by the use of the following analytic continuation rule for arbitrary $Q$:
\begin{equation}
\label{eq:magic}
\prod_{m=1}^{\infty}\left(1-Q\fq^{-m+\frac{1}{2}}\right)^{-1}\rightarrow\prod_{m=1}^{\infty}\left(1-Q\fq^{m-\frac{1}{2}}\right).
\end{equation}
In order to compute the superconformal index, we need in addition to have a ``complex conjugation'', which for an arbitrary function $f$ of the variables $\ft$, $\fq$ and the K\"ahler parameters is given by
\begin{equation}
\label{eq:complexconjugation}
\bar{f}(Q_1,\ldots,Q_M;\ft,\fq)\defeq f(Q_1^{-1},\ldots,Q_M^{-1};\ft^{-1},\fq^{-1}).
\end{equation}
Of course, the above is not a complex conjugation for the case in which the variables are not of unit modulus, and in particular not in the case we are interested in, for which $\ft\rightarrow \infty$ and $\fq\rightarrow 0$. Thus, it is better to simply view \eqref{eq:complexconjugation} as the definition of a new function $\bar{f}$.

Applying the replacement rule \eqref{eq:magic} to $\calM$ and $\overline{\calM}$, we obtain
\begin{equation}
\label{eq:sigmaxy}
\calM(Q;x,y)=\prod_{i,j=1}^{\infty}(1-Qx^{i+j}y^{i-j}), \qquad \overline{\calM}(Q;x,y)=\prod_{i,j=1}^{\infty}(1-Q^{-1}x^{i+j-2}y^{i-j}).
\end{equation}
In computing the index, we also need to expand the refined MacMahon function $M(\ft,\fq)=\calM(1;\ft,\fq)$ and its conjugate $\overline{M}(\ft,\fq)$ for small $x$. Applying \eqref{eq:magic} to $\overline{M}$ naively leads to a trivial zero that has to be removed as shown in \cite{Iqbal:2012xm}. We will then obtain the result
\begin{equation}
\overline{M}(x,y)=\prod_{\substack{i,j=1\\(i,j)\neq(1,1)}}^{\infty}\left(1-x^{i+j-2}y^{i-j}\right).
\end{equation}
After these careful preparations, we can finally spell out the procedure of computing the index. For any $N_f$, we parametrize the K\"ahler moduli $\boldsymbol{Q}=(Q_F,Q_B,Q_1,Q_2, \ldots)$ as
\begin{equation}
\label{eq:Kahlerparametrization}
Q_F=\tilde{A}^2, \qquad Q_B=x_0 \tilde{A}^2,\qquad Q_i=x_i\tilde{A}^{-1},
\end{equation}
where $\tilde{A}$ is the Coulomb modulus and $x_i$ will be determined later so that the symmetries become manifest. Then, we take the topological string partition functions $Z_a$ and their conjugate $\overline{Z}_a$, replace the functions $\calR_{\lambda\mu}$ by $\calM$ and $\calN_{\lambda\mu}$ as determined by \eqref{eq:mainfunctionsdefinitions} and expand in a power series in $x$ by using \eqref{eq:sigmaxy}. The index is defined as the contour integral over the Coulomb moduli \cite{Iqbal:2012xm}
\begin{equation}
\label{eq:defindex}
\mathcal{I}=\frac{1}{2}\oint_{|\tilde{A}|=\epsilon}\frac{d\tilde{A}}{2\pi i \tilde{A}}M(x,y)\overline{M}(x,y) Z(\boldsymbol{Q};x,y)\overline{Z}(\boldsymbol{Q};x,y),
\end{equation}
where $\epsilon>0$ is small enough so that the contour integral in \eqref{eq:defindex} only picks up the residue at zero\footnote{This statement is correct only after expanding in terms of $x$.}.

For the $N_f=0$ and $N_f=1$ cases, application of \eqref{eq:defindex} leads\footnote{In our normalization,  the su(2) characters $\chi_l(u)$ of the $l+1$ dimensional representation are given by $\chi_l(u)=\sum_{m=0}^lu^{l-2m}=u^{l}+u^{l-2}+\cdots +u^{-l}$.} to
\begin{eqnarray}
\label{eq:indexNf01}
\mathcal{I}_0&=&1 + \chi_2(u_1) x^2 + \chi_1(y)(1 +  \chi_2(u_1)) x^3 + (1 +  \chi_4(u_1) +  \chi_2(y)(1+ \chi_2(u_1))) x^4+\mathcal{O}(x^5),\nonumber\\
\mathcal{I}_1&=&1 + (1 + \chi_2(u_1)) x^2 + \chi_1(y)(2  + \chi_2(u_1)) x^3 \\&&+ (1 + \chi_4(u_1) - \chi_1(u_1)(u_2+u_2^{-1})+
    \chi_2(y)(2 + \chi_2(u_1))) x^4+\mathcal{O}(x^5), \nonumber
\end{eqnarray}
where we have set $x_0=u_1^2$ and $x_1=u_1^{-\frac{1}{2}}u_2^{\frac{1}{2}}$ in order to get full agreement with \cite{Kim:2012gu}. For the pure $SU(2)$ case, the global $U(1)$ symmetry is enhanced to $E_1=SU(2)$ and the index $\mathcal{I}_0$ is organized in representations of $SU(2)$ that depend on the fugacity $u_1$. In the $N_f=1$ case, the resulting enhanced global symmetry is $E_2=SU(2) \times U(1)$ and the variable $u_2$ that takes care of the $U(1)$ factor is to be identified with $e^{i\rho/2}$ appearing in equation (4.12) in \cite{Kim:2012gu}.

Starting with two flavors, we need to remove certain problematic factors from the partition functions in order to obtain the index. Specifically, for $N_f=2$, we set
\begin{equation}
Z_2^{\text{ren}}\defeq Z_2^{\prime}=\calM\big(Q_1 Q_2 Q_F\frac{\ft}{\fq}\big)^{-1}Z_{2}=\calM\big(Q_1 Q_2 Q_B\big)^{-1}Z_{2}^{\prime\prime}
\end{equation}
and compute the index $\mathcal{I}_2$ using $Z_2^{\text{ren}}$. Setting the parameter $x_i$ of \eqref{eq:Kahlerparametrization} equal to $x_0=u_1^2$, $x_1=u_1^{-\frac{1}{2}}u_2u_3^{\frac{1}{2}}$ and $x_2=u_1^{-\frac{1}{2}}u_2^{-1}u_3^{\frac{1}{2}}$, we obtain
\begin{equation}
\label{eq:indexNf2}
\begin{split}
\mathcal{I}_2=&1+\Big(1+\chi_2(u_1) +\chi_1(u_1)(u_3+u_3^{-1})  + \chi_2(u_2)\Big) x^2 \\ &+\Big(\chi_1(y)(2+\chi_2(u_1) +\chi_1(u_1)(u_3+u_3^{-1})  + \chi_2(u_2)\Big) x^3+\mathcal{O}(x^4),
\end{split}
\end{equation}
with $u_1$, $u_2$ and $u_3$ counting the charges of the Cartan generators of $SU(2)\times SU(2) \times U(1) \subset  SU(3) \times SU(2)=E_3$, as in \cite{Kim:2012gu}, equation (4.12) et seqq. To make the full $E_3=SU(3) \times  SU(2)$ symmetry manifest however, it is preferable to instead parametrize as
\begin{equation}
x_0=\frac{y_1}{y_2},\qquad  x_1=\sqrt{\frac{y_2}{y_3}} u, \qquad  x_2=\sqrt{\frac{y_2}{y_3}} u^{-1},
\end{equation}
 so that the fundamental representation of $SU(3)$ has the character $y_1+y_2+y_3$. The fugacity $u$ is the variable associated to $SU(2)$ and we also have $\prod_{i=1}^3y_i=1$

For three flavors, we claim that the renormalized partition function that gives the correct index is
\begin{equation}
Z_{3}^{\text{ren}}=\Big[\calM\big(Q_1Q_2Q_F\frac{\ft}{\fq}\big)\calM\big(Q_1Q_3Q_B\big)\Big]^{-1}Z_{3}  .
\end{equation}
Parametrizing the fugacities as
\begin{equation}
x_0=\frac{y_1}{y_2}, \quad x_1=\sqrt{\frac{y_2y_5}{y_3y_4}}, \quad  x_2=\sqrt{\frac{y_2y_4}{y_3y_5}}, \quad x_3=\sqrt{\frac{y_2y_3}{y_4y_5}},
\end{equation}
we recover the global symmetry $E_4=SU(5)$ of the index which reads
\begin{equation}
\begin{split}
\mathcal{I}_3=&1+\chi_{[1,0,0,1]}^{E_4}(\boldsymbol{y}) x^2 +\chi_1(y)\Big(1+\chi_{[1,0,0,1]}^{E_4}(\boldsymbol{y})\Big) x^3\\&+\Big(\chi_2(y)(1+\chi_{[1,0,0,1]}^{E_4}(\boldsymbol{y}))+1+\chi_{[2,0,0,2]}^{E_4}(\boldsymbol{y})\Big)+\mathcal{O}(x^5).
\end{split}
\end{equation}
In the above, $[a_1,a_2,a_3,a_4]$ are the Dynkin labels of $SU(5)$, with $[1,0,0,1]$ denoting the adjoint and $[2,0,0,2]$ the 200-dimensional self-dual representation.

In the four flavor case,  the renormalized partition function is
\begin{equation}
\label{eq:renNf4partitionfunction}
Z_{4}^{\text{ren}}=\Big[\calM\big(Q_1Q_3Q_B\big)\calM\big(Q_3Q_4Q_F\big)\calM\big(Q_2Q_4Q_B\frac{\ft}{\fq}\big)\calM\big(Q_1Q_2Q_F\frac{\ft}{\fq}\big)\Big]^{-1}Z_{4} ,
\end{equation}
as we shall justify in the next subsection by equation \eqref{eq:fourflavornormalization}. Parametrizing the fugacities as
\begin{equation}
x_0=\frac{y_1}{y_2}, \quad x_1=\sqrt{\frac{y_2y_5}{y_3y_4}}, \quad  x_2=\sqrt{\frac{y_2y_4}{y_3y_5}}, \quad x_3=\sqrt{\frac{y_2y_3}{y_4y_5}},\quad x_4=\sqrt{y_2y_3y_4y_5} ,
\end{equation}
we make the enhanced $E_5=Spin(10)$ symmetry of the index manifest. Specifically, for this choice of the fugacities, the character of the vector representation of $Spin(10)$ takes the standard form $\sum_{i=1}^5(y_i+y_i^{-1})$. Plugging \eqref{eq:renNf4partitionfunction} into \eqref{eq:defindex} leads to the following expression for the index
\begin{equation}
\mathcal{I}_4=1+\chi_{[0,1,0,0,0]}^{E_5}(\boldsymbol{y}) x^2 +\chi_1(y)\Big(1+\chi_{[0,1,0,0,0]}^{E_5}(\boldsymbol{y})\Big) x^3+\mathcal{O}(x^4),
\end{equation}
where $[a_1,\ldots, a_5]$ are the Dynkin labels of $Spin(10)$, with $[0,1,0,0,0]$ being the adjoint.

\subsection{Normalizing the partition functions}
\label{subsec:Normalizing}

As we have discovered experimentally a procedure to obtain the correct superconformal indices of the $E_{1,\cdots,5}$ SCFTs,
we will move on to a theoretical formulation of our findings.
We saw an extra factor appearing in the $N_f\geq 2$ theories ($E_{3,4,5}$), a phenomena deeply related to the stringy construction of the five dimensional SCFTs. Let us now give a brief review on the relevant background materials.

\begin{figure}[ht]
 \begin{center}
  \includegraphics[width=150mm,clip]{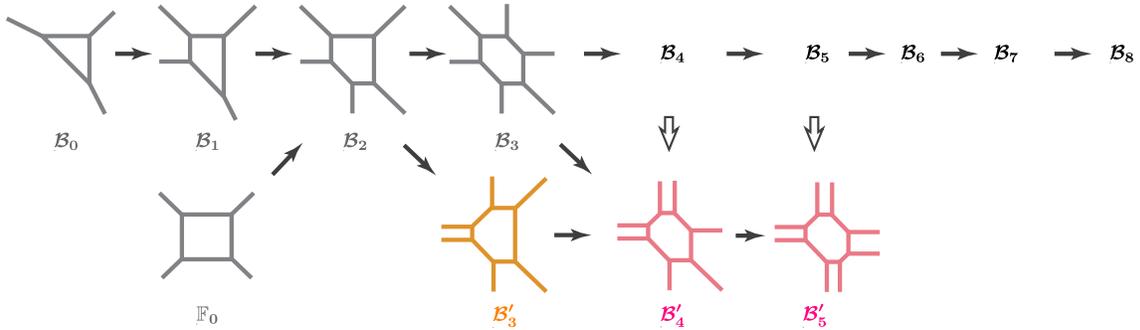}
 \end{center}
 \caption{\it The first line is the sequence of the del Pezzo surface $\mathcal{B}_k$, which is the blow-up of $\,\mathbb{CP}^2$ at $k$ generic points ($\,k=0,1,\cdots,8\,$). $\mathcal{B}_{0,1,2,3}$ and $\,\mathbb{F}_0$ (the five gray diagrams) are the toric del Pezzo surfaces which are the smooth toric Fano varieties in four dimensions. The gray, orange and red diagrams are toric surfaces. For $k\geq 4$, the del Pezzo surface is non-toric, but by setting the blow-up points on special positions we can get toric surfaces (the red diagrams). We can also obtain the two red diagrams from the $\,\mathbb{F}_2$ surface blown up at two points (the orange diagram).}
 \label{fig:delPezzo}
\end{figure}

Originally \cite{Ganor:1996mu,Seiberg:1996vs,Witten:1996qb,Morrison:1996na,Morrison:1996pp,Ganor:1996gu,Klemm:1996hh,Ganor:1996xd}, the 5D $E_n$ SCFTs were studied by considering the small $E_8$ instantons coming from M-theory on $S^1/\mathbb{Z}_2$, which arise when an M5-brane comes near the 9-brane at a boundary of the interval $S^1/\mathbb{Z}_2$.
By further compactifying this M-theory set-up on an additional circle,
we find Type IIA theory on the orientifold $S^1/\mathbb{Z}_2$ with D8-branes, which is called the Type I' superstring theory.
Seiberg in \cite{Seiberg:1996bd} derived certain properties of the $E_n$ SCFTs using this Type I' realization.

Let us consider a point on the moduli space of an elliptically fibered Calabi-Yau threefold where a compact four-cycle shrinks to a point. It is conjectured that F-theory compactified on the Calabi-Yau leads to the same $E_n$ theories \cite{DeWolfe:1999hj}. The $E_n$ theory appears when the del Pezzo surface $\mathcal{B}_n$ vanishes as the four-cycle. By compactifying this F-theory on a circle, we obtain M-theory on the same Calabi-Yau with an elliptic fiber of infinite size.

The 5D $E_n$ SCFT is thus obtained by compactifying M-theory on a Calabi-Yau which includes the del Pezzo surface $\mathcal{B}_n$ as a compact four-cycle. At the singular limit, the del Pezzo surface collapses to a point, the SCFT appears and the finite size four-cycle describes the deformation from the strongly-coupled fixed point.

For some of the $E_n$ theories, we can employ $(p,q)$ 5-brane construction and dual toric Calabi-Yau compactification to investigate the corresponding 5D fixed point theory. Actually, the 5D $E_n$ SCFTs with $n=0,1,2,3$ are constructed by these web configurations ($\mathbb{F}_0$ and $\mathcal{B}_n$), and we can compute their superconformal indices
by using the toric Calabi-Yau compactification \cite{Iqbal:2012xm}. At first sight, the blow-up points of $\mathcal{B}_{1,2,3}$ seem to be placed in a non-generic fashion, because they are located at the three corners of the toric diagram of $\mathbb{P}^2$. However, the $SL(3,\mathbb{C})$ symmetry of $\mathbb{P}^2$ enables us to place three points to any positions and we can therefore arrange the blow-up points to the three corners. The first four del Pezzo surfaces are thus toric varieties \cite{Hori:2003ic}.
This fact does not hold for higher del Pezzo surfaces $\mathcal{B}_{n\geq4}$ since we have exhausted the $SL(3,\mathbb{C})$ transformations, so that these del Pezzo surfaces do not have any toric description. The naive toric descriptions of \textit{del Pezzo-like surfaces} $\mathcal{B}_{n}'$ are given as the red diagrams in figure~\ref{fig:delPezzo}, but they are not genuine del Pezzo surfaces because the blown-up points are specially tuned. As we will see in the following subsections, the 5D theories arising from the diagrams colored in orange and red cause problem because they contain parallel external lines (branes), and in higher cases some of the external lines cross each other. This situation is a reflection of the non-toric nature of the corresponding del Pezzo surfaces. These external configurations then lead to extra states (which are stringy in general).

\begin{figure}[tbp]
 \begin{center}
  \includegraphics[width=140mm,clip]{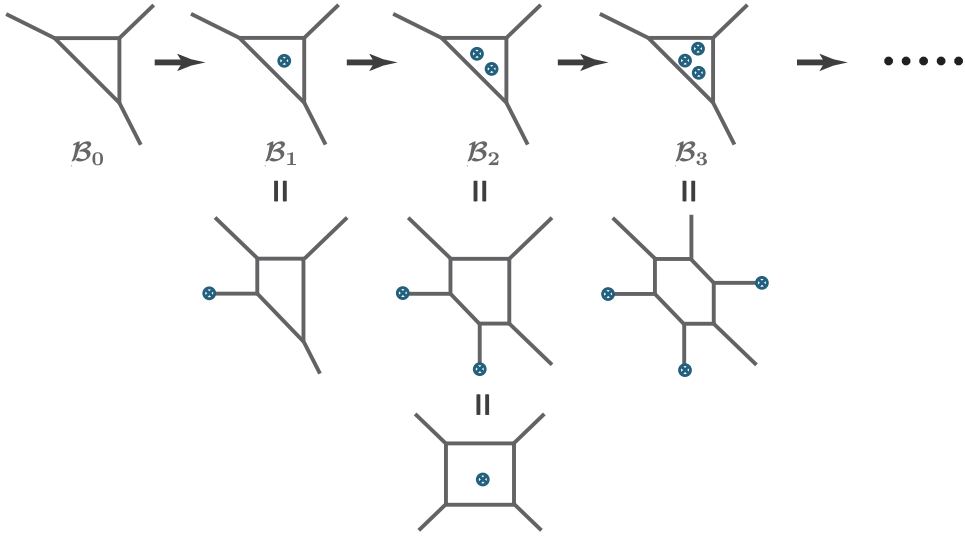}
 \end{center}
 \caption{\it This figure shows the 7-brane realization of blow-ups of the local $\mathbb{CP}^2$. By pulling a 7-brane out of the 5-brane loop, the positions of the corresponding blown up points will be tuned into special locations of the toric geometry $\mathbb{CP}^2$: the corners of the toric diagram.}
 \label{fig:blowup}
\end{figure}

This subtle situation has been resolved by introducing 7-branes into the 5-brane webs, so that the semi-infinite 5-branes become finite, ending on the 7-branes. See also section \ref{sec:7brane}. By moving a 7-brane inside the 5-brane loop, an 5-brane disappears and a 7-brane insertion appears in this new configuration. We can therefore blow up a del Pezzo surface at a generic point by adding a single 7-brane inside the 5-brane loop that describes a compact four-cycle. Figure~\ref{fig:blowup} illustrates the construction of the del Pezzo surfaces as the blow-up of the surface $\mathbb{CP}^2$.

\subsubsection{Our conjecture}

In this paper we compute the superconformal indices of $E_{3,4,5,6}$ theories by using a toric description. As we mentioned above, toric formulation of $\mathcal{B}_{3}'$ and $\mathcal{B}_{4,5,6}$ surfaces causes problems and we have to employ the alternative formulation based on 7-branes. Unfortunately we do not know any efficient way to compute the topological string partition functions of toric Calabi-Yau manifolds on 7-brane backgrounds. This means that we cannot compute the partition functions and the superconformal indices of the corresponding $E_{3,4,5,6}$ SCFTs based on the Iqbal-Vafa scheme \cite{Iqbal:2012xm}. However, we did find an efficient way to compute these partition functions and superconformal indices.
\textit{Our claim is that the partition function of a 5-brane web blown up by 7-branes is equal to that of the naive toric description after eliminating the so-called \textit{non-full spin contents}.
}
Schematically our proposal is
\begin{align}
\label{eq:mainconj}
Z_{\,\textrm{5 \& 7-brane web}}
=\frac{Z_{\textrm{toric}}}{Z_{\textrm{non-full spin}}}.
\end{align}
A 5-brane web describes a five dimensional minimal supersymmetric gauge theory, and its massive BPS states are characterized by the Wigner little group $SO(4)\simeq SU(2)_L\times SU(2)_R$. The label $(j_L,j_R)$ denotes the spin content. In the context of Calabi-Yau compactification, these BPS states come from M2-branes wrapping a two cycle inside the Calabi-Yau $C\in H(\textrm{CY},\mathbb{Z})$. The multiplicities $N_C^{(j_L,j_R)}$ of these states with labels $(j_L,j_R)$ and $C$ provide the Nekrasov instanton partition function of the resulting five dimensional theory
\begin{align}
Z(Q_C;\ft,\fq) &=\exp\Big[ \sum_{C\in H(\textrm{CY},\mathbb{Z})} \sum_{j_L,j_R} Q_C \frac{ (-1)^{2(j_L+j_R)}N_C^{(j_L,j_R)}
\textrm{Tr}_{j_R}\left( \frac{\fq}{\ft}\right)^{\sigma_{R,3}}
\textrm{Tr}_{j_L}\left({\fq}{\ft}\right)^{\sigma_{L,3}}
}{(\ft^{1/2}-\ft^{-1/2})(\fq^{1/2}-\fq^{-1/2})}\Big]\nonumber\\
&=Z(Q_C;\ft^{-1},\fq^{-1}) .
\end{align}
For various Calabi-Yau manifolds and their cycles $C$, this property is satisfied because we have the full spin content
for each spin $(j_L,j_R)$. This is however not always the case. The Wigner little group is mathematically the Lefschetz action \cite{Gopakumar:1998jq, Hori:2003ic} onto the fiber and base direction of (the cohomology class of) the moduli space $\tilde{\mathcal{M}}_C$ of the stable sheaves of the Calabi-Yau, which is a flat vector bundle over the moduli space ${\mathcal{M}}_C$ of the curves in class $C$. The spin content of the BPS state is then given by the decomposition of the cohomology as the representation space of the $SU(2)_L\times SU(2)_R$ Lefschetz action
\begin{align}
H^*(\tilde{\mathcal{M}}_C)
=\bigoplus_{j_L,j_R} \,N_C^{(j_L,j_R)}\big[ (j_L,j_R)\big].
\end{align}
However the existence of the Lefschetz action is not guaranteed in general, and in fact the cohomology is not a full representation space for some curves $C$ on which the moduli space of M2-brane $\mathcal{M}$ is non-compact.

\begin{figure}[ht]
 \begin{center}
  \includegraphics[width=10cm,clip]{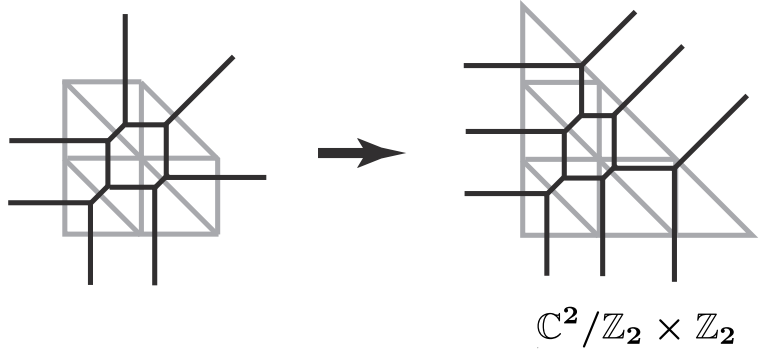}
 \end{center}
 \caption{\it The toric analogue of the del Pezzo surface $\mathcal{B}_6$ as a two point blow-up of the toric $\mathcal{B}_4$-like surface. This surface is the minimal resolution of the orbifold $\mathbb{C}^2/\mathbb{Z}_2\times \mathbb{Z}_2$.}
\label{fig:orbifold}
\end{figure}
In our scheme we can compute a typical non-toric del Pezzo surface $\mathcal{B}_6$ based on a 5-brane web and toric compactification. The key is the toric description of the $\mathbb{CP}^2$ blow-up at six specially placed points, see figure~\ref{fig:orbifold}. This geometry is the resolution of the supersymmetric orbifold $\mathbb{C}^2/\mathbb{Z}_2\times\mathbb{Z}_2$. The genuine del Pezzo surface $\mathcal{B}_6$ compactification is obtained by attaching the 7-branes to the ends of the external 5-branes in the dual 5-brane system and moving these 7-branes inside the 5-brane loop. It is difficult to study this compactification, but we have a nice toric description \eqref{eq:mainconj} and can compute the partition functions explicitly. In this paper, we thus utilize the simple toric description of $\mathcal{B}_6$ to compute the superconformal index of the worldvolume theory, and we find the appearance of the non-trivial $E_6$ SCFT on the five-dimensional space-time.

\subsubsection{Example: the toric description of $\mathcal{B}_5$ }

\begin{figure}[t]
 \begin{center}
  \includegraphics[height=5cm]{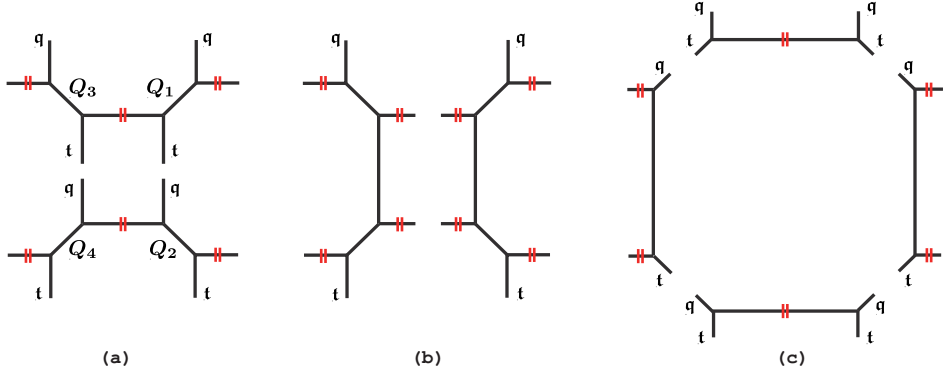}
 \end{center}
 \caption{\it In pictures (a) and (b), we show the four diagrams contributing to the non-full spin content. In (c), we have the flopped diagram, which is precisely equal to the non-full spin content partition function $Z^{\textrm{non-full}}$.}
\label{fig:nonful}
\end{figure}

Let us check our conjecture \eqref{eq:mainconj} by using the typical example $N_f=4$. In the previous subsection we obtained the superconformal index by dividing the topological string partition function of the del Pezzo-like $\mathcal{B}'_5$ geometry (the red colored toric diagram below $\mathcal{B}_5$ in figure~\ref{fig:delPezzo}) by the partition function of extra contributions. Our conjecture is that this extra factor comes from the non-full spin content of the resulting five dimensional theory.

The non-full spin contents arise from M2-branes wrapping a $\mathbb{CP}^2$ whose local structure forms a $\mathcal{O}(0)\oplus \mathcal{O}(-2)$ bundle. In the language of the 5-brane web, such a two-cycle corresponds to a stack of parallel 5-branes. We have four such stacks in the web configuration $\mathcal{B}'_5$, as illustrated by (a) and (b) in figure~\ref{fig:nonful}. Therefore, the partition function of these four partial web diagrams explains the multiplicities of the non-full spin contents of this geometry. The preferred direction for the two diagrams in (a) is horizontal, which makes the computation slightly technical. The refined topological vertex for the upper diagram in (a) gives
\begin{multline}
Z^{\textrm{(a)}}_{\textrm{upper}}(Q_{1,3},Q_B;\ft,\fq)=\sum_{\boldsymbol{\mu}}(-Q_{3})^{\vert \mu_1\vert}
(-Q_{B})^{\vert \mu_2\vert}f_{\mu_2}(\ft,\fq)(-Q_{1})^{\vert \mu_3\vert}\\\times C_{\emptyset \mu_1 \emptyset}(\fq,\ft) C_{\emptyset \mu_1^t \mu_2}(\ft,\fq) C_{\mu_3\emptyset \mu_2^t}(\fq,\ft) C_{\mu_3^t\emptyset \emptyset}(t,q),
\end{multline}
which after using the resummation formulas \eqref{eq:schuridentity1} and \eqref{eq:schuridentity2} leads to
\begin{align}
\label{eq:question1}
Z^{\textrm{(a)}}_{\textrm{upper}}
&=\sum_{\mu}\ft^{||\mu||^2}Q_B^{|\mu|}\tilde{Z}_{\mu^t}(\ft,\fq)\tilde{Z}_\mu(\fq,\ft)\calR_{\emptyset \mu}(Q_3)\calR_{\emptyset \mu}(Q_1)\nonumber\\&=\Big[\calM\big(Q_1\sqrt{\frac{\ft}{\fq}}\big)\calM\big(Q_3\sqrt{\frac{\ft}{\fq}}\big)\Big]^{-1}\sum_{\mu}
\Big(-Q_{B}\sqrt{\frac{\ft}{\fq}}\Big)^{\vert \mu\vert}P_{\mu}(\fq^\rho; \ft,\fq) P_{\mu^t}(\ft^{-\rho}; \fq,\ft).\nonumber\\
&\qquad\times
\prod_{(i,j)\in \mu}
(1-Q_{1}\fq^{i-\frac{1}{2}}\ft^{-j+\frac{1}{2}} )
\prod_{(i,j)\in \mu}
(1-Q_{3}\fq^{i-\frac{1}{2}}\ft^{-j+\frac{1}{2}} ).
\end{align}
In order to express the $\tilde{Z}$ as functions of the Macdonald polynomials, we have made use of \eqref{eq:MacdonaldZ}.
To perform this summation, we use formulae \eqref{eq:Macdonaldextension} and \eqref{eq:MacdonaldCauchy} to find the closed expression
\begin{equation}
Z^{\textrm{(a)}}_{\textrm{upper}}(Q_{1,3},Q_B;\ft,\fq)
=\frac{\calM\big(Q_B\frac{\ft}{\fq}\big)\calM\big(Q_1Q_3Q_B\big)}{\calM\big(Q_1\sqrt{\frac{\ft}{\fq}}\big)\calM\big(Q_3\sqrt{\frac{\ft}{\fq}}\big)\calM\big(Q_1Q_B\sqrt{\frac{\ft}{\fq}}\big)\calM\big(Q_3Q_B\sqrt{\frac{\ft}{\fq}}\big)}.
\end{equation}
In order to obtain the non-full spin content of the above partition function, we have to isolate the parts that change under the transformation $(\ft,\fq)\to (1/\ft, 1/\fq)$. Since by the analytic continuation rule \eqref{eq:magic} we have $\calM(Q;\ft^{-1},\fq^{-1})=\calM(Q\frac{\ft}{\fq};\ft,\fq)$, we see that the corresponding non-full spin partition function is\footnote{The factor
$\calM(Q_B\frac{\ft}{\fq})=\prod (1-Q_B\ft^{i}\fq^{j-1})^{-1}$ forms a full spin content by combining with the contribution coming from the lower diagram of (a).}
\begin{align}
Z^{\textrm{(a),non-full}}_{\textrm{upper}}(Q_{1,3},Q_B;\ft,\fq)=\calM\big(Q_1Q_3Q_B\big).
\end{align}
The lower diagram in (a) is just $Z^{\textrm{(a)}}_{\textrm{upper}}(Q_{2,4},Q_B;\fq,\ft)$,
and by combining it with the upper contribution we find the non-full spin partition function
\begin{align}
Z^{\textrm{(a)}}_{\textrm{non-full}}(Q_{1,\cdots,4},Q_B;\ft,\fq)=\calM\big(Q_1Q_3Q_B\big)\calM\big(Q_2Q_4Q_B\frac{\ft}{\fq}\big).
\end{align}
The partition function of (b) can be derived just as easily by using the same rules as before. The result for the left part of (b) is then
\begin{align}
Z^{\textrm{(b)}}_{\textrm{left}}(Q_{3,4},Q_F;\ft,\fq)=\frac{\calM\big(Q_F\frac{\ft}{\fq}\big)\calM\big(Q_3Q_4Q_F\big)}{\calM\big(Q_3\sqrt{\frac{\ft}{\fq}}\big)\calM\big(Q_4\sqrt{\frac{\ft}{\fq}}\big)\calM\big(Q_3Q_F\sqrt{\frac{\ft}{\fq}}\big)\calM\big(Q_4Q_F\sqrt{\frac{\ft}{\fq}}\big)}.
\end{align}
The right part is obtained by exchanging $\ft$ with $\fq$ and $Q_3$, $Q_4$ with $Q_1$, $Q_2$.We therefore obtain the following partition function for the non-full spin content of the stacks (b):
\begin{align}
Z^{\textrm{(b)}}_{\textrm{non-full}}(Q_{1,\ldots,4},Q_F;\ft,\fq)
=\calM(Q_3Q_4Q_F)\calM\big(Q_1Q_2Q_F\frac{\ft}{\fq}\big).
\end{align}
Therefore, the total partition function of the non-full spin contents is given by combining (a) and (b) and equals
\begin{equation}
\label{eq:fourflavornormalization}
Z_{\textrm{non-full}}(\boldsymbol{Q};\ft,\fq)=\calM(Q_1Q_3Q_B)\calM(Q_3Q_4Q_F)\calM\big(Q_2Q_4Q_B\frac{\ft}{\fq}\big)\calM\big(Q_1Q_2Q_F\frac{\ft}{\fq}\big).
\end{equation}
This function is precisely equal to the prefactor that we divided by in equation \eqref{eq:renNf4partitionfunction} in order to get the properly normalized partition function for the computation of the 5D superconformal index of the $N_f=4$ theory.
These four factors come from four stacks of parallel 5-branes, and in the flopped diagram of figure~\ref{fig:nonful} part (c), these stacks are manifest. In fact, we can even compute the non-full spin partition function $Z_{\textrm{non-full}}$ directly from the four diagrams in (c).

\subsection{Five flavors and the E$_6$ index}
\label{subsec:E6index}

\begin{figure}[ht]
 \centering
  \includegraphics[height=6.5cm]{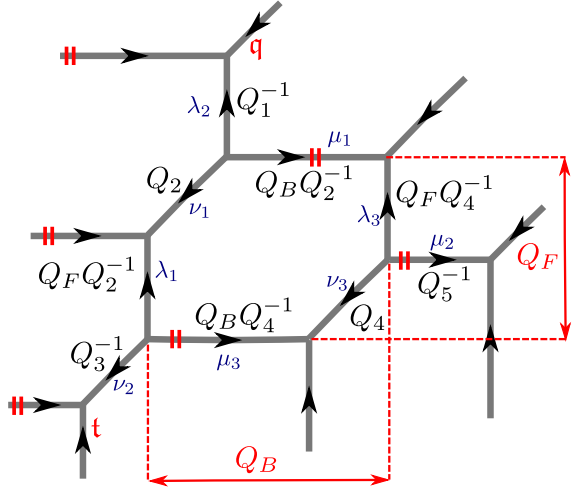}
  \caption{\it The toric diagram for SU(2) with five flavors. In this case, we also label the partitions associated to the internal edges. There are no non-trivial framing factors.}
\label{fig:e6toricdiagram}
\end{figure}

The toric diagram for the SU(2) five flavors case is significantly different from the previous cases.
As we argued in section \ref{sec:7brane}, it has the form given in figure~\ref{fig:e6toricdiagram}.
The inversions appearing in $Q_1$, $Q_3$ and $Q_5$ are purely for convenience. All the framing factors are trivial. The topological string partition function for this toric diagram is
\begin{equation}
\label{eq:toppartZ}
\begin{split}
Z_5=&\sum_{\boldsymbol{\lambda},\boldsymbol{\mu},\boldsymbol{\nu}}(-Q_{1}^{-1})^{|\lambda_2|}(-Q_{3}^{-1})^{|\nu_2|}(-Q_{5}^{-1})^{|\mu_2|}(-Q_{2})^{|\nu_1|}(-Q_FQ_{2}^{-1})^{|\lambda_1|}(-Q_BQ_{4}^{-1})^{|\mu_3|}\\
&\times (-Q_{4})^{|\nu_3|}(-Q_FQ_{4}^{-1})^{|\lambda_3|}(-Q_BQ_{2}^{-1})^{|\mu_1|}C_{\emptyset \lambda_2^t \emptyset }(\fq,\ft)C_{ \nu_2^t\emptyset\emptyset}(\fq,\ft)C_{\emptyset \emptyset \mu_2^t}(\fq,\ft)\\&\times  C_{\nu_1\lambda_2\mu_1}(\ft,\fq)C_{\nu_1^t\lambda_1^t \emptyset }(\fq,\ft)C_{\nu_2\lambda_1\mu_3}(\ft,\fq)C_{ \nu_3^t\emptyset \mu_3^t}(\fq,\ft)C_{\nu_3\lambda_3\mu_2}(\ft,\fq)C_{\emptyset \lambda_3^t \mu_1^t}(\fq,\ft).
\end{split}
\end{equation}
Applying now the resummation formulas \eqref{eq:schuridentity1} and \eqref{eq:schuridentity2} to $Z_5$ leads to
\begin{multline}
Z_5=\sum_{\boldsymbol{\mu}}(-Q_2^{-1}Q_B)^{|\mu_1|}(-Q_5^{-1})^{|\mu_2|}(-Q_4^{-1}Q_B)^{|\mu_3|}\prod_{i=1}^3\fq^{\frac{||\mu_i||^2}{2}}\ft^{\frac{||\mu_i^t||^2}{2}}\tilde{Z}_{\mu_i}(\ft,\fq)\tilde{Z}_{\mu_i^t}(\fq,\ft)\\\times\frac{\calR_{\mu_1^t\emptyset}(Q_1^{-1})\calR_{\emptyset\mu_1}(Q_2)\calR_{\emptyset\mu_3}(Q_3^{-1})\calR_{\mu_3^t\mu_2}(Q_4)\calR_{\mu_3^t\emptyset}(\frac{Q_F}{Q_1})\calR_{\mu_3^t\emptyset}(\frac{Q_F}{Q_2})\calR_{\emptyset\mu_1}(\frac{Q_F}{Q_3})\calR_{\mu_2^t\mu_1}(\frac{Q_F}{Q_4})}{\calR_{\emptyset\emptyset}(\frac{Q_2}{Q_1}\sqrt{\frac{\fq}{\ft}})\calR_{\emptyset\emptyset}(\frac{Q_F}{Q_1Q_3}\sqrt{\frac{\fq}{\ft}})\calR_{\emptyset\emptyset}(\frac{Q_F}{Q_2Q_3}\sqrt{\frac{\fq}{\ft}})\calR_{\mu_3^t\mu_1}(Q_F\sqrt{\frac{\fq}{\ft}})\calR_{\mu_3^t\mu_1}(Q_F\sqrt{\frac{\ft}{\fq}})}.
\end{multline}
Using the experience acquired experimentally in the computation of the indices for $N_f=0,\ldots,4$ as well as the theoretical justifications of subsection \ref{subsec:Normalizing}, we claim that the properly formalized partition function for $N_f=5$ reads
\begin{equation}
\label{eq:renormalizede6partfunction}
\begin{split}
Z_5^{\text{ren}}=&\Big[\calM\big(\frac{Q_2}{Q_1}\big)\calM\big(\frac{Q_F}{Q_1Q_3}\big)\calM\big(\frac{Q_F}{Q_2Q_3}\big)\calM\big(\frac{Q_B}{Q_1Q_2}\big)\calM\big(\frac{Q_F}{Q_4Q_5}\big)\calM\big(\frac{Q_BQ_F}{Q_1Q_2Q_4Q_5}\big)\\&\times \calM\big(\frac{Q_B}{Q_3Q_4}\frac{\ft}{\fq}\big)\calM\big(\frac{Q_4}{Q_5}\frac{\ft}{\fq}\big)\calM\big(\frac{Q_B}{Q_3Q_5}\frac{\ft}{\fq}\big)\Big]^{-1}Z_5.
\end{split}
\end{equation}
In order to get the index from the above equation, we introduce the Coulomb modulus $\tilde{A}$, see \eqref{eq:twoparam1} and \eqref{eq:twoparam2}. Computing the index using formula \eqref{eq:renormalizede6partfunction} is numerically difficult, to obtain even the $x^2$ order term requires summing over all partitions $\mu_i$ whose total number of boxes is smaller or equal to 8. We nevertheless managed to check numerically the superconformal index given in \cite{Kim:2012gu}
\begin{equation}
\label{eq:index5flavor}
\mathcal{I}_5=1+\chi_{\mathbf{ 78}}^{E_6}x^2+(1+\chi_{\mathbf{78}}^{E_6})\chi_1(y)x^3+\Big(\chi_2(y)\big(1+\chi_{\mathbf{78}}^{E_6}\big)
+1+\chi_{\mathbf{2430}}^{E_6}\Big) x^4
+ \mathcal{O}(x^5),
\end{equation}
for several sets of particular values for the fugacities, which makes us confident that \eqref{eq:renormalizede6partfunction} is correct. We refer the reader to appendix \ref{app:E6} for the details involving the different characters and parametrizations.

We finish this section by discussing the non-full spin contributions that were removed in \eqref{eq:renormalizede6partfunction}. In subsection \ref{subsec:Normalizing} we computed the non-full spin part of the four flavors $SU(2)$ theory directly and observed that we obtained a piece of the form $\calM(Q)$ for each pair of external parallel branes, where $Q$ determines the distance between the two branes. We propose that this holds for the $T_3$ junction as well and since we have 9 pairs of external parallel branes, we arrive at \eqref{eq:renormalizede6partfunction}. The only remaining issue is the position of the variables $\ft$ and $\fq$ which tells us whether to divide out $\calM(Q)\equiv \calM(Q;\ft,\fq)$ or $\calM(Q\frac{\ft}{\fq})\equiv \calM(Q\frac{\ft}{\fq};\ft,\fq)=\calM(Q;\fq,\ft)$. We have solved this problem experimentally by computing the index.
\begin{figure}[ht]
 \centering
  \includegraphics[height=5cm]{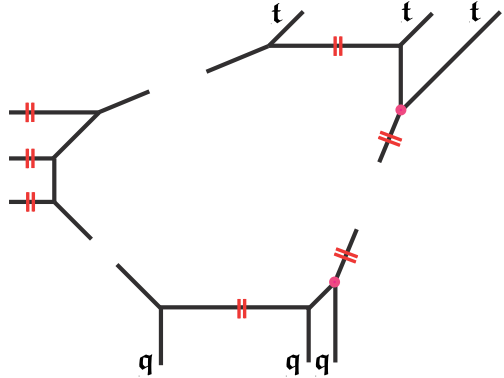}
  \caption{\it This figure depicts the flopped version of the $T_3$ geometry and the parallel 5-brane contributions. The circles denote the new topological string vertex introduced by Iqbal and Koz{\c c}az.}
\label{fig:nonfull}
\end{figure}

Let us finally mention a possible way to actually compute the non-full spin contribution directly. In the brane setup, the three stacks of parallel 5-branes provide this extra contribution, so it is useful to go to the frame which makes these parallel stacks manifest. In the Calabi-Yau setup, we can move into this frame by using a flop transition and the resulting toric diagram is illustrated in figure~\ref{fig:nonfull}. This geometry contains the new vertices $T_{\lambda\mu\nu}(\ft,\fq)$ depicted by the red circles. The vertex of this new type was introduced by Iqbal and Koz{\c c}az in \cite{Iqbal:2012mt}. We can expect that these non-full spin content contributions coincide with the refined topological vertex partition function of this geometry. Since it is non-trivial to compute this new vertex function, we leave the problem open.

\subsection{The $T_{N}$ partition function}
\label{sec:TNpartitionfunction}

In this subsection, we generalize the computation of the $T_3$ partition function to the $T_{N}$ case. In order to do this, it is convenient to cut the toric diagram diagonally into $N$ subdiagrams called \emph{strips}.
\begin{figure}[ht]
 \centering
  \includegraphics[height=6cm]{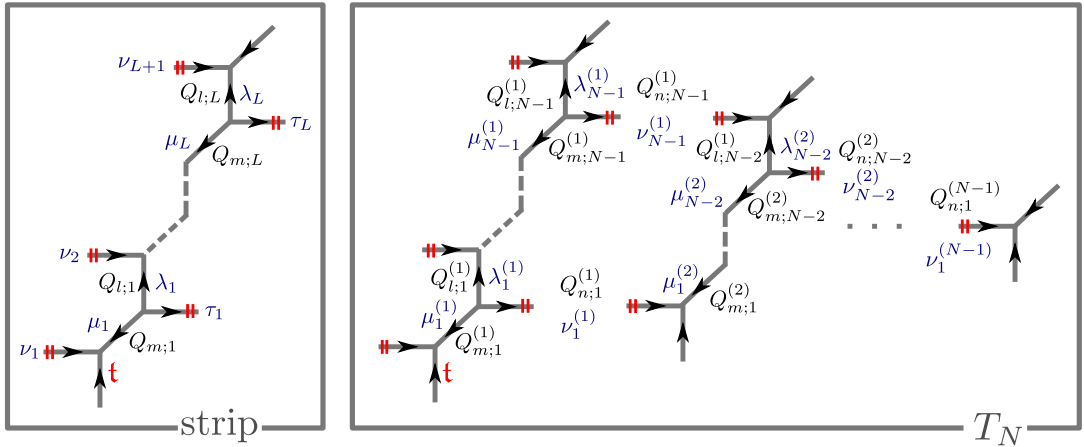}
  \caption{\it The left part of the figure shows the strip diagram, while the right one depicts the dissection of the $T_N$ diagram into $N$ strips. The partitions associated with the horizontal, diagonal and vertical lines are $\nu_{i}^{(j)}$, $\mu_{i}^{(j)}$ and $\lambda_{i}^{(j)}$ with $j=1,\ldots, N-1$, $i=1,\ldots, N-j$ respectively. The K\"ahler parameters of the horizontal, diagonal and vertical lines are $Q_{n;i}^{(j)}$, $Q_{m;i}^{(j)}$, $Q_{l;i}^{(j)}$ respectively with the same range of indices. }
\label{fig:strip}
\end{figure}
We consider the strip diagram of arbitrary length $L\leq 0$, drawn in the left in figure~\ref{fig:strip}. The corresponding partition function depends on the external horizontal partitions $\boldsymbol{\nu}=(\nu_1,\ldots, \nu_{L+1})$, $\boldsymbol{\tau}=(\tau_1,\ldots, \tau_{L})$ as well as the parameters $\boldsymbol{Q}_m=(Q_{m;1},\ldots, Q_{m;L})$ and $\boldsymbol{Q}_l=(Q_{l;1},\ldots, Q_{l;L})$. It takes the form
\begin{equation}
\label{eq:strippartitionfunction1}
Z_{\boldsymbol{\nu}\boldsymbol{\tau}}^{\text{strip}}(\boldsymbol{Q}_m, \boldsymbol{Q}_l;\ft,\fq)\defeq \sum_{\boldsymbol{\lambda},\boldsymbol{\mu}}\prod_{i=1}^L(-Q_{m;i})^{|\mu_i|}(-Q_{l;i})^{|\lambda_i|}\prod_{j=1}^{L+1}C_{\mu_j^t\lambda_{j-1}^t\nu_j^t}(\fq,\ft)\prod_{k=1}^LC_{\mu_k\lambda_k\tau_k}(\ft,\fq),
\end{equation}
where $\mu_{L+1}=\lambda_0=\emptyset$. We claim that this expression can be resummed to the following result:
\begin{align}
\label{eq:strippartitionfunction2}
Z_{\boldsymbol{\nu}\boldsymbol{\tau}}^{\text{strip}}(\boldsymbol{Q}_m, \boldsymbol{Q}_l;\ft,\fq)=&\prod_{j=1}^{L+1}t^{\frac{||\nu_j^t||^2}{2}}\tilde{Z}_{\nu_j^t}(\fq,\ft)\prod_{j=1}^L\fq^{\frac{||\tau_j||^2}{2}}\tilde{Z}_{\tau_j}(\ft,\fq)\nonumber\\&\times\prod_{i\leq j=1}^L\frac{\calR_{\nu_i^t\tau_j}\Big(Q_{m;j}\prod_{k=i}^{j-1}Q_{m;k}Q_{l;k}\Big)
\calR_{\tau_i^t\nu_{j+1}}\Big(Q_{l;i}\prod_{k=i+1}^jQ_{m;k}Q_{l;k}\Big)}{\calR_{\nu_i^t\nu_{j+1}}\Big(\prod_{k=i}^jQ_{l;k}Q_{m;k}\sqrt{\frac{\fq}{\ft}}\Big)}\nonumber\\&\times \prod_{i\leq j=1}^{L-1}\calR_{\tau_i^t\tau_{j+1}}\Big(\prod_{k=i}^jQ_{l;k}Q_{m;k+1}\sqrt{\frac{\ft}{\fq}}\Big)^{-1}.
\end{align}
The complete $T_{N}$ diagram is made out of $N$ such strip diagrams as depicted in figure~\ref{fig:strip}. The choice of the K\"ahler parameters is explained in detail in appendix \ref{app:parametrizationgeneralTN}. Using the result \eqref{eq:strippartitionfunction2} and renormalizing the outcome as in the previous subsections, we arrive at the following expression for the topological string partition function of the $T_{N}$ toric diagram:
\begin{align}
\label{eq:TNjunctionpartitionfunction}
Z_{T_{N}}^{\text{ren}}=&\sum_{\boldsymbol{\nu}}\prod_{r=1}^{N}\Big(-\boldsymbol{Q}_n^{(r)}\Big)^{|\boldsymbol{\nu}^{(r)}|}Z^{\text{strip}}_{\boldsymbol{\nu}^{(r-1)}\boldsymbol{\nu}^{(r)}}(\boldsymbol{Q}_m^{(r)}, \boldsymbol{Q}_l^{(r)};\ft,\fq)\nonumber\\&\times\prod_{i\leq j=1}^{N-1}\Big[\calM\big(\prod_{k=i}^jQ_{l;k}^{(1)}Q_{m;k}^{(1)}\big)\calM\big(\prod_{k=i}^jQ_{m;1}^{(k)}Q_{n;1}^{(k)}\frac{\ft}{\fq}\big)\calM\big(\prod_{k=i}^jQ_{l;N-k}^{(k)}Q_{n;N-k}^{(k)}\big)\Big]^{-1}.
\end{align}
In the above formula, we have set $\nu_{j}^{(0)}=\emptyset$ for all $j$, see the right part of figure~\ref{fig:strip}. This partition function depends on the parameters $\boldsymbol{Q}_m^{(j)}$, $\boldsymbol{Q}_n^{(j)}$ and $\boldsymbol{Q}_l^{(j)}$ of which there are $3\sum_{i=1}^{N-1}i$ in total. Since each hexagon in the toric diagram comes with two independent equations that relate the parameters to each other and there are $\sum_{i=1}^{N-2}i$ hexagons in the $T_{N}$ toric diagram, we arrive at the conclusion that there are a total of
\begin{equation}
\label{eq:independentmoduli}
3\sum_{i=1}^{N-1}i-2\sum_{i=1}^{N-2}i=\frac{(N+4)(N-1)}{2}
\end{equation}
independent moduli. The result \eqref{eq:independentmoduli} fully agrees with equation (3.36) in \cite{Kozcaz:2010af}. When $N$ is equal to 3, this gives us 7 parameters corresponding to $Q_F,Q_B,Q_1,\ldots, Q_5$ from subsection \ref{subsec:E6index}.

\subsection{The $T_2$ partition function revisited}
\begin{figure}[thbp]
 \begin{center}
  \includegraphics[width=80mm,clip]{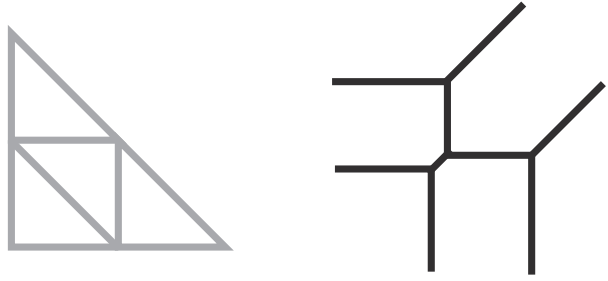}
 \end{center}
 \caption{\it The toric and web diagram of the $T_2$ geometry.  }
 \label{fig:t2}
\end{figure}
In this subsection, we shall rederive \eqref{eq:TNjunctionpartitionfunction} for $N=2$ and rewrite it as an infinite product. The $T_2$ theory is given by the intersection of three stacks of two parallel 5-branes as show in figure~\ref{fig:t2}.
The toric diagram of the dual Calabi-Yau manifold has the same shape as this web diagram. We can therefore compute the five dimensional Nekrasov partition function of the deformed $T_2$ theory by using the refined topological vertex formalism.
As shown in subsection \ref{sec:TNpartitionfunction} as well as in appendix \ref{app:parametrizationgeneralTN}, this theory has three independent K\"ahler parameters. In order to simplify the notation, we do not use the notation of subsection \ref{sec:TNpartitionfunction} here, but instead write $Q_1\equiv Q_{m,1}^{(1)}$, $Q_2\equiv Q_{l,1}^{(1)}$ and $Q_3\equiv Q_{n,1}^{(1)}$.  As we will show in the following, the partition function of the $T_2$ theory is given by the infinite product
\begin{equation}
\label{eq:T2partitionfunction}
Z'_{T_2}(\boldsymbol{Q})=
\prod_{i,j=1}^{\infty}\frac{(1-Q_1Q_2Q_3\ft^{i-\frac{1}{2}}\fq^{j-\frac{1}{2}})\prod_{k=1}^3(1-Q_k\ft^{i-\frac{1}{2}}\fq^{j-\frac{1}{2}})}{(1-Q_1Q_2\ft^i\fq^{j-1})(1-Q_1Q_3\ft^{i-1}\fq^{j})(1-Q_2Q_3\ft^{i}\fq^{j-1})}.
\end{equation}
This formula is precisely equal to the one which was conjectured by Koz{\c c}az, Pasquetti and Wyllard in \cite{Kozcaz:2010af}, equation (5.3). In the paragraphs to follow, we shall provide a derivation based on the properties of the Macdonald polynomials.

We begin with the refined partition function of the $T_2$ geometry. The general case was written down in subsection \ref{sec:TNpartitionfunction}, which we repeat\footnote{Note that $\ft$ and $\fq$ are reversed relative to the previous sections.} here:
\begin{equation}
\label{eq:T2partitionfunction1}
Z_{T_2}'(\boldsymbol{Q})= \sum_{\boldsymbol{R}}
\prod_{i=1}^3(-Q_i)^{\vert R_i\vert} C_{ R_1^t\emptyset\emptyset}(\ft,\fq)C_{\emptyset R_2^t \emptyset}(\ft,\fq) C_{\emptyset\emptyset R_3^t}(\ft,\fq) C_{R_1R_2 R_3}(\fq,\ft),
\end{equation}
where $\boldsymbol{Q}=(Q_1,Q_2,Q_3)$.
This vertex is off strip in the sense of Iqbal and Kashani-Poor's paper \cite{Iqbal:2004ne}. It is very difficult to compute such an off strip diagram in general. However this $T_2$ geometry is special and the closed expression for the unrefined partition function was given in \cite{Sulkowski:2006jp}. By using the usual identities \eqref{eq:schuridentity1} and \eqref{eq:schuridentity2}, we can easily obtain from \eqref{eq:T2partitionfunction1} the following expression
\begin{equation}
\label{eq:T2ver1}
Z_{T_2}'(\boldsymbol{Q})=\sum_{R_3}(-Q_3)^{\vert R_3\vert}\fq^{\frac{\parallel R_3^t\parallel^2}{2}}\ft^{\frac{\parallel R_3\parallel^2}{2}}\tilde{Z}_{R_3^t}(\ft,\fq)\tilde{Z}_{R_3}(\fq,\ft)\calM\big(Q_1Q_2\frac{\ft}{\fq}\big)\calR_{R_3\emptyset}(Q_1)\calR_{\emptyset R_3^t}(Q_2).
\end{equation}
As we did for the computations in the $SU(2)$ gauge theory case,
let us normalize this partition function with the following one-loop like factor
\begin{equation}
Z^0_{T_2}(Q_1,Q_2)=\calM\big(Q_1Q_2\frac{\ft}{\fq}\big)\calM\big(Q_1\sqrt{\frac{\ft}{\fq}}\big)^{-1}\calM\big(Q_2\sqrt{\frac{\ft}{\fq}}\big)^{-1}
\end{equation}
so that \eqref{eq:T2ver1} becomes
\begin{equation}
\label{eq:T2middle}
Z'_{T_2}(\boldsymbol{Q})=Z^0_{T_2}\sum_R (-Q_3)^{\vert R\vert}
\ft^{\frac{\parallel R\parallel^2}{2}}\fq^{\frac{\parallel R^t\parallel^2}{2}}\prod_{(i,j)\in R}\frac{(1-Q_1\sqrt{\frac{\fq}{\ft}}\ft^{1-j}\fq^{i-1})(1-Q_2\sqrt{\frac{\ft}{\fq}}\ft^{j-1}\fq^{1-i}) }{(1-\ft^{R_i-j+1}\fq^{R^t_j-i})(1-\ft^{R_i-j}\fq^{R^t_j-i+1})}.
\end{equation}
In the above, we have used a trick that allowed us to replace  in the numerator $\fq^{R_j^t-i}$ by $\fq^{i-1}$.  We can further rewrite this expression by using a specialization of the Macdonald symmetric polynomials
\begin{align}
P_R(\ft^{N-\frac{1}{2}},\ft^{N-\frac{3}{2}},\cdots,\ft^{\frac{1}{2}};\fq,\ft)
=\ft^{\frac{\parallel R^t\parallel^2}{2}}\prod_{(i,j)\in R}
\frac{(1-\fq^{j-1}\ft^{N+1-i})}{(1-\fq^{R_i-j}\ft^{R_j^t-i+1})}
\end{align}
taken from \cite{MacDonnaldSymmetric}, equation (6.11') on page 337. By using the Cauchy formula \eqref{eq:MacdonaldCauchy} for the Macdonald symmetric polynomials, we find
\begin{align}
\label{eq:ZT2temp}
Z'_{T_2}(\boldsymbol{Q})
=&Z^0_{T_2}\sum_{R}
(-Q_3)^{\vert R\vert}
P_{R^t}(\ft^{N-\frac{1}{2}},\ft^{N-\frac{3}{2}},\cdots,\ft^{\frac{1}{2}};\fq,\ft)P_R(\fq^{M-\frac{1}{2}},\fq^{M-\frac{3}{2}},\cdots,\fq^{\frac{1}{2}};\ft,\fq)\nonumber\\
=&
Z^0_{T_2}\prod_{i=1}^N\prod_{j=1}^M\big(1-Q_3\ft^{i-\frac{1}{2}}\fq^{j-\frac{1}{2}}\big).
\end{align}
Here, we have specialized $Q_1$ and $Q_2$, so that $N$ and $M$, defined via the equations
\begin{align}
\ft^N=Q_1\ft^{-\frac{1}{2}}\fq^{\frac{1}{2}},
\quad
\fq^M=Q_2\ft^{\frac{1}{2}}\fq^{-\frac{1}{2}},
\end{align}
become integers. Due to this specialization, it is possible to rewrite the finite product as an infinite one
\begin{equation}
\prod_{i=1}^N\prod_{j=1}^M\big(1-Q_3\ft^{i-\frac{1}{2}}\fq^{j-\frac{1}{2}}\big)=\prod_{i,j=1}^\infty\frac{(1-Q_3\ft^{i-\frac{1}{2}}\fq^{j-\frac{1}{2}})(1-Q_1Q_2Q_3\ft^{i-\frac{1}{2}}\fq^{j-\frac{1}{2}})}{(1-Q_1Q_3\ft^{i-1}\fq^{j})(1-Q_2Q_3\ft^{i}\fq^{j-1})}.
\end{equation}
By ``analytic continuation'', the right hand side should give the correct result for the partition function \eqref{eq:ZT2temp} with generic $Q_1$ and $Q_2$. Putting everything together, we arrive at the following compact product expression:
\begin{equation}
\label{eq:T2partitionfunctionfinal}
Z'_{T_2}(\boldsymbol{Q})=\frac{\calM(Q_1Q_3)\calM\big(Q_1Q_2\frac{\ft}{\fq}\big)\calM\big(Q_2Q_3\frac{\ft}{\fq}\big)}{\calM\big(Q_1\sqrt{\frac{\ft}{\fq}}\big)\calM\big(Q_2\sqrt{\frac{\ft}{\fq}}\big)\calM\big(Q_3\sqrt{\frac{\ft}{\fq}}\big)\calM\big(Q_1Q_2Q_3\sqrt{\frac{\ft}{\fq}}\big)}.
\end{equation}
Due to the definition of the function $\calM$ in \eqref{eq:mainfunctionsdefinitions}, this is exactly the desired partition function \eqref{eq:T2partitionfunction} for the $T_2$ geometry.
Since there is no compact four-cycle in  this geometry, there is no constant map contribution. This formula is therefore complete up to the well-known classical partition function which is irrelevant for the superconformal index.

\newpage

\section{$W_N$ 3-point functions}
\label{sec:correlationfunctions}

In this section, we interpret the topological string partition functions that we obtained in the previous section through the lens of the 5D AGTW relation and relate them to the correlation functions of the 2D $q$-deformed Toda CFTs. We explore the generalization of the 5D/2D AGTW dictionary, originally studied in \cite{Nieri:2013yra}, namely the correspondence between the 5D superconformal index (of a generalized quiver) and the $n$-point $q$-deformed Toda correlation function\footnote{The quiver diagram drawn \`a la Gaiotto looks identical to the diagram associated with the conformal block.}. 
This is in strict distinction to the 4D/2D case, where the index corresponds to topological QFT correlators \cite{Gadde:2009kb,Gadde:2011ik} and it stems from the fact that the 5D superconformal index is not as simple as the 4D index, since apart from the plethystic exponential part, it includes an instanton contribution.

In the context of the AGTW relation between 4D theories of class $\mathcal{S}$ and 2D conformal field theories, the $T_N$ theory corresponds to the three punctured Riemann sphere and is a basic building block in the construction of the Sicilian gauge theories. In the original paper by Gaiotto, these theories were called ``generalized quiver theories".
The 5D $T_N$ theories, i.e. 5-brane junctions, also lead to ``Sicilian gauge theories" in 5D by gluing the edges of the junctions\footnote{An edge of a stack of 5-branes in a junction describes a subgroup of the flavor symmetry of the $T_N$ theory, the gluing procedure of two junctions is realized by gauging two such subgroups and coupling them with the corresponding vector multiplet.}, and some of them provide well-defined UV fixed point theories. Therefore, the 5D AGTW relation suggests that the Nekrasov partition functions of these gauge theories can be recast\footnote{For an M-theory based derivation of this statement, see \cite{Bonelli:2009zp,Alday:2009qq,Tan:2013xba}.} into the conformal blocks of $q$-deformed Toda theories with $q$-deformed $W_N$ symmetry \cite{Awata:2009ur,Awata:2010yy}. By generalizing the results of \cite{Nieri:2013yra}, one obtains the following relation between the 5D superconformal index, which is the partition function on $S^4\times S^1$, and the correlation function of the corresponding $q$-Toda field theory:
\begin{align}
\mathcal{I}^{5D}(x,y)=\int [da] \Big|
Z_{\textrm{Nek}}^{\textrm{5D}}(a,m,\beta,\epsilon_{1,2})\Big|^2
\propto \langle
V_{\boldsymbol{\alpha}_1}(z_1)\cdots V_{\boldsymbol{\alpha}_n}(z_n)
\rangle_{q\textrm{-Toda}}.
\end{align}
Strictly speaking, the generalization to $N>2$ is straightforward only for the degenerate\footnote{See \cite{Bonelli:2011fq,Bonelli:2011wx} for a topological string description of the semi-degenerate multipoint function of $q$-Toda theory.} cases, but we  conjecture that the relation holds for the others as well.
It is a vast project to properly investigate this 5D AGTW correspondence, we therefore focus on the $T_N$ theory which is a basic building block of the uplift of the class $\mathcal{S}$ theories. In the following we will study the partition functions of the $T_N$ brane junctions from the perspective of the DOZZ formula.

\subsection{The $T_2$ theory}

In \eqref{eq:T2partitionfunction}, we have obtained the partition function of the $T_2$ theory in 5D. This result is expected to be related to the $q$-deformed Liouville theory through the AGTW relation.
 On the 2D side, the only results available so far \cite{Nieri:2013yra} are obtained by employing the symmetry constraints of the $q$-deformed algebra. This method gives the DOZZ functions of the $q$-deformed  Liouville theory  up to an overall unknown function.

\subsubsection{The $q$-Liouville DOZZ formula}

The $q$-Liouville correlation functions are expressed with the help of the $\fq$-deformed $\Upsilon$-function
\begin{align}
\Upsilon_{\,\ft,\fq} (x)
\defeq\frac{1}{\Gamma_{\ft,\fq}(x)\Gamma_{\ft,\fq}(\epsilon-x)},
\end{align}
where $\epsilon=\epsilon_1+\epsilon_2$ and the $\fq$-deformed Gamma function is defined, via a zeta function regularization, by the following infinite product:
\begin{equation}
\Gamma_{\ft,\fq}(x)\defeq\exp \frac{d}{ds}\zeta_{\ft,\fq}(s;x)\Bigm|_{s=0},
\qquad\zeta_{\ft,\fq}(s;x)\defeq\sum_{i,j=1}^\infty
 \left(2 \sinh \frac{\beta}{2}(x+\epsilon_1 i+\epsilon_2j) \right)^{-s}.
\end{equation}
Here $\beta$ is the circumference of the fifth dimension, and the quantum-deformation parameters are given by the $\Omega$-backgrounds as
\begin{align}
\fq=e^{-\beta\epsilon_1},\qquad
\ft=e^{\beta\epsilon_2}.
\end{align}
We can recast the functions $\calM$ of \eqref{eq:mainfunctionsdefinitions} by using the $\Gamma_{\ft,\fq}$-function as follows
\begin{align}
\mathcal{M}\Big(Q\Big(\frac{\ft}{\fq}\Big)^{\frac{\alpha}{2}} \Big)
=\frac{\prod_{i,j=1}^{\infty}e^{-\frac{\beta}{2}\left(m +\epsilon_1(i-\alpha\frac{1}{2})+\epsilon_2(j-\alpha\frac{1}{2})\right)}}
{\Gamma_{\ft,\fq}\left(m-\alpha\frac{\epsilon}{2}\right)}
\end{align}
with $Q=e^{-\beta m}$. The partition function $Z'_{T_2}$ of  \eqref{eq:T2partitionfunctionfinal} can then be rewritten as
\begin{multline}
Z'_{T_2}(\boldsymbol{Q})=\Big(\prod_{i,j=1}^{\infty}e^{\frac{\beta\epsilon}{4}}\Big)
\prod_{i,j=1}^{\infty}(1-\ft^{i-\frac{1}{2}}\fq^{j-\frac{1}{2}})\\
\times
\frac{\Gamma_{\ft,\fq}\left(m_1-\frac{\epsilon}{2}\right) \Gamma_{\ft,\fq}\left(m_2-\frac{\epsilon}{2}\right)
\Gamma_{\ft,\fq}\left(m_3-\frac{\epsilon}{2}\right) \Gamma_{\ft,\fq}\left(m_1+m_2+m_3-\frac{\epsilon}{2}\right)}{
\Gamma_{\ft,\fq}\left(-\frac{\epsilon}{2}\right)
\Gamma_{\ft,\fq}\left(m_1+m_2-{\epsilon}\right)
\Gamma_{\ft,\fq}\left(m_2+m_3-{\epsilon}\right)
\Gamma_{\ft,\fq}\left(m_1+m_3\right)
}.
\end{multline}
By introducing the Liouville parameters
\begin{align}
m_1=-\alpha_1- \alpha_2+ \alpha_3-\frac{\epsilon}{2},\quad
m_2=\alpha_1+ \alpha_2+ \alpha_3+\frac{3\epsilon}{2},\quad
m_3=\alpha_1- \alpha_2- \alpha_3-\frac{\epsilon}{2},
\end{align}
we can rewrite it into the following form similar to the DOZZ formula for the $q$-Liouville correlation function, up to a divergent prefactor
\begin{multline}
Z'_{T_2}(\boldsymbol{Q})\propto\prod_{i,j=1}^{\infty}(1-\ft^{i-\frac{1}{2}}\fq^{j-\frac{1}{2}})\\
\times
\frac{
\Gamma_{\ft,\fq}\left(-\alpha_1- \alpha_2+ \alpha_3-\epsilon\right)
\Gamma_{\ft,\fq}\left(\sum_{k=1}^3\alpha_k+\epsilon\right)
\Gamma_{\ft,\fq}\left(\alpha_1- \alpha_2- \alpha_3-\epsilon\right)
\Gamma_{\ft,\fq}\left(\alpha_1-\alpha_2+\alpha_3 \right)}{
\Gamma_{\ft,\fq}\left(-\frac{\epsilon}{2}\right)
\Gamma_{\ft,\fq}\left(2\alpha_1 \right)
\Gamma_{\ft,\fq}\left(-2\alpha_2-\epsilon \right)
\Gamma_{\ft,\fq}\left(2\alpha_3 \right)
}.
\end{multline}
The combination of $Z'_{T_2}$ and $\bar{Z}'_{T_2}$ appears in the superconformal index of the 5D theory, which leads to the DOZZ formula for the $q$-Liouville three-point functions \cite{Nieri:2013yra, Kozcaz:2010af, Bao:2011rc}
\begin{align}
C^{\ft,\fq}_{\textrm{DOZZ}} \propto 
\Big\vert \prod_{i,j=1}^{\infty}(1-\ft^{i-\frac{1}{2}}\fq^{j-\frac{1}{2}})^{-1}Z'_{T_2}(\boldsymbol{Q})\Big\vert^2.
\end{align}
The Liouville parameter is given by $-\epsilon$, which is opposite to the usual conventions \cite{Alday:2009aq}. We thus obtain the $q$-DOZZ three-point function up to an unknown coefficient, and this situation is completely the same as in \cite{Nieri:2013yra}. In that paper, the authors determined the DOZZ partially without referring to any Lagrangian formulation
of the $q$-Liouville theory. The lack of the Lagrangian description makes the determination of the proportionality coefficient difficult.

\subsubsection{The superconformal index}

We can apply the expression \eqref{eq:T2partitionfunction} to compute the superconformal index of the $T_2$ SCFT after extracting the full spin content. Since the non-full spin content breaks the symmetry $(\ft,\fq)\to (1/\ft,1/\fq)$, we find the genuine contribution from the south pole of $S^4$, after expressing the result using \eqref{eq:sigmaxy},
\begin{equation}
Z_{T_2}(\boldsymbol{Q})=\prod_{i,j=1}^{\infty}
(1-Q_1Q_2Q_3x^{i+j-1}y^{i-j})^{-1}\prod_{k=1}^3(1-Q_kx^{i+j-1}y^{i-j})^{-1}.
\end{equation}

Since the ``Euler number"\footnote{Twice the number of the compact four-cycles.} of this Calabi-Yau manifold is zero,
there is no constant map contribution.
Moreover the $T_2$ theory does not have any Coulomb branch deformation,
so the superconformal index is simply the product of the north and south pole contributions
\begin{multline}
\label{free-hyper}
\mathcal{I}_{\,T_2}=Z_{T_2}(\boldsymbol{Q};\ft,\fq)\,Z_{T_2}(\boldsymbol{Q}^{-1};\ft^{-1},\fq^{-1})
=1+\left(\chi^{SU(4)}_{[1,0,0]}+\chi^{SU(4)}_{[0,0,1]}\right)x\\+
\bigg[\,
\chi_2(y)\left(\chi^{SU(4)}_{[1,0,0]}+\chi^{SU(4)}_{[0,0,1]}
\right)
+\chi^{SU(4)}_{[2,0,0]}+1+\chi^{SU(4)}_{[0,0,2]}+\chi^{SU(4)}_{[1,0,1]}\bigg]x^2+\cdots,
\end{multline}
where the $SU(4)$ characters labeled by Dynkin labels are given by
\begin{align}
\chi^{SU(4)}_{[1,0,0]}=\sum_{i=1}^4Q_i,\qquad
\chi^{SU(4)}_{[0,0,1]}=\sum_{i=1}^4Q_i^{-1},\quad \cdots,
\end{align}
for fugacities $Q_i$ satisfying $Q_1Q_2Q_3Q_4=1$. We can thus see that the flavor symmetry is enhanced as $U(1)^3 \rightarrow SU(4)\simeq Spin(6)$. The index \eqref{free-hyper} is the index of four free hypermultiplets and $SU(4)$ is the flavor symmetry that rotates them.

\subsection{The $T_N$ theory and the $A_{N-1}$ $q$-Toda CFT}

The refined topological vertex formalism enables us to write down the partition function of the $T_N$ geometry, since we can find a consistent choice of the preferred direction. In fact, this partition function does not correspond to a conventional gauge theory, and so far no closed expression, as for the $T_2$ case, has been found. This, however, does not mean that there is no nice expression mathematically, and actually we have the expression as a summation over Young diagrams. So if we can perform the summation exactly, a closed expression will immediately follow from our expression. Though this might be an interesting problem for mathematicians, we leave it open and assume the existence of a closed expression in the following. 

Recall that the superconformal index, can be obtained from the corresponding Nekrasov partition function, i.e. the topological string partition function, through the following formula \textit{\`a la }localization
\begin{align}
\label{eq:5dAGTW}
\mathcal{I}^{\,\textrm{5D}}(x,y)=\int da_1\cdots da_F\ \Big|
Z_{\textrm{Nek}}^{\textrm{5d}}(a,m,\beta,\epsilon_{1,2})\Big|^2.
\end{align}
The integral is taken over the loop variables $Q_{Fg}=e^{-\beta a_g}$, which parametrize the local deformation of the brane web diagram. Let $F$ be the number of these variables, namely the number of the elementary faces in the web-diagram. 
In \cite{Nieri:2013yra} the relation between 4-point functions of 2D $q$-deformed Liouville and the superconformal index of $SU(2)$ gauge theory with four flavors was discussed.
Generalizing, we expect the superconformal index of an $SU(N)$ quiver to be related to a correlation function of the 2D $q$-deformed Toda just like for the AGTW relation in 4D. This is because the (irregular) conformal blocks of this 2D CFT coincide with the corresponding Nekrasov partition functions.
We therefore expect the following 5D uplift of the AGTW relation:
\begin{table}[htb]
\begin{center}
\begin{tabular}{r|c}
\quad gauge theory on $S^1\times S^4$& $q$-Toda CFT\quad\\
\hline
\quad superconformal index $\mathcal{I}^{\,\textrm{5D}}$& correlation function $\langle V_{\boldsymbol{\alpha}}\cdots V_{\boldsymbol{\alpha}}\rangle$ \quad \\
\end{tabular}
\end{center}
\end{table}

To fit the web diagram into the Gaiotto construction of SCFTs, we need to decompose the diagram into $T_N$ geometries and stacks of parallel (color) 5-branes connecting the $T_N$ blocks. These $T_N$ blocks are the analogues of the three punctured spheres in the Gaiotto construction, and the connecting procedure by the stacks represents gauging (the subgroup of) the flavor symmetries of $T_N$s. The parameters associated with the gauged external lines of the $T_N$s then become the internal momenta of the 2D CFT, and the non-gauged external leg parameters are the momenta of the vertex operators inserted in the $q$-Toda correlator. The claims made thus far are completely parallel to the original AGT story \cite{Alday:2009aq} on 4D $SU(2)$ theories.

\subsubsection{Counting $W_N$ three-point functions}

A subtlety appears in the $T_{N\geq3}$ cases as pointed out by Wyllard \cite{Wyllard:2009hg}, due to a characteristic of the higher spin $W_N$-algebra. We review this subtlety in the well known non $q$-deformed case, expecting the situation to be essentially the same for the $q$-deformed one. Usually, one can reduce a CFT three-point function of descendant operators to that of three primary operators by using the Ward identities of the conformal algebra. However, in the $A_{N\geq2}$ Toda CFT, one \textit{cannot} derive the correlation functions of the descendant fields purely from the correlation functions of the $W_N$ primary fields. This fact can by roughly explained \cite{Kozcaz:2010af} by counting the available Ward identities. The chiral $A_{N-1}$ Toda CFT has $N(N-1)/2$ basic descendant operators,
\begin{align}
\left(W^{(s)}_{-l}\right)^nV_{\boldsymbol{\alpha}},\quad \textrm{for}\quad
l=1,2,\cdots,s-1,\quad
s=2,3,\cdots,N.
\end{align}
The number of basic three-point functions is hence $3N(N-1)/2$ after taking into account the three primaries $V_{\boldsymbol{\alpha}_{1,2,3}}(z=\infty,1,0)$. On the other side, the $W_N$-algebra provides Ward identities for the following generators
\begin{align}
W_{l}^{(s)},\qquad \textrm{for}\quad
l=-s+1,-s+2,\cdots,s-1,\quad
s=2,3,\cdots,N,
\end{align}
leading to only $N^2-1$ constraints. Therefore, we cannot reduce
\begin{align}
\frac{3N(N-1)}{2}-(N^2-1)=\frac{(N-1)(N-2)}{2}
\end{align}
types of basic three-point functions. This means that {\it the three-point functions of three $W_N$ primary operators are not enough to write down the generic Toda correlation function}, and we need to introduce additional ${(N-1)(N-2)}/{2}$ types of three-point functions for the descendant operators. Moreover, we have to specify $3(N-1)$ values for the Toda momenta of the three basic primary operators $V_{\boldsymbol{\alpha}_{1,2,3}}$. Remember that a Toda momentum is an $(N-1)$-vector $\boldsymbol{\alpha}=(\alpha_1,\cdots,\alpha_{N-1})$. Let us consider the $W_3$ symmetry as an example \cite{Kozcaz:2010af}. The additional building blocks are the three-point functions of the following type:
\begin{align}
\langle
V_{\boldsymbol{\alpha}_1}(\infty)\,
\left((W_{-1}^{(3)})^{n}V_{\boldsymbol{\alpha}_2}(1)\right)\,
V_{\boldsymbol{\alpha}_3}(0)
\rangle.
\end{align}
They are labeled by the positive integers $n=1,2,\cdots$, and therefore we need an infinite amount of additional data.
The choices of the original correlation function and its degenerate channel determine which three-point functions appear in the construction. This situation makes the study of the Toda three-point functions very difficult\footnote{The building block of the $SU(3)$ gauge theory with $6$ fundamentals is the three-point function with the insertion of a semi-degenerate operator, namely the simple punctures in the Gaiotto's construction. This case is very simple, because the transformation $W_{-1}^{(3)}$ on the semi-degenerate operator can be replaced with the well-behaved Virasoro transformation $W_{-1}^{(2)}=L_{-1}$ and the primary three-point functions are enough to compute the correlation functions as in the Liouville theory.}.

For the $W_3$ case, we can recast this situation explicitly in terms of the three- and four-point functions \cite{Bowcock:1993wq}.
The $W_3$ three-point functions can be associated to maps from the Verma module to the complex numbers:
\begin{align}
\vert \Psi\rangle\quad \mapsto \quad \langle \boldsymbol{\alpha_1}\vert V_{\boldsymbol{\alpha_2}}(1) \vert \Psi\rangle.
\end{align}
Following \cite{Bowcock:1993wq}, we introduce the following combinations of the generators
\begin{align}
&e_m(z)=L_m-2zL_{m-1}+z^2L_{m-2},& &f_m(z)=W_m-3zW_{m-1}+3z^2W_{m-2}-z^3W_{m-3},& \nonumber\\
&e'_0(z)=L_{0}-2zL_{-1}+z^2L_{-2},& &f'_0(z)=W_0-3zW_{-1}+3z^2W_{-2}-z^3W_{-3},& \nonumber\\
&e''_0(z)=-zL_{-1}+z^2L_{-2},& &f''_0(z)=-zW_{-1}+2z^2W_{-2}-z^3W_{-3}.&
\end{align}
The OPEs between a primary field and a symmetry generator then imply
\begin{align}
&\langle \boldsymbol{\alpha_1}\vert V_{\boldsymbol{\alpha_2}}(z) e_m(z)=0,\quad m=-1,-2,-3,\cdots,\nonumber\\
&\langle \boldsymbol{\alpha_1}\vert V_{\boldsymbol{\alpha_2}}(z) \left(e'_0(z)-\Delta_1\right)=0,\nonumber\\
&\langle \boldsymbol{\alpha_1}\vert V_{\boldsymbol{\alpha_2}}(z) \left(e''_0(z)-\Delta_2\right)=0,\nonumber\\
&\langle \boldsymbol{\alpha_1}\vert V_{\boldsymbol{\alpha_2}}(z) f_m(z)=0,\quad m=-1,-2,-3,\cdots,\nonumber\\
&\langle \boldsymbol{\alpha_1}\vert V_{\boldsymbol{\alpha_2}}(z) \left(f'_0(z)-w_1\right)=0,\nonumber\\
&\langle \boldsymbol{\alpha_1}\vert V_{\boldsymbol{\alpha_2}}(z) \left(f''_0(z)-w_2\right)=0,
\end{align}
where $w$ is the eigenvalue of $W_0$ on a primary state $W_0\vert \boldsymbol{\alpha}\rangle=w( \boldsymbol{\alpha})\vert \boldsymbol{\alpha}\rangle$. In the following we specialize to $z=1$. However, the Verma module cannot be spanned by the generators $\{e_m,e'_0,e''_0,f_m,f'_0,f''_0\}$ only. We need to introduce an additional generator, $W_{-1}$, in order to construct the full Verma module. A generic state then takes the form
\begin{align}
\vert \Psi\rangle&=
\sum_{\ell_1\geq \cdots\geq1}\sum_{m_1\geq \cdots\geq1}\sum_{p,q,r,s,n}
C\,e_{-\ell_1}e_{-\ell_2}\cdots f_{-m_1}f_{-m_2}\cdots
\left( e'_0\right)^p\left( e''_0\right)^q
\left( f'_0\right)^r\left( f''_0\right)^s
\left( W_{-1}\right)^n\vert \boldsymbol{\alpha}\rangle\nonumber\\
&=\sum_{n=0}^\infty
P^{\Psi}_n(e_{-1},e_{-2},\cdots;f_{-1},f_{-2},\cdots;e'_0,e''_0,f'_0,f''_0)\left( W_{-1}\right)^n \vert \boldsymbol{\alpha}\rangle,
\end{align}
where $P^{\Psi}_n$ is a polynomial. The three-point function then becomes
\begin{align}
 \langle \boldsymbol{\alpha_1}\vert V_{\boldsymbol{\alpha_2}}(1) \vert \Psi\rangle
 =\sum_{n=0}^\infty
P^{\Psi}_n(0,\cdots;0,\cdots;\Delta_1,\Delta_2,w_1,w_2)\,
 \langle \boldsymbol{\alpha_1}\vert V_{\boldsymbol{\alpha_2}}(1)
\left( W_{-1}\right)^n \vert \boldsymbol{\alpha}\rangle.
\end{align}
The infinite number of independent $\langle \boldsymbol{\alpha_1}\vert V_{\boldsymbol{\alpha_2}}(1)
\left( W_{-1}\right)^n \vert \boldsymbol{\alpha}\rangle$ leads to infinitely many fusion rules, which complicates the situation.

Next, we can apply the result to the four-point functions. Let $\{\vert \Psi_Y\rangle\}$ be an orthonormal basis of the Verma module with highest weight state $|\boldsymbol{\alpha} \rangle$.  The four-point function, after inserting the projector $\mathbb{P}_{\boldsymbol{\alpha}}=\sum_Y\vert \Psi_Y\rangle\langle \Psi_Y\vert$  is given by
\begin{multline}
 \langle \boldsymbol{\alpha_1}\vert V_{\boldsymbol{\alpha_2}}(1) \mathbb{P}_{\boldsymbol{\alpha}} V_{\boldsymbol{\alpha_3}}(z) \vert \boldsymbol{\alpha_4}\rangle
 =\sum_{n_1,n_2=0}^\infty
 \langle \boldsymbol{\alpha_1}\vert V_{\boldsymbol{\alpha_2}}(1)
\left( W_{-1}\right)^{n_1} \vert \boldsymbol{\alpha}\rangle
\langle
\boldsymbol{\alpha}\vert
\left( W_{1}\right)^{n_2} \vert V_{\boldsymbol{\alpha_3}}(z) \vert \boldsymbol{\alpha_4}\rangle
\\
\times
\sum_Y
P^{\Psi_Y}_{n_1}(0,\cdots;\Delta_1,\Delta_2,w_1,w_2)
P^{\Psi_Y}_{n_2}(0,\cdots;\Delta_4,\Delta_3,w_4,w_3) \, 
\end{multline}
with fixed the internal momentum $\boldsymbol{\alpha}$.
The product of two polynomials $P$ is the four-point conformal block, and the coefficient, which corresponds to two DOZZ factors\footnote{Here, we have only discussed the chiral part, and in order to compute the full correlation function we need to introduce the anti-chiral part as well.}, are labeled by a pair of positive integers $\{n_{1},n_2\}$
\begin{align}
 \langle \boldsymbol{\alpha_1}\vert V_{\boldsymbol{\alpha_2}}(1,1) V_{\boldsymbol{\alpha_3}}(z,\bar{z}) \vert \boldsymbol{\alpha_4}\rangle
 =
 \int \left[d\boldsymbol{\alpha}\right]
& \sum_{n_1,n_2=0}^\infty
 C_{\textrm{DOZZ}}^{n_1}(\boldsymbol{\alpha_1},\boldsymbol{\alpha_2},\boldsymbol{\alpha})
  C_{\textrm{DOZZ}}^{n_2}(\boldsymbol{\alpha},\boldsymbol{\alpha_3},\boldsymbol{\alpha_4})\nonumber\\
  &\times\big|
\mathcal{F}^{n_1,n_2}(\boldsymbol{\alpha_1},\boldsymbol{\alpha_2};\boldsymbol{\alpha};\boldsymbol{\alpha_3},\boldsymbol{\alpha_4})\big|^2.
\end{align}
We would like to find the analogous object on the gauge theory side.

\subsubsection{Counting parameters from the topological string side}

There is a corresponding situation in our web construction of Sicilian gauge theories. Let us thus use $T_N$ as an illustration. We can easily count the number of possible deformations of the $T_N$ theory using the toric web diagram. The global deformations of the theory, which change the physical parameters of the theory, are associated with the external lines of the web such that
\begin{align}
\#(\textrm{global deformations})&=\#(\textrm{external lines})-3
=3(N-1).
\end{align}
They are the relevant deformations of the fixed point theory associated with the kinetic terms of the gauged Cartan flavor symmetry. In particular, they provide the fugacities of the superconformal index. The breathing modes which correspond to the Coulomb branch direction are associated with the faces of the web, so the counting goes as
\begin{align}
\#(\textrm{local deformations})=\#(\textrm{internal faces})
=\frac{(N-1)(N-2)}{2}.
\end{align}
These deformations turn on the VEVs of the scalar components of the associated local $U(1)$ vector multiplets, and their values are the loop variables $Q_{Fg}$ in the $T_N$ theory. The $T_N$ theory then has $(N-1)(N+4)/2$ parameters in total, and the same counting was given in \cite{Benini:2009gi} based on the 5-brane construction.

The $3(N-1)$ global parameters agree nicely with the three types of external momenta $\boldsymbol{\alpha}_{1,2,3}$. We can thus expect that the fugacities of the global deformations coincide with the Toda momenta. Then, what are the variables associated with the local deformations? The number of the local deformations is $(N-1)(N-2)/2$, and it is equal to the types of Toda three-point functions. We can therefore expect these parameters to specify the different types.

\begin{table}[t]
\begin{center}
  \begin{tabular}{|| m{4.9cm} | m{4.4cm}  | m{4.1cm} ||}
        \hline
    Topological strings (5-web)   & Superconformal Index & $q$-Toda \\   \hline   \hline
    $\mathfrak{q},     \mathfrak{t}$ & $x,y$ & ``Central charge'' $c$
     \\   \hline   \hline
Global deformations $e^{-\beta m}$ \,  (distance between the external legs)     & Flavor fugacities $u_i , x_i , y_i$  & External momenta $\boldsymbol{\alpha}_i$ 
 \\
    \hline
Local deformations $Q_{Fg''}$ for gluing pants (breathing modes for gluing pants)
  & Gauge group fugacities  & Internal  momenta $\boldsymbol{\alpha}$ \\
      \hline  \hline
   Local deformations $\tilde{A}$ or   $Q_{Fg'}$    inside $T_N$    (breathing modes inside $T_N$)  & Residues $(r,s)$  of the contour integral (of $\int d a_{g'} \propto \int d\log Q_{Fg'}$)  & Labels $n$ of  the different three-point functions \\ 
	\hline
  \end{tabular}
\end{center}
  \caption{{\it In this table we summarize all the different parameters/notations that appear when we consider the correspondence between the different objects: partition functions, superconformal indices and Toda correlation functions. The superconformal index is the bridge between the 5-web picture and the 2D $q$-Toda theory.}}
\label{table:conventions}
\end{table}

By expanding this idea, we can generalize the parametrization to the theories obtained by gluing the Triskelion blocks $T_N$. It is useful, while following this argument to also keep an eye on table~\ref{table:conventions} which summarizes the different parameters appearing in this article. In the language of the toric web description, all the parameters originate from the K\"ahler parameters of the Calabi-Yau. After the decomposition as a Sicilian gauge theory, these parameters fall into three categories: global deformations, local deformations, and gauged flavor symmetries. The first are the masses of the matters multiplets, i.e. the fugacities for the flavor symmetry of the Sicilian theory. They are precisely the Toda momenta of the vertex operators. The parameters in the second class specify the choices of the descendant operators $n_{g'}$ in the three-point functions associated with the $T_N$ blocks. The final parameters are the Coulomb branch parameters $a_{g''}$ of the Sicilian theory over which we have to integrate. These parameters are just the internal momenta of the Toda correlation function in a fixed channel. The corresponding (undeformed) Toda correlation functions then take the form
\begin{multline}
\label{eq:Toda-npt}
\langle
V_{\boldsymbol{\alpha}_1}(z_1,\bar{z}_1)\cdots V_{\boldsymbol{\alpha}_n}(z_n,\bar{z}_n)
\rangle_{\textrm{Toda}}
\\=
\prod_{g''} \int [da_{g''}] \left(
\sum_{ \{n_{g'}\} } \prod C_{DOZZ}(m,n_{g'})
\Big|
\mathcal{F}(a,m,\beta,\epsilon_{1,2})\Big|^2 \right),
\end{multline}
where $\mathcal{F}$ is the corresponding Toda conformal block. By applying the Ward identities to the internal lines, where all the descendant operators run, we can reduce the correlation function into a combination of three-point functions, and then a special choice of three-point functions labeled by $\{n_{g'}\}$ appears. This result is illustrated on the right hand side of the above formula. We expect that exactly the same structure holds for the $q$-deformed case.

\subsubsection{Discrete versus continuous parameters}

Going back to the $q$-deformed Toda theory, we find that formula \eqref{eq:Toda-npt} is very similar to the expression \eqref{eq:5dAGTW} of the superconformal index, once we make use of the 5D AGTW identification between the conformal block and the Nekrasov partition function. However, there is an important difference. In the 5D superconformal index we take the integral over all the breathing modes $\log (Q_{Fg})$, but in the Toda correlation function these parameters are split into $n_{g'}$ and $a_{g''}$. Recall that the $n_{g'}$s are the breathing modes of the $T_N$ blocks, and $a_{g''}$ are that of the web-loops generated by the gluing procedure. In contrast to the superconformal index, the $n_{g'}$ parameters take integral values as explained above, and therefore we have to consider a summation over the parameters in the Toda correlation function, not an integral. Nevertheless, we would like to expect that the integral procedure on the superconformal index side corresponds to the summation procedure on the 2D CFT side. We will consider the situation with integral values of the K\"ahler parameters, like $Q_{Fg'}\propto \fq^{N}$ appear automatically because of certain special mechanism. Integral K\"ahler parameters are also very natural from the perspective of the topological string, because the closed topological string comes from the open topological string theory through the geometric transition and in this context a K\"ahler parameter is equal to the number of D-branes times the topological string coupling constant, that is to say the $\Omega$-background. We can expand this idea to the following possible AGTW relation between a partition function on $S^1\times S^4$ and a $q$-deformed Toda correlation function. Let us consider a gauge theory obtained from $T_N$ junctions\footnote{We can generalize to theories obtained from all types of three punctured spheres. We do not know the forms of these three-point functions except for free hypermultiplets $\tilde{T}_N$ and hence skip discussing details on these generalized cases.} whose flavor symmetries are partially gauged and coupled. Its superconformal index after evaluating the residue leads to a certain combination of the three-point functions as the coefficient of the instanton partition function, that is the conformal block
\begin{align}
\mathcal{I}^{\,\textrm{5D}}\propto& \prod_{g}\oint \frac{dQ_{Fg}}{2\pi iQ_{Fg}} \prod_j  \vert Z^{(j)}_{T_{N}}(Q_{Fg'})
\,Z_{\textrm{1-loop}}\vert^2 \vert
{Z}_{\textrm{inst.}}(Q_{Fg''})\vert^2
 \nonumber\\
 =&
  \prod_{g''}\oint \frac{dQ_{Fg''}}{2\pi iQ_{Fg''}}
\left(\sum_{\boldsymbol{n}_j}
  \prod_j C^{\ft,\fq}_{\textrm{DOZZ}}(\boldsymbol{n}_j) \prod C^{\ft,\fq}_{\textrm{DOZZ}}\,
\vert
\mathcal{F}(Q_{Fg''})\vert^2\right)\\
   =&\langle
V_{\boldsymbol{\alpha}_1}(z_1,\bar{z}_1)\cdots V_{\boldsymbol{\alpha}_n}(z_n,\bar{z}_n)
\rangle_{q\textrm{-Toda}},\nonumber
\end{align}
where $\boldsymbol{n}_j=\{{n}_{g'(j)}\}$ labels the choices of combining the three-point functions for the $j$-th $T_N$ block.

The coefficients of the superconformal index are written in terms of the residues, and heuristically we know that the poles are labeled by integers as\footnote{The denominator of the instanton partition function for a given instanton number is the Kac determinant, so the corresponding topological string partition function as a function of $Q_F$ has such poles. Since this pole structure comes from the Cauchy formula computation of the refined topological vertex, we expect this to be the case for a generic partition function.}
\begin{align}
Q_F\propto \ft^{r}\fq^s,\quad r,s \in \frac{1}{2}\mathbb{Z}.
\end{align}
This observation implies that the residue of the absolute square of the topological string partition function is the ``building block'' with discrete labels $(r,s)$ in the superconformal index, which is analogous to that any correlation function can be written in terms of the DOZZ three-point function with discrete label $n_{g'}$. This observation is expected to give a clue to relate the degrees of freedom of the breathing modes in the topological string partition function with the degree label $n_{g'}$ on the 2D CFT side. Furthermore, we might be able to find the transformation from the set of residues to the set of three-point functions. It might be highly non-trivial and complicated, and so we leave it for further studies.

Thus, the 5D uplift of the AGTW relation naturally leads to the expected DOZZ formula for the $q$-deformed Toda three-point functions, and they are automatically derived from the $T_N$ Nekrasov partition function via the structure of the superconformal indices. Once we can compute the closed form of the three-point functions, the DOZZ functions of ordinary Toda CFTs automatically follow from them by using the 4D limit of the gauge theory side. The 5D viewpoint thus provides a powerful approach to investigate an, yet-undiscovered, explicit expression of the Toda three-point functions.

\subsubsection{Example: the $T_3$-$\tilde{T}_3$ system}

\begin{figure}[tbp]
 \begin{center}
  \includegraphics[width=120mm,clip]{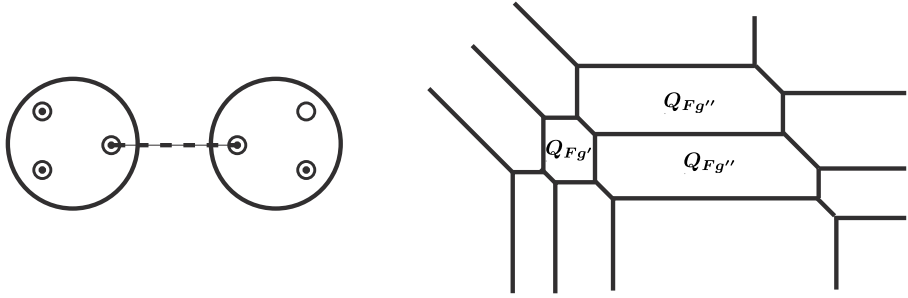}
 \end{center}
 \caption{Sicilian theory obtained by gluing $T_3$ and $\tilde{T}_3$ through an $SU(3)$ vector multiplet.  }
\label{fig:T3T3}
\end{figure}

Figure~\ref{fig:T3T3} illustrates the 5D theory of $T_3$ and nine free hypermultiplets $\tilde{T}_3$ coupled via an $SU(3)$ vector multiplet. In this figure, the parameters (see table~\ref{table:conventions}) $Q_{Fg'}$ and $Q_{Fg''}$ corresponding to the local deformations\footnote{In appendix \ref{app:parametrizationgeneralTN}, these parameters are denoted $\tilde{A}$.} inside $T_N$ and the local deformations for gluing pairs of pants respectively are depicted on the web diagram of the $SU(3)$ quiver. The left part of the figure is the analogue of the Gaiotto curve of the theory, and the 5-brane web is depicted on the right. The breathing mode inside the $T_3$ block is the parameter $Q_{Fg'}$,. The residue integral over this parameter leads to a summation over integers. We expect that this part provides the $A_2$ Toda three-point functions. The three horizontal legs connecting the $T_3$ and $\tilde{T}_3$ webs correspond to the $SU(3)$ vector multiplet. Therefore, the breathing modes for this part are the parameters $Q_{Fg''}$, and they are associated with the internal momenta of the corresponding conformal block.

\section{Conclusions}
\label{sec:conclusions}

In this article, we have studied the 5D $T_N$ theories realized on the 5-brane junctions with two different tools. On one hand, we computed their SW curves for any $N$ such that the $SU(N)^3$ symmetry is manifest. We were also able to reproduce these curves using a certain limit  in the parameter space of the SW curve of the 5D $SU(N)^{N-2}$ linear quiver theory obtained in \cite{Bao:2011rc}. Furthermore, this curves yield the SW curves for the 4D $T_N$ SCFTs. Our results are consistent with those given in \cite{Gaiotto:2009we}. For the simplest non-trivial case $T_3$, we explicitly demonstrated how the $SU(3)^3$ symmetry enhances to $E_6$ and reproduced the curves given in \cite{Eguchi:2002fc,Eguchi:2002nx,Minahan:1997ch}.

On the other hand, we used topological string theory techniques to compute the superconformal index of the $T_3$ junction from the Nekrasov partition function. We first carried out the same computation for the 5D $SU(2)$ $N_f=0,\ldots, 4$ gauge theories, where we found that the contribution with \textit{non-full spin content} has to be removed from the partition function in order to reproduce the results of \cite{Kim:2012gu}. We thus claim that this contribution corresponds to the extra degrees of freedom which appear when we describe (by choosing special blow-up points) the local, generically non-toric, del Pezzo surface using toric geometry. The non-full spin content contributions arise from BPS M2 branes with a non-compact moduli space that are wrapping a two-cycle. In the language of the web diagram, whenever external parallel 5-branes appear, multiplets with an incomplete spin content show up. Such degrees of freedom do not exist in the original 5D $SU(2)$ gauge theory and have to be removed from the topological partition function in order to obtain the Nekrasov partition function of the theory. We therefore conjectured a procedure to remove the extra degrees of freedom for generic toric diagrams, which needs further study in the future. Armed with this idea, we proceed to the $T_3$ case and checked that it gives the same index as in \cite{Kim:2012gu}. We also generalized to the $T_N$ case and provided the formula to the superconformal index from the topological string partition function.

It is a long standing problem to compute the generic 3-point functions in Toda CFTs. We proposed in section \ref{sec:correlationfunctions} a way to obtain, via the 5D AGTW correspondence, the 3-point functions for primary operators of the 2D $W_N$ $q$-deformed Toda field theories from the 5D $T_N$ Nekrasov partition functions, up to an overall factor. Any $n$-point function of the $q$-deformed Toda theory can be obtained by computing the superconformal index of the appropriate 5D gauge theory. Furthermore, the relation between the labels of different $q$-Toda 3-point function and the $T_N$ breathing modes was addressed in section~\ref{sec:correlationfunctions}. From the gauge theory point of view, calculating the index, although technical, is a fully algorithmic procedure. A first step in this direction would be to explicitly compute the 5D superconformal index for the $T_N$ junction and/or the corresponding 4D partition function.

Another important open problem is to perform the summation of the Young diagrams in order to get a product formula for the partition function of the $T_N$ junction, as is possible to do for $T_2$. Once it is done, we expect to be able to take a well defined $q\rightarrow 1$ limit and obtain the 3-point functions of the undeformed Toda CFT. Furthermore, by generalizing the methods of \cite{Fateev:2007ab, Fateev:2008bm}, one should be able to get differential equations, whose solutions should be given by the 3-point functions obtained via our Nekrasov partition functions. Moreover, a bootstrap approach like the one used in \cite{Nieri:2013yra} should provide an alternative way of calculating the 3-point functions.

Finally, 7-branes played a major role in this paper. At the same time, the topological vertex formalism provided a powerful computational tool. It is natural to expect that a topological vertex formalism also should exist in the presence of 7-branes. If we let 5-branes end on 7-branes at finite distance, the partition function might be different from the setup where all the external 5-branes extend to infinity. Related to this problem, it is also important to study how to deal with the \textit{jumping} of the 5-branes described in \cite{Benini:2009gi, Vafa:2012fi}. Deeper understanding of these questions will make it possible to calculate the topological string partition functions for more generic theories including 5D $E_7$ and $E_8$ CFTs.

\section*{Acknowledgments}
We are very grateful to Can Koz{\c c}az for helpful discussions and suggestions, and for providing us with a useful Mathematica code. We would also like to thank Albrecht Klemm, Kazunobu Maruyoshi, Yuji Tachikawa, J\"org Teschner and Dan Xie for insightful comments and discussions. VM, EP and MT are grateful to the Kavli Institute for the Physics and Mathematics of the Universe for their kind hospitality during the finishing stage of this work. FY thanks the Yukawa Institute for Theoretical Physics at Kyoto University, where the discussions during the YITP workshop on ``Field theory and string theory'' (YITP-W-13-12 ) were useful to complete this work. FY also thanks the organizers and the participants of ``Workshop on Geometric Correspondences of Gauge Theories'' held at ICTP for fruitful discussions. EP thanks the Initial Training Network GATIS for support. The research of FY was partially supported by the INFN project TV12 and the grant ``Competition for Young SISSA Scientists''.

\appendix

\section{Special functions}
\label{app:special}

For the reader's convenience, we collect here the definitions of the special functions used in the main text. The most used ones are:
\begin{equation}
\label{eq:mainfunctionsdefinitions}
\begin{split}
\tilde{Z}_{\nu}(\ft,\fq)\defeq& \prod_{i=1}^{\ell(\nu)}\prod_{j=1}^{\nu_i}\left(1-\ft^{\nu^t_j-i+1}\fq^{\nu_i-j}\right)^{-1},\\
\calR_{\lambda\mu}(Q;\ft,\fq)\defeq &\prod_{i,j=1}^{\infty}\left(1-Q \ft^{i-\frac{1}{2}-\lambda_j}\fq^{j-\frac{1}{2}-\mu_i}\right)=\calM(Q\sqrt{\frac{\ft}{\fq}};\ft,\fq)^{-1}\calN_{\lambda^t\mu}(Q\sqrt{\frac{\ft}{\fq}};\ft,\fq),\\
\calM(Q;\ft,\fq)\defeq&\prod_{i,j=1}^{\infty}(1-Q \ft^{i-1}\fq^{j})^{-1},\\
\calN_{\lambda\mu}(Q;\ft,\fq)\defeq &\prod_{i,j=1}^{\infty}\frac{1-Q\ft^{i-1-\lambda_j^t}\fq^{j-\mu_i}}{1-Q\ft^{i-1}\fq^j}\\=&\prod_{(i,j)\in\lambda}(1-Q\ft^{\mu_j^t-i}\fq^{\lambda_i-j+1})\prod_{(i,j)\in\mu}(1-Q\ft^{-\lambda_j^t+i-1}\fq^{-\mu_i+j}).
\end{split}
\end{equation}
We also occasionally use the Macdonald polynomials $P_{\nu}(\textbf{x};t,q)$. Our main reference is appendix B in \cite{Awata:2005fa}. We reproduce here the formulas that we need. We can relate the Macdonald polynomials, at specific values of the vector $\textbf{x}$, to the $\tilde{Z}_{\nu}$ functions as
\begin{equation}
\label{eq:MacdonaldZ}
\ft^{\frac{||\nu^t||^2}{2}}\tilde{Z}_{\nu}(\ft,\fq)=P_{\nu}(\ft^{-\rho};\fq,\ft), \qquad \fq^{\frac{||\nu^t||^2}{2}}\tilde{Z}_{\nu}(\fq,\ft)=P_{\nu}(\fq^{-\rho};\ft,\fq).
\end{equation}
Furthermore, we will need the following two exchange relations
\begin{equation}
\label{eq:Macdonaldexchange}
P_{\nu}(\ft^{\rho};\fq,\ft)=\Big(-\sqrt{\frac{\ft}{\fq}}\Big)^{|\nu|}\fq^{\frac{||\nu||^2}{2}}\ft^{-\frac{||\nu^t||^2}{2}}P_{\nu}(\ft^{-\rho};\fq,\ft), \qquad P_{\nu}(\textbf{x};\fq,\ft)=P_{\nu}(\textbf{x};\fq^{-1},\ft^{-1}),
\end{equation}
as well as the ``extension" identity
\begin{equation}
\label{eq:Macdonaldextension}
P_{\nu}(\ft^{-\rho},Q\ft^{\rho};\fq,\ft)=P_{\nu}(\ft^{-\rho};\fq,\ft)\prod_{(i,j)\in \nu}\big(1-Q\ft^{1-i}\fq^{j-1}\big).
\end{equation}
Lastly, there is a variant of the Cauchy identity for the Macdonald polynomials 
\begin{equation}
\label{eq:MacdonaldCauchy}
\sum_{\nu}Q^{|\nu|}P_{\nu}(\textbf{x};\fq,\ft)P_{\nu^t}(\textbf{y};\ft,\fq)=\prod_{i,j=1}^{\infty}(1+Q x_iy_j).
\end{equation}

\section{Comparison with the $E_6$ curve of Minahan-Nemeschanski type}
\label{sec:Minahan-Nemeshanski}

In this section, we check that our curve investigated in section \ref{sec:seibergwitten} is
equivalent to the curve which was known previously \cite{Minahan:1997ch,Eguchi:2002fc,Eguchi:2002nx}.

On the one hand, our curve \eqref{eq:su6xsu2_manifest} is rewritten in terms of the quartic polynomial
\begin{align}
\label{eq:quartic}
y^2 = a w'{}^4 + b w'{}^3 + c w'{}^2 + d w' + e
\end{align}
 by shifting and rescaling the coordinate $t'$ as
\begin{align}
y \equiv 2 t' + \left[
- 2 {w^{\prime}}^3
+ \left( \sum_{i=1}^6 \tilde{m}_i^{\prime} \right) {w^{\prime}}^2
+ U_1^{\prime} w^{\prime}
+ \left( \tilde{M}^{\frac{1}{2}}+\tilde{M}^{-\frac{1}{2}}\right)
\right]
\end{align}
so that no linear term appears. The coefficients in \eqref{eq:quartic} are given by
\begin{align}
&a \equiv  - 4U_1^{\prime} + (\chi^{SU(6)}_{6})^2 - 4 \chi^{SU(6)}_{15},&
&b \equiv  2 U_1^{\prime} \chi_6^{SU(6)} - 4 \chi_2^{SU(2)} + 4 \chi_{20}^{SU(6)},&
\nonumber\\
&c \equiv  U_1^{\prime}{}^2 + 2\chi^{SU(6)}_{6} \chi^{SU(2)}_2 - 4 \chi^{SU(6)}_{\overline{15}},&
&d \equiv 2 U_1^{\prime} \chi^{SU(2)}_2 + 4 \chi^{SU(6)}_{\overline{6}},&
\\
&e \equiv  (\chi^{SU(2)}_{2})^2 - 4,& &&\nonumber
\end{align}
where we have introduced the characters of $SU(6)$ and $SU(2)$ as
\begin{align}
\chi^{SU(6)}_{6} = \sum_{i=1}^6 \tilde{m}'_i,
\qquad
\chi^{SU(6)}_{15} = \sum_{1\le i < j \le 6}  \tilde{m}'_i \tilde{m}'_j,
\qquad
\chi^{SU(6)}_{20} = \sum_{1\le i < j < k \le 6}  \tilde{m}'_i \tilde{m}'_j \tilde{m}'_k,
\cr
\chi^{SU(6)}_{\overline{15}} = \sum_{1\le i < j \le 6}  \tilde{m}'_i{}^{-1} \tilde{m}'_j{}^{-1},
\qquad
\chi^{SU(6)}_{\overline{6}} = \sum_{i=1}^6 \tilde{m}'_i{}^{-1},
\qquad
\chi^{SU(2)}_{2} = M^{\frac{1}{2}} + M^{-\frac{1}{2}}.
\label{ch2-6}
\end{align}
From the Seiberg-Witten one-form
\begin{equation}
\lambda_{\rm SW} \propto \log t' d \log w',
\end{equation}
we can explicitly show that
the holomorphic one-form of this curve is given by
\begin{align}
\label{eq:holo1}
\omega
\propto \left. \frac{\partial}{\partial U_1'} \lambda_{\rm SW} \right|_{w'}
\propto \frac{dw'}{y}.
\end{align}

On the other hand, the 5D $E_6$ curve has been
studied in \cite{Minahan:1997ch,Eguchi:2002fc,Eguchi:2002nx},
which provides the 5D uplift of the Minahan-Nemeschanski theory \cite{Minahan:1996cj, Minahan:1996fg}.
It is given in terms of a cubic polynomial
\begin{align}
y^2
= & A x^3 + B x^2 + C x + D ,
\label{eq:ES}
\end{align}
where
\begin{align}
&
A \equiv 4,
\qquad
B \equiv -u^2+4\chi^{E_6}_{27},
\qquad
C \equiv
(2 \chi_{78} - 12 )u +
\left( 4 \chi_{\overline{351}}^{E_6} - 4 \chi^{E_6}_{\overline{27}} \right),
\cr
&D \equiv
4 u^3
+ 4 \chi^{E_6}_{\overline{27}} u^2
+ (4 \chi^{E_6}_{351} - 4 \chi^{E_6}_{27} ) u
+  4 \chi^{E_6}_{2925} - (\chi^{E_6}_{78})^2
\end{align}
with the holomorphic one-form being
\begin{align}
\label{eq:holo2}
\omega = \frac{dx}{y}.
\end{align}

Since \eqref{eq:ES} is written in terms of the characters of $E_6$,
we need to rewrite them in terms of the characters of $SU(6)$ and $SU(2)$
to be able to compare with our curve \eqref{eq:quartic}. The decompositions of the relevant representations of $E_6$
into $SU(6) \times SU(2)$ representations are the following:
\begin{align}
27 & = (6,2) + (\overline{15},1),
\nonumber\\
2925 &
= (1,3) + (20,2) + (20,4) + (35,1) + (35,3) + (70,2) + (\overline{70},2)
\nonumber\\
& \qquad\qquad\qquad\quad
 + (175,1) + (189,3)
+ (280,1) + (\overline{280},1) + (540,2),
\\
351 &
= (6,2) + (\overline{15},3) + (\overline{21},1) + (84,2) + (\overline{105},1),\nonumber\\
78 & = (1,3) + (20,2) + (35,1).\nonumber
\end{align}
Furthermore, by using the relations derived from the decompositions of the product representations of the fundamental representation of $SU(6)$ into irreducible representation, we can rewrite the characters of any irreducible representation in terms of the characters of the fundamental representations:
\begin{align}
&21 = 6  \otimes 6 - 15,&
&35= 6 \otimes \bar{6} - 1,&\nonumber\\
&70 = 6 \otimes 15 - 20,&
&84 = \bar{6} \otimes 15 - 6,&\nonumber\\
&105 = \bar{6} \otimes 20 - 15,&
&175= 20 \otimes 20 - 15 \otimes \overline{15}&\\
&189 = 15 \otimes \overline{15} - 1 - 35,&
&280 =
6 \otimes 6 \otimes \overline{15}
- 15 \otimes \overline{15}
- 6 \otimes \overline{6}
+ 1,&\nonumber\\
&540 = 6 \otimes \overline{6} \otimes 20
- 6 \otimes 15 - \overline{6} \otimes \overline{15}.& &&\nonumber
\end{align}
The above expressions should only be understood on the level of the characters. We have also suppressed the corresponding relations for the dual representations. For completeness, we also give the decompositions for $SU(2)$:
\begin{align}
3 = 2 \otimes 2 - 1,
\qquad
4 = 2 \otimes 2 \otimes 2 - 2(2).
\end{align}
Combining these sets of identities, we find that the characters of $E_6$
can be rewritten in terms of the characters of the fundamental representations of $SU(6)$ and $SU(2)$.

Taking into account that the holomorphic one-forms for both curves are given in the standard form (see \eqref{eq:holo1} and \eqref{eq:holo2}), we find it straightforward to compare their modular function called ``$j$-invariant'' in order to show the equivalence of the curves \eqref{eq:quartic} and \eqref{eq:ES}. Agreement of the $j$-invariant indicates that the periods of the curves are identical up to $SL(2,\mathbb{Z})$ modular transformations,  which means that the curves coincide with each other.

For the elliptic curve in the standard Weierstrass form
\begin{align}
y^2 = 4 z^3 - g_2 z - g_3,
\end{align}
the $j$-invariant is defined as
\begin{align}
J(\tau) = \frac{g_2{}^3}{g_2{}^3 - 27 g_3{}^2}.
\end{align}
As for elliptic curves with more generic expressions, the $j$-invariant is derived by first going to the standard Weierstrass form by proper coordinate transformation and then expressing $g_2$ and $g_3$ in terms of the coefficients of the original polynomial.
For the quartic polynomial \eqref{eq:quartic}, the denominator of the $j$-invariant is given by its discriminant multiplied by $16 a^{-6}$, while for the cubic polynomial \eqref{eq:ES}, it is given by its discriminant\footnote{The discriminant is given by
\begin{align}
\Delta
=&
-128 a^2 c^2 e^2 + 256 a^3 e^3 - 4 b^3 d^3
+ 16 a c^4 e - 4 a c^3 d^2 - 6 a b^2 d^2 e
+144 a b^2 c e^2 + 144 a^2 c d^2 e
\cr
& + 18 a b c d^3 + b^2 c^2 d^2 - 4 b^2 c^3 e - 192 a^2 b d e^2
-80 a b c^2 d e + 18 b^3 c d e - 27 b^4 e^2 - 27 a^2 d^4
\end{align}
for the quadratic polynomial \eqref{eq:quartic} and given by
\begin{align}
\Delta
= -27 A^2 D^2 + 18 A B C D + B^2 C^2 - 4 B^3 D - 4 A C^3
\end{align}
for the cubic polynomial \eqref{eq:ES}.
} multiplied by $16A^{-4}$.

The numerator is given for the quartic polynomial \eqref{eq:quartic} as
\begin{align}
g_2 = \frac{4}{3 a^2} (c^2 - 3 bd + 12 ae)
\end{align}
while for the cubic case \eqref{eq:ES} as
\begin{align}
g_2 = \frac{4}{3 A^2} (B^2 - 3 AC).
\end{align}

By explicit calculation, we have checked that
the $j$-invariants for the both curve coincide,
under the identification of the Coulomb moduli parameters
\begin{align}
U_1' = u, 
\end{align}
although the explicit expression is too long and complicated to write down here.

\section{$E_6$ characters and parametrization}
\label{app:E6}

The purpose of this appendix is two-fold. First, we would like to connect the parametrization of section \ref{subsec:E6index} to the one appearing in section \ref{sec:seibergwitten}. Second, we will provide some additional details concerning the embedding of the group $SU(3)^3$ in $E_6$ and the information this provides us for formula \eqref{eq:index5flavor}. 

\begin{figure}[ht]
 \centering
  \includegraphics[height=6cm]{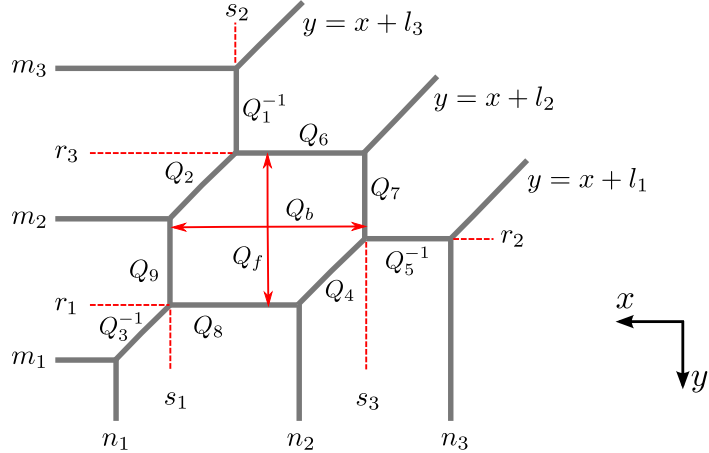}
  \caption{\it This figure illustrates the parametrization of the K\"ahler moduli as functions of the toric geometry. The equations $y=x+l_i$ come from the conditions $\tilde{W}=-\tilde{L}_i\tilde{T}$..}
\label{fig:e6distance}
\end{figure}

Let us begin by relating the parametrizations. In order to do so, we need to connect the K\"ahler moduli $Q_B,Q_F,Q_1,\ldots, Q_5$ to the distances appearing in the toric diagram using \eqref{eq:Qlength}. For this purpose, we draw the $T_3$ diagram again and parametrize the positions (in the unit of $2 \pi \alpha'$)
of the relevant branes as depicted in figure~\ref{fig:e6distance}.  In order for the toric geometry to have the proper angles, we need to have the following constraints
\begin{equation}
\label{eq:e6distance1}
m_2-r_3=s_1-s_2,\qquad m_1-r_1=n_1-s_1,\qquad r_1-r_2=n_2-s_3.
\end{equation}
Furthermore, from the intersection points of the exterior diagonal lines, we read
\begin{equation}
\label{eq:e6distance2}
m_3=s_2+l_3, \qquad r_3=s_3+l_2,\qquad r_2=n_3+l_1.
\end{equation}
We would now like to solve for the $r_i$ and $s_i$ in terms of the exterior parameters $m_i$, $n_i$ and $l_i$, for which \eqref{eq:const_LMN} gives us
\begin{equation}
\sum_{i=1}^3m_i=\sum_{i=1}^3n_i=\sum_{i=1}^3l_i=0.
\end{equation}
The system of equations \eqref{eq:e6distance1} and \eqref{eq:e6distance2} has rank 5, so that we need to use an additional equation to determine the size of the inner cycle. We dispose of a certain amount of freedom in parametrizing the size, but in order to get parameters with good behavior with respect to the $\sigma$ and $\tau$ transformations of \eqref{ST_trans}, it turns out to be useful to introduce the parameter $a$ by the relation
\begin{equation}
r_1-r_3=2a-n_1+l_3. 
\end{equation}
Then, by equation \eqref{eq:Qlength}, we find 
\begin{equation}
\label{eq:twoparam1}
Q_F=e^{- \beta (r_1-r_3) }= \tilde{A}^2 \tilde{N}_1^{-1}\tilde{L}_3, \qquad 
Q_B=e^{- \beta (s_1-s_3) }= \tilde{A}^2 \tilde{M}_1^{-1}\tilde{L}_1^{-1},
\end{equation}
and similarly
\begin{equation}
\label{eq:twoparam2}
Q_1= \tilde{A}\tilde{M}_3\tilde{L}_3, \  
Q_2= \tilde{A}\tilde{M}_2\tilde{L}_3,\  
Q_3= \tilde{A}\tilde{M}_1^{-1}\tilde{N}_1^{-1},\   
Q_4= \tilde{A}\tilde{N}_2\tilde{L}_1^{-1},\  
Q_5= \tilde{A}\tilde{N}_3\tilde{L}_1^{-1}.
\end{equation}
Furthermore, we find the obvious relations $Q_6=Q_BQ_2^{-1}$, $Q_7=Q_FQ_4^{-1}$, $Q_8=Q_BQ_4^{-1}$, $Q_9=Q_FQ_2^{-1}$. In the previous equations, we have set $ \tilde{A}=e^{- \beta a}$.

The 78-dimensional adjoint representation of $E_6$ 
is decomposed into the representation of its subgroup 
$SU(3)_1 \times SU(3)_2 \times SU(3)_3$ as
\begin{equation}
78 = (\overline{3},3,3)+(3,\overline{3},\overline{3})+(8,1,1)+(1,8,1)+(1,1,8).
\label{decomp}
\end{equation}
We note that in the above decomposition, the $SU(3)_1$ appears differently from the other two. The problem is to find out, which parameters among $\tilde{M}$, $\tilde{N}$ or $\tilde{L}$ correspond to  $SU(3)_1$. This can be seen from the discrete symmetry \eqref{ST_trans}, from which we read that in order for the $E_6$ character to be invariant, we must relate $SU(3)_1$ to the $\tilde{M}_i$. Therefore, the characters of the fundamental representation of the different $SU(3)$ are written as
\begin{equation}
\chi^{SU(3)_1}_{3} =\sum_{i=1}^3\tilde{M}_i,\qquad \chi^{SU(3)_2}_{3} =\sum_{i=1}^3\tilde{N}_i,\qquad \chi^{SU(3)_3}_{3} =\sum_{i=1}^3\tilde{L}_i.
\end{equation}
The characters of the dual representations $\overline{3}$ are obtained by sending the parameters $\tilde{M}$, $\tilde{N}$ and $\tilde{L}$ to their inverses and the characters of the adjoints are of course given by $\chi^{SU(3)}_{8} =  \chi^{SU(3)}_{3} \chi^{SU(3)}_{\overline{3}} - 1$. In the index formula \eqref{eq:index5flavor},  the ${\bf 2430}$-dimensional representation of $E_6$ 
labeled by the Dynkin index $[0,0,0,0,0,2]$
(the adjoint is labeled by $[0,0,0,0,0,1]$) appears. It decomposes as follows:
\begin{align}
\label{decomp2}
2430 =\, &(1,1,1) + (\overline{3},3,3)+(3,\overline{3},\overline{3})+(8,1,1)+(1,8,1)+(1,1,8)+ 
\nonumber \\
&+ (8,8,1) + (8,1,8) + (1,8,8)+(8,8,8) + (27,1,1) + (1,27,1) + (1,1,27)\\
&+ (6,3,3) + (\overline{3},3,\overline{6}) + (\overline{3},\overline{6},3) + (3,6,\overline{3})
+ (3,\overline{3},6) + (\overline{6},\overline{3},\overline{3})+ (6,\overline{6},\overline{6}) + (\overline{6},6,6)
\nonumber \\
&+ (\overline{15},3,3) + (\overline{3},15,3) + (\overline{3},3,15) 
+ (15,\overline{3},\overline{3}) + (3,\overline{15},\overline{3}) + (3,\overline{3},\overline{15}).\nonumber
\end{align} 
By using the following formulas
\begin{equation}
6 = 3 \otimes 3 - \overline{3},\quad 
15 
  = 6 \otimes \overline{3} - 3,\quad 
27 
  = 6 \otimes \overline{6} - 8 - 1,
\end{equation}
where we suppressed the symbol ``$\chi^{SU(3)_i}$'' $(i=1,2,3)$ to denote the character,
we can express the character of the 2430-dimensional representation of $E_6$ entirely in terms of the parameters $\tilde{M}$, $\tilde{N}$ and $\tilde{L}$. Plugging the parametrizations \eqref{eq:twoparam1} and \eqref{eq:twoparam2} into the expression for the normalized topological partition function \eqref{eq:renormalizede6partfunction} and computing numerically the index using generic values for the parameters, we have been able to check our final expression \eqref{eq:index5flavor} for the $E_6$ index.

\section{Parametrization for $T_N$}
\label{app:parametrizationgeneralTN}

\begin{figure}[htbp]
  \centering
\includegraphics[height=12cm]{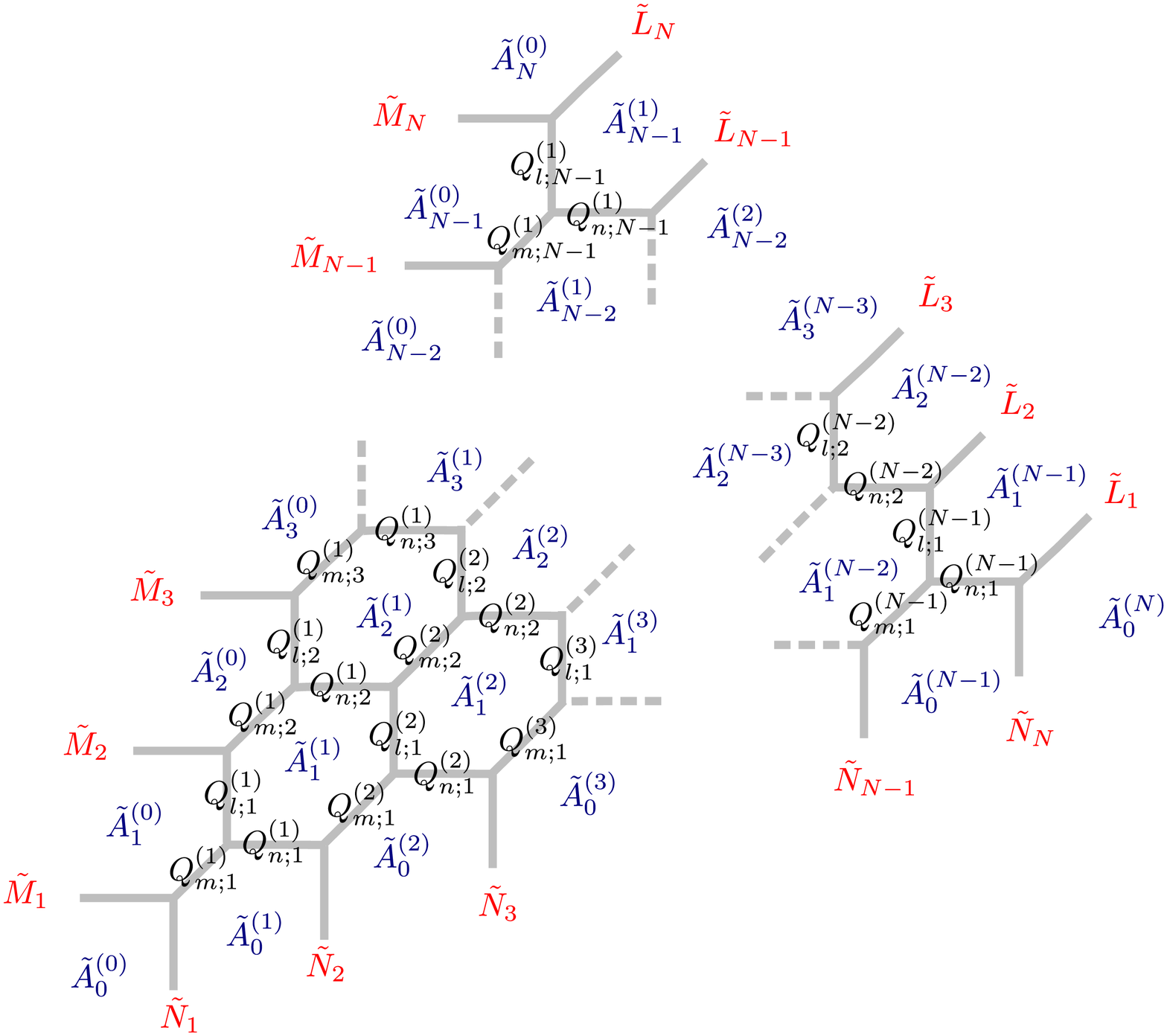}
  \caption{Parametrization for $T_N$. We denote the K\"ahler moduli parameters corresponding to the horizontal lines as $Q_{n;i}^{(j)}$, to the vertical lines as $Q_{l;i}^{(j)}$, and to tilted lines as $Q_{m;i}^{(j)}$. We denote the breathing modes as $\tilde{A}^{(j)}_i$. The index $j$ labels the strips in which the diagram can be decomposed.}
  \label{Brane-gen}
\end{figure}

The purpose of this appendix is to relate the K\"ahler parameters $Q_m$, $Q_n$ and $Q_l$ for the general junction $T_N$ of section \ref{subsec:curvegeneralN} with the parametrization of the SW curves of section \ref{sec:TNpartitionfunction}.  The situation is depicted in Figure \ref{Brane-gen}. The K\"ahler moduli parameters of the 2-cycles are denoted as $Q_{n;i}^{(j)}$, $Q_{l;i}^{(j)}$, $Q_{m;i}^{(j)}$ ($i,j \ge 1$, $i+j \le N$) and the breathing modes corresponding to the 4-cycles are denoted as $\tilde{A}_{i}^{(j)}$ ($i,j \ge 1$, $i+j \le N-1$). We interpret the breathing modes as Coulomb moduli parameters. For later convenience, we also introduce parameters
$\tilde{A}_{0}^{(j)}$, $\tilde{A}_{i}^{(0)}$, and $\tilde{A}_{i}^{(N-i)}$ corresponding to the non-compact 4-cycles, which should be determined once we fix the parameters of the SW curve $\tilde{M}_n$, $\tilde{N}_n$, and $\tilde{L}_n$.

In figure~\ref{Brane-gen}, we see that each two cycle is
shared by two 4-cycles and attached to two more 4-cycles.
We propose that the K\"ahler moduli parameters for the 2-cycles are determined only in terms of the breathing modes of these four neighboring 4-cycles as
\begin{align}
\label{eq:PQR}
Q_{n;i}^{(j)}
 = \frac{\tilde{A}_{i}^{(j)} \tilde{A}_{i-1}^{(j)}}%
    {\tilde{A}_{i}^{(j-1)} \tilde{A}_{i-1}^{(j+1)}},
\qquad
Q_{l;i}^{(j)}
 = \frac{\tilde{A}_{i}^{(j)} \tilde{A}_{i}^{(j-1)}}%
    {\tilde{A}_{i-1}^{(j)} \tilde{A}_{i+1}^{(j-1)}},
\qquad
Q_{m;i}^{(j)}
 = \frac{\tilde{A}_{i}^{(j-1)} \tilde{A}_{i-1}^{(j)}}%
    {\tilde{A}_{i}^{(j)} \tilde{A}_{i-1}^{(j-1)}}.
\end{align}
\begin{figure}[htbp]
  \centering
\includegraphics[height=3.5cm]{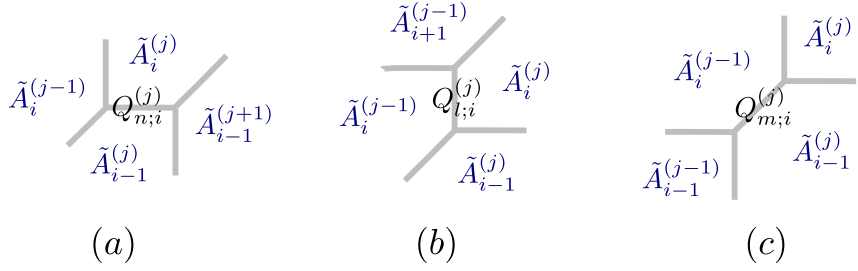}
  \caption{The pictures (a), (b) and (c) illustrate the relation between the Coulomb moduli and the K\"ahler parameters for the horizontal, vertical and diagonal edges respectively.}
  \label{fig:PQR}
\end{figure}
Here, we see as in figure \ref{fig:PQR} that the breathing modes of the two 4-cycle which share the considered two cycle appear in the numerator while for other come in the denominator. We further assume that the relations \eqref{eq:PQR} are valid also for $m=0$, for $n=0$ and for $n+m=N$. Furthermore, we explicitly check that for each "hexagon" in the diagram, the following expected identities are satisfied:
\begin{equation}
Q_{l;i}^{(j)} Q_{m;i+1}^{(j)} = Q_{m;i}^{(j+1)} Q_{l;i}^{(j+1)},
\qquad
Q_{n;i}^{(j)} Q_{m;i}^{(j+1)} = Q_{m;i+1}^{(j)} Q_{n;i+1}^{(j)} .
\end{equation}
At the border of the diagram, we can relate the K\"ahler to the positions of the branes as
\begin{align}
Q_{m;i}^{(1)} Q_{l;i}^{(1)} = \frac{\tilde{M}_i}{\tilde{M}_{i+1}},
\qquad
Q_{m;1}^{(i)} Q_{n;1}^{(i)} = \frac{\tilde{N}_j}{\tilde{N}_{j+1}},
\qquad
Q_{n;i}^{(N-i)} Q_{l;i}^{(N-i)} = \frac{\tilde{L}_i}{\tilde{L}_{i+1}}.
\end{align}
In order to fully determine the relations between the SW curve and the K\"ahler parameters, we assume that the breathing modes of the non-compact cycles are given by 
\begin{align}
\tilde{A}_{i}^{(0)} = \prod_{k=1}^i \tilde{M}_k,
\qquad
\tilde{A}_{0}^{(j)} = \prod_{k=1}^j \tilde{N}_k,
\qquad
\tilde{A}_{i}^{(N-i)}  = \prod_{k=1}^i \tilde{L}_k.
\label{asum_A}
\end{align}
Now, we can see that the $T_N$ generalization of the discrete symmetry $\sigma$ in equation \eqref{ST_trans} acts on the Coulomb moduli on the border of the diagram as
\begin{align}
\tilde{A}_{i}^{(0)} \leftrightarrow \tilde{A}_{0}^{(i)},
\qquad
\tilde{A}_{i}^{(N-i)} \leftrightarrow \tilde{A}_{N-i}^{(i)}.
\end{align}
If we further assume that it acts on the Coulomb moduli parameters in the interior as
\begin{align}
\tilde{A}_{i}^{(j)} \leftrightarrow \tilde{A}_{j}^{(i)},
\end{align}
we find by \eqref{eq:PQR} that it action on the K\"ahler moduli parameters is
\begin{align}
Q_{n;i}^{(j)} \leftrightarrow Q_{l;j}^{(i)},
\qquad
Q_{m;i}^{(j)} \leftrightarrow Q_{m;j}^{(i)}.
\label{sym_PQR1}
\end{align}
Similarly, the generalization of the discrete symmetry $\tau$ of equation \eqref{ST_trans}
acts as follows on the moduli on the border of the diagram
\begin{align}
\tilde{A}_{i}^{(0)} \leftrightarrow \tilde{A}_{i}^{(N-i)},
\qquad
\tilde{A}_{0}^{(j)} \leftrightarrow \tilde{A}_{0}^{(N-j)}.
\end{align}
Again, assuming that it acts on the Coulomb moduli parameters of the interior as
\begin{align}
\tilde{A}_{i}^{(j)} \leftrightarrow \tilde{A}_{i}^{(N-i-j)},
\end{align}
we find that its action on the K\"ahler moduli parameters is
\begin{align}
Q_{n;i}^{(j)} \leftrightarrow Q_{m;i}^{(N-i-j+1)},
\qquad
Q_{l;i}^{(j)} \leftrightarrow Q_{l;i}^{(N-i-j+1)}.
\label{sym_PQR2}
\end{align}
The transformations \eqref{sym_PQR1} and \eqref{sym_PQR2}
 are natural from the point of view of the toric diagram. They further imply that the original assumption of \eqref{asum_A} was natural as well. We can now express the K\"ahler parameters only as functions of the independent elements of the toric diagram. We obtain the explicit formulas
\begin{align}
& Q_{n;1}^{(1)} = \tilde{A}_{1}^{(1)}
               (\tilde{M}_{1})^{-1} (\tilde{N}_{2})^{-1},& &&
\nonumber\\ 
& Q_{n;1}^{(N-1)}
= (\tilde{A}_{1}^{(N-2)})^{-1}
   \tilde{L}_{1} (\tilde{N}_{N})^{-1},& &&
\nonumber\\ 
& Q_{n;N-1}^{(1)}
= \tilde{A}_{N-2}^{(1)}
   \tilde{M}_{N} \tilde{L}_{N-1},& &&
\nonumber\\ 
& Q_{n;1}^{(j)}
=  \tilde{A}_{1}^{(j)} (\tilde{A}_{1}^{(j-1)})^{-1}
    (\tilde{N}_{j+1})^{-1},&
   &(2 \le j \le N-2)&
\\ 
& Q_{n;i}^{(1)}
= \tilde{A}_{i}^{(1)} \tilde{A}_{i-1}^{(1)}
    (\tilde{A}_{i-1}^{(2)})^{-1} \prod_{k=1}^i (\tilde{M}_k)^{-1},&
    &(2 \le i \le N-2)&
\nonumber\\ 
& Q_{n;i}^{(N-i)}
= \tilde{A}_{i-1}^{(N-i)}
(\tilde{A}_{i}^{(N-i-1)})^{-1} \tilde{L}_{i},&
    &(2 \le i \le N-2)& 
\nonumber\\ 
& Q_{n;i}^{(j)}
 = \tilde{A}_{i}^{(j)} \tilde{A}_{i-1}^{(j)}
    (\tilde{A}_{i}^{(j-1)})^{-1} (\tilde{A}_{i-1}^{(j+1)})^{-1},& 
& (i,j \ge 2, \quad i+j \le N-1)& \nonumber
\end{align}
for the K\"ahler parameters of the horizontal lines,
\begin{align}
& Q_{l;1}^{(1)} = \tilde{A}_{1}^{(1)} (\tilde{M}_{2})^{-1} (\tilde{N}_{1})^{-1},& &&
\nonumber\\
& Q_{l;1}^{(N-1)}
= \tilde{A}_{1}^{(N-2)} \tilde{N}_{N}  (\tilde{L}_{2})^{-1},& &&
\nonumber\\
& Q_{l;N-1}^{(1)}
= (\tilde{A}_{N-2}^{(1)})^{-1}
   (\tilde{M}_{N})^{-1} (\tilde{L}_{N})^{-1},& &&
\nonumber\\
& Q_{l;1}^{(j)}
= \tilde{A}_{1}^{(j)} \tilde{A}_{1}^{(j-1)} (\tilde{A}_{2}^{(j-1)})^{-1}
    \prod_{k=1}^j (\tilde{N}_{k})^{-1},& 
    &(2 \le j \le N-2)&
\\
& Q_{l;i}^{(1)}
= \tilde{A}_{i}^{(1)} (\tilde{A}_{i-1}^{(1)})^{-1}
   (\tilde{M}_{i+1})^{-1},&
    &(2 \le m \le N-2)&
\nonumber\\
& Q_{l;i}^{(N-i)}
= \tilde{A}_{i}^{(N-i-1)} (\tilde{A}_{i-1}^{(N-i)})^{-1}
  (\tilde{L}_{i+1})^{-1},& &(2 \le i \le N-2)&
\nonumber\\
& Q_{l;i}^{(j)}
 = \tilde{A}_{i}^{(j)} \tilde{A}_{i}^{(j-1)}
    (\tilde{A}_{i-1}^{(j)})^{-1} (\tilde{A}_{i+1}^{(j-1)})^{-1}&
    &(i,j \ge 2, \quad m+n \le N-1)&\nonumber
\end{align}
for the parameters of the vertical lines and
\begin{align}
& Q_{m;1}^{(1)}
= (\tilde{A}_{1}^{(1)})^{-1}\tilde{M}_{1} \tilde{N}_{1},& &&
\nonumber\\
& Q_{m;1}^{(N-1)}
= \tilde{A}_{1}^{(N-2)}(\tilde{L}_{1})^{-1} \tilde{N}_{N-1}, & &&
\nonumber\\
& Q_{m;N-1}^{(1)}
= \tilde{A}_{N-2}^{(1)}
     \tilde{M}_{N-1} \tilde{L}_{N},& &&
\nonumber\\
& Q_{m;1}^{(j)}
= \tilde{A}_{1}^{(j-1)} (\tilde{A}_{1}^{(j)})^{-1}
     \tilde{N}_{j},& 
    &(2 \le j \le N-2)&
\\
& Q_{m;i}^{(1)}
= \tilde{A}_{i-1}^{(1)} (\tilde{A}_{i}^{(1)})^{-1}
     \tilde{M}_{i},&  &(2 \le m \le N-2)&
\nonumber\\
& Q_{m;i}^{(N-i)}
= \tilde{A}_{i}^{(N-i-1)} \tilde{A}_{i-1}^{(N-i)}
    (\tilde{A}_{i-1}^{(N-i-1)})^{-1}
    \prod_{k=1}^i (\tilde{L}_{k})^{-1} ,&
    &(2 \le m \le N-2)&
\nonumber\\
& Q_{m;i}^{(j)}
 = \tilde{A}_{i}^{(j-1)} \tilde{A}_{i-1}^{(j)}
    (\tilde{A}_{i}^{(j)})^{-1} (\tilde{A}_{i-1}^{(j-1)})^{-1},&
&(m,n \ge 2, \quad m+n \le N-1)&\nonumber
\end{align}
for the parameters of the diagonal lines.
This reproduce the parametrization for $N=3$.



\providecommand{\href}[2]{#2}\begingroup\raggedright\endgroup

\end{document}